\documentclass[%
 reprint, 
 nofootinbib,
 amsmath,amssymb,
 aps,
]{revtex4-2}

\usepackage{graphicx}
\usepackage{dcolumn}
\usepackage{bm}
\usepackage{hyperref}
\usepackage{mhchem}
\usepackage{siunitx}

\usepackage{float}
\usepackage{booktabs}
\usepackage{natbib}
\usepackage{subfigure}

\defcitealias{LopezHonorez2016}{LH16}

\begin{document}

\preprint{APS/123-QED}

\title{The Effects of Dark Matter Annihilation and Dark Matter-Baryon Velocity Offsets at Cosmic Dawn}

\author{Liqiang Hou}
 \email{lhou2@ncsu.edu}
 \affiliation{
 North Carolina State University
}%
\affiliation{
 Perimeter Institute of Theoretical Physics
}%
\affiliation{
 University of Waterloo
}%

\author{Katherine J. Mack}%
 \email{kmack@perimeterinstitute.ca}
\affiliation{%
 Perimeter Institute of Theoretical Physics
}%
\affiliation{
 University of Waterloo
}%
\affiliation{
 York University
}%

\date{\today}
\begin{abstract}
Dark matter annihilation has the potential to leave an imprint on the properties of the first luminous structures at Cosmic Dawn as well as the overall evolution of the intergalactic medium (IGM).
In this work, we employ a semi-analytic method to model dark matter annihilation during Cosmic Dawn (approximately redshift $z=20$ to $40$), examining potential modifications to IGM evolution as well as gas collapse, cooling, and star formation in mini-halos. Our analysis takes into account the effects of dark matter-baryon velocity offsets, utilizing the public \texttt{21cmvFAST} code, and producing predictions for the 21cm global signal.
The results from our simplified model suggest that dark matter annihilation can suppress the gas fraction in small halos and alter the molecular cooling process, while the impact on star formation might be positive or negative depending on parameters of the dark matter model as well as the redshift and assumptions about velocity offsets. This underscores the need for more comprehensive simulations of the effects of exotic energy injection at Cosmic Dawn as observational probes are providing us new insights into the process of reionization and the formation of first stars and galaxies.

\end{abstract}

\maketitle


\section{\label{sec:intro}Introduction}

Cosmic dawn marks the end of the universe's Dark Ages, following recombination and preceding the formation of the first galaxies and stars. 
While it is challenging to directly observe, this epoch is a promising target for investigations of the impact of Beyond the Standard Model physics such as dark matter (DM) particle interactions. In particular, dark matter annihilation, which can inject energy into primordial gas prior to and during the epoch of the formation of the first stars, has the potential to substantially alter the evolution of early galaxies and the intergalactic medium (IGM) during Cosmic Dawn.

In the local universe, researchers have investigated various potential signals of dark matter annihilation, including the excess gamma ray emission from the direction of the Galactic center~\cite{Hooper2013, collaboration2015, daylan2016characterization}, but uncertainties in astrophysical contributions make the interpretation of these excesses challenging. The high-redshift universe presents an alternative target, trading the complexity of disentangling a signal from mature astrophysical sources for new challenges in observation and interpretation.

Previous studies~\cite{Padmanabhan2005, Galli2013} have found that dark matter annihilation can alter the recombination history, and imprint an observable signal on the anisotropy of the cosmic microwave background (CMB). Furthermore, the injection of the energy from dark matter annihilation into the primordial gas could imprint distinctive features on the cosmological 21cm signal from neutral hydrogen during cosmic dawn~\cite{Ghara2015, Valds2012, Evoli2014}. 
The 21-cm signal corresponds to the hyperfine transition of neutral hydrogen and arises due to the energy difference between the spin-up and spin-down states of the hydrogen atom's electron. As a probe of the high-redshift universe~\cite{Furlanetto2006, Pritchard2008, pritchard2012}, the 21-cm neutral hydrogen line presents a promising avenue for exploring various aspects of early structure formation, as it is sensitive to the state of the neutral hydrogen gas from which early stars and galaxies form.

Injected energy from dark matter annihilation can significantly impact baryonic processes and, consequently, the formation of the first galaxies and stars~\cite{Cirelli2009, Ripamonti2010}. Due to the lack of heavy elements, the collapse of the gas that formed the first stars (known as Population III stars) required molecular hydrogen cooling, which is highly sensitive to gas properties \cite{Bromm2002, Bromm2013, Hirano2015}. These stars formed within dark matter minihalos with typical masses in the range of \(M \sim 10^5 - 10^6 M_\odot\), at redshift \(z \sim 20 - 40\). The first galaxies and stars subsequently coupled the spin temperature of the primordial gas to the kinetic temperature through Wouthuysen-Field coupling, via Lyman-$\alpha$ photon emission, and caused heating through X-ray emission. Both of these processes contribute to the strength and timing of 21cm absorption against the CMB. It is therefore important to determine the interplay between possible exotic energy injection and the formation of the first stars and galaxies to robustly interpret the 21cm signal from Cosmic Dawn.

In our analysis, we consider the impact of dark matter annihilation on both the spatially averaged IGM and on the formation of the first stars in primordial mini-halos. To connect our studies to future observations, we employ a semi-numerical model to obtain the global 21cm signal. We focus on MeV to GeV dark matter, following \cite{LopezHonorez2016} (hereafter \citetalias{LopezHonorez2016}), where the distinctive signature is significant. Specifically, in this mass range, DM annihilation significantly heats and ionizes the IGM. This extra heating alters the timing and depth of the 21cm absorption trough and modifies the spatial power spectrum, producing a distinctive signature that can help distinguish DM effects from standard astrophysical processes. We incorporate recent calculations from \cite{Slatyer2016} on the energy deposition fraction to estimate the impact on the thermal history. We examine the effect of stellar feedback from DM annihilation by introducing a Jeans analysis and examining the relevant gas cooling processes. 

We also consider the effect of dark matter–baryon velocity offsets, also known as streaming. This phenomenon originates from the relative velocity between the two components that results due to the coupling of baryons and radiation prior to the recombination epoch, and can suppress structure formation, particularly at high redshifts and on small scales \cite{tseliakhovich2010relative, Tseliakhovich2011}. This effect is significant in influencing gas accretion and star formation, which therefore alters the 21cm signal, as highlighted in recent studies \cite{Muoz2019, Schauer2022Dwarf}. We include the streaming effect in our analysis and discuss its role in shaping the 21cm signal in the presence of dark matter annihilation.

The organization of this paper is as follows: Section~\ref{sec:previous} compares recent works considering the 21cm signal in the presence of dark matter annihilation. Section~\ref{sec:methods} describes the methods used in this work, with a review of the relevant halo mass scales and collapse criteria. Section~\ref{sec:model} describes our model, including our approaches to dark matter annihilation and the baryons physics within dark matter halos. Section~\ref{sec:simulation} details the numerical simulations used to derive the 21cm signal and presents the results of our analysis. In Section~\ref{sec:discuss}, we discuss the role of molecular cooling and streaming velocities in our model. Finally, we conclude the work in Sec.~\ref{sec:conclusion}. We also explore the potential contribution of local DM annihilation effects within dark matter halos, as shown in appendix.~\ref{sec:model.local}. DM heating and ionization could reshape the galaxies' formation in small halos, when local halo deposition efficiency is large.

\section{Previous Studies}\label{sec:previous}

The effect of dark matter annihilation on the 21cm signal has been investigated in numerous studies. \citet{Natarajan2009} explored the potential for DM annihilation to influence the 21cm signal, suggesting that such interactions could produce a distinctive signature detectable in observations. \citet{Evoli2014} focused on the calculation of the 21cm brightness temperature at cosmic dawn, utilizing the 21cmFAST code to examine the effects of $\chi\chi \to \mu^+\mu^-$ annihilation. \citetalias{LopezHonorez2016} constructed a comprehensive brightness temperature history for $\chi\chi\to e^+e^-$ annihilation, highlighting that certain ranges of dark matter masses could leave significant imprints on the 21cm signal. \citet{Cheung2019} utilized spin temperature data from the EDGES experiment to constrain various dark matter models. \citet{basu2020lower} examined the impact of dark matter annihilation with DM-baryon scattering on the 21cm signal, uncovering that such interactions could significantly alter the expected signal. \citet{Cang2023} calculated the inhomogeneous effects of DM annihilation, demonstrating that these effects could enhance the 21cm power spectrum by orders of magnitude, thus providing a potential method for distinguishing between different DM models. \citet{Qin2023} investigated the impact of DM annihilation on molecular hydrogen (\ce{H2}) formation for star formation, utilizing the DarkHistory code. This study found that the exotic energy injection from DM annihilation could either accelerate or delay the formation of the first stars, underscoring the complex interplay between DM annihilation and star formation.

In \citetalias{LopezHonorez2016}, the authors focus on the DM annihilation of particles with masses in the MeV to GeV range into electron-positron pairs. They employ an energy deposition fraction calculation from~\cite{Liu2016} and find enhancements in both temperature and ionization fraction during the epoch of reionization. The dark matter annihilation contribution to the thermal history was calculated in the \texttt{CosmoRec} code for the early universe, and \texttt{21cmFAST} was used for the period from cosmic dawn to reionization. Several parameters such as the halo mass function and X-ray efficiency affect the calculated spin temperature, as does the DM particle mass. In \citetalias{LopezHonorez2016}, models with $m_\mathrm{DM} \sim 100$ MeV produced the strongest effects.

However, a comprehensive study of the 21cm signal in conjunction with first star formation requires a more detailed treatment. While some studies include stellar Lyman-Werner (LW) feedback, the properties of molecular cooling halos can be significant for the formation of the first stars, and these are not automatically included in studies based on \texttt{21cmFAST}. This omission could lead to a delay in the onset of star formation (e.g.,~\cite{Evoli2014, LopezHonorez2016}). Meanwhile, DM annihilation may impact the star formation rate in those small halos, thereby altering early stellar feedback. The energy deposition fraction, which varies with redshift and dark matter mass, also requires careful consideration. Previous studies \cite{basu2020lower, Cang2023} have utilized energy deposition values derived from numerical results~\citep{shull1985, chen2004particle}(commonly referred to as the SSCK model). Although the SSCK results have been widely adopted and are supported by recent calculations, they are based on 3 keV electrons, and discrepancies increase as the electron energy rises. Thus, an improved treatment is crucial for accurately understanding and constraining the effects of dark matter annihilation~\cite{Slatyer2016}. Additionally, the inclusion of DM-baryon velocity offsets, which cause suppression on small scales and alters the existing dark matter annihilation power, is important for self-consistency. 

Table.~\ref{tab:model-compare} compares our model with several recent studies on dark matter annihilation in the context of the 21cm signal. Our work utilizes the same deposit fraction table as \citetalias{LopezHonorez2016} and includes the effects of molecular cooling and streaming. It is important to note that our model does not fully integrate all aspects of previous studies. For instance, we did not consider the DM-baryon elastic scattering discussed in~\cite{basu2020lower}. Like \cite{Qin2023}, we include molecular cooling to account the star formation in low-mass halos. For Lyman-Werner (LW) feedback, we assume that dark matter (DM) annihilation has a negligible impact, and therefore consider only the regular stellar LW feedback. This is because LW photon deposition data for DM annihilation is unavailable, and accurately modeling it would require detailed energy deposition simulations, which we leave for future work.

\begin{table*}[]
\caption{Models of Previous Dark matter Annihilation Studies in 21cm Astronomy. This table only features key processes in our scope. }
\label{tab:model-compare}
\begin{tabular}{@{}lccccc@{}}
\toprule
Reference & Channels & Deposit Fraction & Gas Cooling & LW Feedback                        & $v_{bc}$ \\ \midrule
\citet{Evoli2014}   & $\chi \chi \to \mu^+\mu^-$            & Analytic Fit     & Atomic Cooling   &    -                & no $v_{bc}$             \\
\citet{LopezHonorez2016}   & $\chi \chi \to e^+e^-$            & Updated Table             & Atomic Cooling         &       -       & no $v_{bc}$            \\
\citet{Cheung2019}   & $\chi \chi \to e^+e^-$            & SSCK             & Atomic Cooling& -  & no $v_{bc}$             \\ 
\citet{basu2020lower} & $\chi \chi \to e^+e^-$            & SSCK             & Atomic Cooling& -  & no $v_{bc}$             \\ 
\citet{Cang2023}   & $\chi \chi \to e^+e^-$            & SSCK             & Atomic Cooling&  - & no $v_{bc}$             \\ 
\citet{Qin2023}   & $\chi \chi \to e^+e^-$   & DarkHistory  & Molecular Cooling  &  Exotic     & no $v_{bc}$             \\
\midrule
\textit{This paper} &
  $\chi \chi \to e^+e^-$ &
  Updated Table &
  Molecular Cooling &
  Regular &
\begin{tabular}[c]{@{}c@{}}no $v_{bc}$\\ $v_{bc}$ included\end{tabular} \\ \bottomrule
\end{tabular}
\end{table*}

\section{\label{sec:methods}Setup}

\subsection{Base model (no dark matter annihilation)}

To model early cosmic structure formation without dark matter annihilation, we follow a sequence of key processes that impact the formation of first stars and the evolution of the IGM (and therefore the 21cm signal).

\begin{enumerate}
    \item \textbf{Gas Collapse:} In the early universe, dark matter halos act as gravitational wells. Those halos accumulate gas, where the process depends critically on the halo's mass. When the halo's mass exceeds the Jeans mass, $M_\mathrm{J}$, the gravitational force is strong enough to overcome the thermal pressure, allowing the gas to collapse. The time-averaged Jeans mass (known as the filtering mass, $M_\mathrm{F}$) then determines the gas fraction that halos can retain. Detailed calculations of the Jeans mass and gas fraction are provided in Section~\ref{sec:methods.JeansMass}.
    \item \textbf{Gas Cooling:} Star formation requires that gas within dark matter halos cool efficiently in order to reach the required density for nuclear ignition. This cooling efficiency is described by the minimum cooling mass, $M_\mathrm{cool}$, a threshold that determines whether a halo can support the cooling rates required for star formation. Halos with masses above $M_\mathrm{cool}$ have sufficient cooling to lower gas temperature, allowing the gas to collapse further to form individual stars. This cooling is dependent on the channel (atomic or molecular hydrogen cooling) and thus also the abundance of molecular hydrogen. A detailed discussion is provided in Section~\ref{sec:methods.Mcool}.
    \item \textbf{Star Formation:} As the gas sufficiently cools within primordial halos, the first stars begin to form. The radiation emitted by these stars includes both Lyman-$\alpha$ and X-ray photons, which are particularly important in altering the IGM in a manner relevant to the 21cm signal. Lyman-$\alpha$ photons couple the gas spin temperature to its kinetic temperature via the Wouthuysen-Field effect. Meanwhile, X-ray photons heat the gas, raising its temperature and modifying the 21cm absorption signal against the CMB. The Pop III star formation rate at a specific redshift is described by the total fraction of star-forming baryons, $f_\mathrm{coll}$, which is introduced in Section~\ref{sec:methods.starformation}.
\end{enumerate}

\subsubsection{Jeans Mass and Gas Fraction}\label{sec:methods.JeansMass}

For a primordial gas cloud associated with a dark matter halo, the thermal pressure of the gas acts counter to gravitational forces, which can prevent the cloud from collapsing. The characteristic scale at which a gas cloud can overcome thermal pressure and undergo gravitational collapse is known as the Jeans scale $k_J$, described as follows

\begin{equation}\label{eq:JeansScale}
    k_J = \frac{a}{c_s}\sqrt{4\pi G \rho}\;,
\end{equation}
where $a \equiv 1/(1+z)$ is the scale factor, $G$ is the gravitational constant, $\rho$ is the total density, and $c_s$ is the sound speed of the gas. 

For the gas in a halo with mass $M_h$ at redshift $z$, we assume that the total density $\rho = \rho_\mathrm{DM} + \rho_\mathrm{B} \approx \rho_\mathrm{DM}$, as the baryonic density is negligible compared to the dark matter density in the early Universe. The sound speed $c_s$ is given by

\begin{equation}\label{Eq:cs_DM}
    c_s(z, M) = \sqrt{\frac{5 k_B T_g(z, M)}{3\mu m_N}}\;,
\end{equation}
where $T_g(z, M)$ is the gas temperature within halo mass $M$ at redshift $z$, $\mu=1.22$ is the mean molecular weight of the neutral primordial gas, $m_N$ is the proton mass, and $k_B$ is the Boltzmann constant. 

The Jeans mass $M_\mathrm{J}$ is defined by the enclosed mass within the Jeans scale,

\begin{equation}
    M_\mathrm{J} = \frac{4}{3} \pi \rho \left(\frac{\lambda_\mathrm{J}}{2}\right)^3\;,
\end{equation}
where $\lambda_\mathrm{J} = 2\pi / k_\mathrm{J}$ is the Jeans wavelength. 

To estimate the gas fraction within dark matter halos, we must first calculate  the time-averaged Jeans mass~\cite{Gnedin2000, Naoz2007, barkana2011scale}, also known as the filtering mass $M_\mathrm{F}$,

\begin{equation}\label{equ.filteringmass}
    M_\mathrm{F}^{3/2}(a) = \frac{3}{a} \int_0^a M_\mathrm{J}^{3/2}(a)da'\left(1-\sqrt{\frac{a'}{a}}\right)\;,
\end{equation}
where $a$ is the scale factor and $M_\mathrm{J}$ is the Jeans mass at a given redshift.

The gas fraction $f_\mathrm{gas}$ inside a halo with mass $M_h$ at redshift $z$ follows the equation

\begin{equation}\label{eq:fgas}
    f_\mathrm{gas}(z, M_h)=f_{\mathrm{b}, 0}\left[1+\left(2^{\alpha / 3}-1\right)\left(\frac{M_\mathrm{F}}{M_h}\right)^\alpha\right]^{-3 / \alpha}\;,
\end{equation}
where $M_\mathrm{F}$ is the filtering mass at redshift $z$ given by Equation~\ref{equ.filteringmass} and where we have taken $\alpha=0.7$ as suggested by \cite{Tseliakhovich2011}. The cosmic baryon fraction $f_{\mathrm{b}, 0}(z)$ is given by

\begin{equation}\label{eq:fgas_0}
    f_{\mathrm{b}, 0}(z) = \frac{\Omega_b(z)}{\Omega_m(z)} ( 1 + 3.2 r_\mathrm{LSS})\;,
\end{equation}
where $\Omega_b(z)$ and $\Omega_m(z)$ are the density parameters for baryons and matter, respectively, at redshift $z$, and $r_\mathrm{LSS}$ is a redshift-dependent function defined in \cite{Naoz2007}.

\subsubsection{Minimum Cooling Mass}\label{sec:methods.Mcool}

The minimum cooling mass $M_\mathrm{cool}$ represents the smallest mass of star-forming halos, below which the gas is unable to cool efficiently and form stars. The cooling of gas at Cosmic Dawn occurs primarily via atomic hydrogen cooling and molecular hydrogen cooling. 

\textit{Atomic cooling} is an important process at high temperatures ($T_\mathrm{vir} > 10^4$ K), corresponding to virial masses $M_{\rm vir} > 10^8 M_{\odot}$. At these temperatures, gas is able to cool through the emission of atomic lines. 

\textit{Molecular cooling} is important at lower temperatures, where gas is able to cool through the emission of rotational and vibrational lines of molecules. Molecular cooling is expected to be important for halos with $M_{\rm vir} \lesssim 10^7 M_{\odot}$, where atomic cooling is less effective.

For the first stars formed in small halos, gas cooling is dominated by molecular (\ce{H2}) cooling. The cooling rate must be large enough to lower the gas temperature and enable collapse and star formation. In this case, the factors determining the cooling criterion include the density of each species involved in \ce{H2} production and cooling (hydrogen, helium, electrons), as well as the gas temperature $T_g$.

We define the minimum cooling mass $M_\mathrm{cool}$ corresponding to the critical virial temperature $T_\mathrm{crit}$ where cooling is sufficient. For atomic cooling, it is given by $T_\mathrm{crit}=10^4$ K. For molecular cooling, the critical virial temperature is determined by several factors involved in the cooling process, which will be provided later in Section~\ref{sec:H2cool}. The relation between the halo mass and virial temperature is given by the formula~\cite{Barkana2001}, 

\begin{equation}\label{eqn:Tvir}
    T_{\mathrm{vir}}=1.98 \times 10^4\left(\frac{M_\mathrm{vir}}{10^8 h^{-1} \mathrm{M}_{\odot}}\right)^{\frac{2}{3}}\left(\frac{\Omega_{m,0}}{\Omega_m}\right)^{\frac{1}{3}}\left(\frac{1+z}{10}\right) \mathrm{K} \;,
\end{equation}
where $\Omega_{m,0}$ is the cosmic matter density today, and $\Omega_{m}$ is the cosmic matter density at redshift $z$.

\subsubsection{Star formation rate}\label{sec:methods.starformation}

For Pop III stars, we relate the star formation rate to the star-forming baryon collapsed fraction, which is the total fraction of star-forming baryons in the cosmos. This can be written as

\begin{equation}\label{eq:fcoll}
    f_\mathrm{coll} = \int_{M_\mathrm{min}}^\infty M \frac{dn}{dM} \frac{f_\mathrm{gas}}{\rho_b} f_\star dM\;,
\end{equation}
where $dn/dM$ is the halo mass function, $\rho_b$ is the comoving density of baryons, $f_\mathrm{gas}$ is the gas fraction within halos, $M_\mathrm{min}$ is the minimum star-forming halo mass, and $f_\star$ is the star formation efficiency. Since molecular cooling is essential for Pop III star formation at cosmic dawn, we enable star formation in halos where molecular cooling occurs and set the minimum star-forming halo mass at $M_\mathrm{min} = M_\mathrm{cool}$. This implies that only halos with a mass greater than $M_\mathrm{cool}$ can initiate star formation. Considering both molecular cooling halos and atomic cooling halos in our calculations significantly accelerates the onset of the 21-cm absorption signal compared to scenarios where only atomic cooling halos are taken into account.

The star formation efficiency, $f_\star$, is taken from \cite{Fialkov2013}. It is assumed to be constant for halos with masses above the atomic cooling mass, while for halos with masses between the molecular cooling mass and the atomic cooling mass, it follows an exponentially decaying function with halo mass. We employ the piecewise function

\begin{equation}
    f_\star(M)= \begin{cases}f_* & \text { if } M \geqslant M_{\text {atom}} \\ f_* \frac{\log \left(M / M_{\text {cool}}\right)}{\log \left(M_{\text {atom}} / M_{\text {cool }}\right)} & \text { if } M_{\text {cool}}<M<M_{\text{atom}} \\ 0 & \text { otherwise }\end{cases}\;,
\end{equation}
where $M_{\text {atom}}$ is the mass threshold of atomic cooling halos, given by $T_\mathrm{vir}>10^4$ K, $M_{\text{cool}}$ is the minimum cooling mass defined in this work, and $f_*$ is the star formation efficiency for atomic cooling halos, which is usually set to $f_* = 0.1$, as adopted in \cite{LopezHonorez2016, Muoz2019}.

\subsection{Effects of dark matter annihilation}

In this section, we describe the effects of dark matter annihilation on early halos and the 21cm signal.

The \textit{thermal history} of the intergalactic medium can be altered by the radiation background from dark matter annihilation, which can increase the IGM temperature, ionization fraction, and Lyman-$\alpha$ photon background. Since the spin temperature of neutral hydrogen is coupled to the gas kinetic temperature during this period, an increase in gas temperature leads to a reduction in the 21cm absorption signal.

The \textit{Jeans mass} can also be suppressed by dark matter annihilation. As the IGM is heated by dark matter annihilation, the gas in dark matter halos experiences increased pressure, which may overcome the gravitational attraction of the halo. 

The Jeans mass $M_J$ then increases, and the gas fraction $f_\mathrm{gas}(z, M_h)$ decreases within low mass halos with $M_h \approx M_J$. Consequently, the formation of Population III stars could be suppressed due to the lower baryon fraction within small halos, leading to a reduced radiation background from stars.

The \textit{minimum cooling mass} can potentially be altered by dark matter annihilation through several mechanisms. On one hand, the gas fraction altered by dark matter annihilation affects the gas density within halos, influencing the cooling process. On the other hand, energy deposition from dark matter annihilation can modify the abundance of molecular hydrogen by affecting the heating and ionization terms in the \ce{H2} production process, and it may also impact the cooling rate. However, the overall effect of annihilation on the minimum cooling mass is complex and remains unclear. 
 
In this work, we develop a simple model to analytically estimate the cooling mass for molecular cooling halos under dark matter annihilation, as described in Section~\ref{sec:H2cool}.

\section{\label{sec:model}Models}

\subsection{\label{sec:model.dm}Dark Matter Annihilation}

For self-annihilating dark matter (DM), the annihilation power per unit volume is described by the equation

\begin{equation}
    \frac{dE}{dV dt} = \frac{\langle \sigma v \rangle}{m_{\mathrm{DM}}} \rho_{\mathrm{DM}}^2,
\end{equation}

where $\langle \sigma v \rangle$ represents the thermally averaged annihilation cross-section, $\rho_{\mathrm{DM}}$ is the DM density, and $m_{\mathrm{DM}}$ denotes the mass of the DM particle. 

For the case of s-wave dark matter annihilation, the cosmic microwave background (CMB) spectrum provides an upper limit on the cross-section, as reported by the Planck collaboration: \( p_\mathrm{ann} \leq 4.1 \times 10^{-28} \, \si{\cubic\centi\metre\per\second\per\giga\electronvolt} \), where \( p_\mathrm{ann} = f_\mathrm{eff} \langle \sigma v \rangle / m_{\mathrm{DM}} \) \cite{collaboration2014planck, Liu2016}. Here, \( f_\mathrm{eff} \) is a constant proxy for the deposited fraction, which depends on the dark matter particles and cosmological parameters. Given the value of \( f_\mathrm{eff} \), one can derive constraints on the dark matter annihilation cross-section.  In this study, we employ a constant mass-weighted cross-section, \( \langle \sigma v \rangle / m_{\mathrm{DM}} = 10^{-27} \, \si{\cubic\centi\metre\per\second\per\giga\electronvolt} \). This value is allowed for \( f_\mathrm{eff} \lesssim 0.1 \) and serves as a useful benchmark to check against current limits. Additionally, we assume annihilation occurs through the channel \( \chi \chi \to e^+ e^- \).

Since the annihilation rate depends on the squared density of dark matter, the formation of collapsed halos results in a boost in the cosmic dark matter annihilation power. We separate the power of dark matter annihilation into two parts: $\frac{dE}{dt} = (\frac{dE}{dt})_{\mathrm{smooth}} + (\frac{dE}{dt})_{\mathrm{struct}}$. The former comes from the smooth dark matter background, and the latter comes from collapsed dark matter structures. In terms of the boost factor $B(z)$, accounting for this structure effect, the annihilation power at redshift $z$ can be written as

\begin{equation}\label{eq:dE_dVdt}
\begin{split}
\frac{dE}{dV dt}\bigg|_\mathrm{injected} &= \frac{dE}{dVdt}\bigg|_\mathrm{smooth} + \frac{dE}{dVdt}\bigg|_\mathrm{struct} \\
    &= \frac{\langle \sigma v \rangle}{m_{\mathrm{DM}}} \rho_{\mathrm{DM,0}}^2 (1+z)^6 \left(1+B(z)\right)\;,
\end{split}
\end{equation}
where $\rho_\mathrm{DM,0}$ indicates the average DM density in the current time. The component of annihilation power from structure is related to the smooth component by

\begin{equation}
\frac{dE}{dVdt}\bigg|_\mathrm{struct} = B(z) \frac{dE}{dVdt}\bigg|_\mathrm{smooth}.
\end{equation}

The boost factor depends on the halo mass function and halo profile, written in \cite{Cirelli2009} as

\begin{equation}\label{eq:boost_factor}
    B(z) = \frac{1}{\rho_{\mathrm{DM,0}}^2 (1+z)^3} \int_{M^\mathrm{min}_\mathrm{h}}^{\infty} \frac{dN}{dM} dM \int_{0}^{R_\mathrm{vir}} \rho_\mathrm{DM}^2(r) 4\pi r^2 dr,
\end{equation}
where $\frac{dN}{dM}$ is the comoving halo mass function (HMF), and $\rho_\mathrm{DM}(r)$ is the dark matter halo density profile. The minimum halo mass, \(M^\mathrm{min}_\mathrm{h}\), representing the smallest mass of halos, depends on the model of dark matter particle. As noted in~\cite{Evoli2014, LopezHonorez2016, Mack2014}, this value affects the intensity of  the structured boost factor of the dark matter annihilation signal, altering the 21cm signal. We use the value \(M^\mathrm{min}_\mathrm{h} = 10^{-9} \mathrm{M}_\odot\) in this paper.

For the halo mass function, we use the Sheth-Tormen model with parameters $a'=0.75$ and $p'=0.3$, identified as optimal fits~\cite{Sheth2001}. We also employ the mass-concentration relation from \cite{diemer2019accurate} and use the Navarro-Frenk-White (NFW) profile to characterize dark matter halos, which provides the total dark matter annihilation for a given halo.

Note that the deposited energy is highly dependent on the chosen cosmological and DM models. These dependencies include the density profile, mass concentration relation, and halo mass function, as detailed in \cite{Mack2014, schon2014dark}. Additionally, our model assumes DM annihilates entirely into $e^+ e^-$ pairs and lies in the approximate 10 MeV to GeV mass range.

The DM energy deposition into gas is modelled as follows: the deposited fraction $f_c(z)$ assumes that power deposition is proportional to the injected power at the same redshift, going into each channel $c$ for heating, HI and HeI ionization, and Ly-$\alpha$ photons. Thus, the deposited energy in channel $c$ per unit time per baryon is given as

\begin{equation}\label{eq.epsilon_c}
\epsilon_c^\mathrm{DM}(z) = \frac{1}{n_\mathrm{B}} f_c(z) \frac{dE}{dV dt}\bigg|_\mathrm{injected},
\end{equation}
where $n_\mathrm{B}$ is the mean baryon number density at redshift $z$.

The deposited energy in Equation~\ref{eq.epsilon_c} represents the energy deposited into the background IGM. For gas in overdense regions, such as dark matter halos, the energy deposit rate is given by the total DM energy from both local DM and the global DM background: $\epsilon^\mathrm{DM}_c = \epsilon^\mathrm{DM}_\mathrm{c, local} + \epsilon^\mathrm{DM}_\mathrm{c, BG}$. The local deposit rate originates from local dark matter annihilation, which could heat and ionize the surrounding gas~\cite{Schn2017, Clark2017}. However, the local effect depends heavily on the interaction between annihilation products and baryonic particles in the gas environment, which remains poorly understood. In this study, we focus on the global DM energy deposition, leaving the discussion of local effects to the Appendix. For the global annihilation background, we assume the energy deposit rate per baryon in halos is the same as that in the IGM. Therefore, we consistently use $\epsilon_c^\mathrm{DM}(z)$ from Equation~\ref{eq.epsilon_c} throughout this work.

We derive the deposition fractions using the transfer functions from \cite{Slatyer2016}, which calculate these fractions for high-energy photons and $e^+e^-$ pairs resulting from DM annihilation. The DM annihilation emission power deposited per baryon is $\epsilon^\mathrm{DM}_c$, where $c$ represents the deposition channel, as defined above (heating, HI and HeI ionization, and Ly-$\alpha$ photons).

The energy injection of DM annihilation into the IGM can be described by following terms. For the effect on the gas temperature,

\begin{equation}
    \left.\frac{d T_{K}}{d z}\right|_{\mathrm{DM}}=\frac{d t}{d z} \frac{2}{3 k_{B}\left(1+x_{e}\right)} \epsilon_{\mathrm{heat}}^{\mathrm{DM}}\label{Equ.evo_T}\;,
\end{equation}
where $k_B$ is the Boltzmann constant and $x_e$ is the electron fraction. 

The ionization and Ly-$\alpha$ excitation due to dark matter annihilation are given by

\begin{align}
\left.\Lambda_{\mathrm{ion}}\right|_{\mathrm{DM}}&=\mathfrak{f}_{\mathrm{H}} \frac{\epsilon_{\mathrm{HI}}^{\mathrm{DM}}}{E_{\mathrm{HI}}}+\mathfrak{f}_{\mathrm{He}} \frac{\epsilon_{\mathrm{HeI}}^{\mathrm{DM}}}{E_{\mathrm{HeI}}}\label{Equ.evo_ion} \;,\\
J_{\alpha}|_{\mathrm{DM}}&=\frac{ n_\mathrm{B}}{4 \pi} \frac{\epsilon_{\mathrm{Ly} \alpha}^{\mathrm{DM}}}{h \nu_{\alpha}} \frac{1}{H(z) \nu_{\alpha}}\label{Equ.evo_lymana}\;,
\end{align}
where $E_\mathrm{HI}$ and $E_\mathrm{HeI}$ are the ionization energies for hydrogen and helium, $\mathfrak{f}_\mathrm{H}$ and $\mathfrak{f}_\mathrm{He}$ represent the number fractions of hydrogen and helium, $n_\mathrm{B}$ is the number density of baryons, $h$ is the Planck constant, $H(z)$ is the Hubble constant at redshift $z$, and $\nu_{\alpha}$ is the emission frequency of a Ly-$\alpha$ photon.

\subsection{\label{sec:H2cool}Molecular Cooling}

For the birth of the first stars, the gas within dark matter halos requires cooling for further collapse, and that cooling is primarily achieved via molecular hydrogen~\cite{Galli1998, Glover2006, Yoshida2007, Glover2008, Glover2009}. Molecular cooling occurs only if the halo mass is larger than the minimum cooling mass, $M_\mathrm{cool}$. In this section, we build an analytical cooling model to calculate the influence of the minimum cooling mass in the presence of dark matter annihilation.

\subsubsection{Gas Density}

To calculate gas cooling in molecular cooling halos, we must first estimate the gas profile, which can be approximated by following the distribution of the dark matter (DM) profile. In this approximation, the gas density in a halo of mass $M_h$, denoted as $\rho_\mathrm{gas}(M_h, r)$, is expressed as $\rho_\mathrm{gas}(M_h, r) = f_\mathrm{gas}(M_h) \rho_\mathrm{DM}(M_h, r)$, where $f_\mathrm{gas}(M_h)$ is the gas fraction.

In accordance with cosmological hydrodynamic simulations and theoretical arguments~\cite{Machacek2001, Visbal2014}, prior to efficient cooling, the gas in the halo is expected to settle into a cored density profile. The core is characterized by a nearly constant density within a radius of approximately $R_\mathrm{core} \approx 0.1 R_\mathrm{vir}$. In this work, we assume efficient gas cooling in the core region, defined as $R_\mathrm{core} = 0.1 R_\mathrm{vir}$. Consequently, the gas density in the core of halo $M_h$ is given by

\begin{equation}\label{eqn:rho_core_HM}
    \rho^\mathrm{HM}_\mathrm{gas, core}(M_h) = f_\mathrm{gas}(z, M_h) \rho_\mathrm{DM}(M_h, 0.1 R_\mathrm{vir})\;,
\end{equation}
where the gas fraction $f_\mathrm{gas}(z, M_h)$ is derived from the filtering mass in Equation~\ref{eq:fgas}.

However, this assumption breaks down on small scales, particularly when $k \gtrsim k_J$, as the gas density cannot form a cusp-like central density  like the dark matter in a standard NFW profile, due to thermal pressure. For these minihalos, we apply constraints from adiabatic compression and use the core density derived from hydrostatic equilibrium, given by \cite{Tegmark1997, Barkana2001, Visbal2014}

\begin{equation}\label{eqn:rho_core_LM}
    \rho^\mathrm{LM}_\mathrm{gas, core}(M_h) = \bar{\rho}_{b} \left(1+ \frac{6}{5}\frac{T_\mathrm{vir}(M_h)}{T_\mathrm{IGM}}\right)^{3/2}\;,
\end{equation}
where $\bar{\rho}_{b}$ represents the average baryonic density, and $T_\mathrm{IGM}$ denotes the background IGM temperature. Thus, the final core gas density of halo mass $M_h$ is given by the minimum of the two values: $\rho_\mathrm{gas, core}(M_h) = \min\{\rho^\mathrm{HM}_\mathrm{gas, core}, \rho^\mathrm{LM}_\mathrm{gas, core}\}$. In high-mass halos, $\rho^\mathrm{HM}_\mathrm{gas, core}$ dominates, while in low-mass halos, $\rho^\mathrm{LM}_\mathrm{gas, core}$ dominates. Importantly, the IGM temperature affects the gas density in both cases, and as a result, dark matter annihilation also influences gas cooling by altering the gas density.

\subsubsection{$\ce{H2}$ Production}

Molecular hydrogen production in primordial gas is primarily through the \ce{H^-} mechanism:

\begin{align}
\ce{H + e- &\to H- + \gamma}\label{equ.He-} \qquad (\text{rate } k_1) \;,\\
    \ce{H- + H &\to H2 + e^-}\label{equ:H-H} \qquad (\text{rate } k_2)\;.
\end{align}

We also consider the mutual neutralization and radiative recombination reactions:

\begin{equation}\label{equ:H+H-}
    \ce{H+ + H- \to H + H}  \qquad (\text{rate } k_3)\;.
\end{equation}
\begin{equation}\label{equ:H+e-}
    \ce{H+ + e- -> H} \qquad (\text{rate } k_4)\;.
\end{equation}

The corresponding reaction rates for Equations~\ref{equ.He-}-\ref{equ:H+e-} are denoted as \( k_1, k_2, k_3, k_4 \), and they are functions of the gas temperature \( T \). From reference~\cite{hutchins1976thermal}, we have \( k_1 = 1.83\times 10^{-18} T^{0.8779}\ \mathrm{cm^3 s^{-1}} \). The reaction rates \( k_2 \) and \( k_3 \) have relatively weak temperature dependences, and are calculated in \cite{kreckel2010experimental, stenrup2009mutual}; we set $k_2 / k_3 \approx 0.03$. For the recombination rate \( k_4 \), instead of using the coefficient adopted by \cite{Tegmark1997}, \( k_4 = 1.88\times 10^{-10} T^{-0.644}\ \mathrm{cm^3 s^{-1}} \), we follow \cite{Nebrin2023}, and use the case-B recombination rate: \( k_4 = 2.11\times 10^{-10} T^{-0.72}\ \mathrm{cm^3 s^{-1}} \), because the neutral gas in the minihalo is optically thick to Lyman-continuum photons. 

In addition, we assume that $\ce{H-}$ reaches its equilibrium abundance, where $n_{\ce{H-}} = k_1 n_{\ce{e-}} / k_2$, and that the abundance of ionized hydrogen is equal to free electrons, $n_{\ce{e-}} = n_{\ce{H+}}$.

The abundance of electrons is written as

\begin{equation}\label{Equ:dxe}
    \frac{d x_e}{dt} = - k_{4} n_{\ce{H}} x_e^2\;.
\end{equation}
The fraction of molecular hydrogen can be calculated via

\begin{equation}\label{Equ:dxH2}
    \frac{d x_{\ce{H2}}}{dt} = k_{1} x_e n_\mathrm{H} \left( 1 + \frac{k_{3}}{k_{2}} x_e \right)^{-1} \;.
\end{equation}
It is possible to solve the above equations and obtain the abundance of $x_e$ and $\ce{H2}$ as a function of time $t$:

\begin{align}
x_e &= \frac{x_0}{1 + x_0 n_\mathrm{H} k_{4} t} \;,\\
    x_{\ce{H2}} &= x_{\ce{H2}, 0} + \frac{k_{1}}{k_{4}} \ln\left(1 + \frac{k_{4} n_\mathrm{H} x_0 t}{1+k_3/k_2 x_0} \right)\label{eqn:xH2_solution}\;,
\end{align}
where $x_0$ and $x_{\ce{H2}, 0}$ are the initial abundances of electrons and \ce{H2}. The logarithmic term implies that the abundance of $\ce{H2}$ is growing rapidly at the beginning, and then becomes slow.

Considering the impact of dark matter annihilation ionizing gas during the cooling process alters Equation.~\ref{Equ:dxe}:

\begin{equation}\label{eqn:xe_DM}
    \frac{d x_e}{dt} = - k_{4} n_{\ce{H}} x_e^2 + \left.\Lambda_{\mathrm{ion}}\right|_{\mathrm{DM}}\;,
\end{equation}
where $\left.\Lambda_{\mathrm{ion}}\right|_{\mathrm{DM}}$ is the ionization rate per hydrogen atom due to dark matter annihilation given by Equation~\ref{Equ.evo_ion}. The first term on the right-hand side of the equation represents the recombination rate as a function of the electron fraction. When the electron fraction is high, recombination dominates over annihilation-induced ionization, causing the electron fraction to decrease. In contrast, if $\left.\Lambda_{\mathrm{ion}}\right|_{\mathrm{DM}}$ exceeds the recombination rate, the electron fraction will increase over time.

\begin{figure*}
    \centering
    \includegraphics[width=0.85\linewidth]{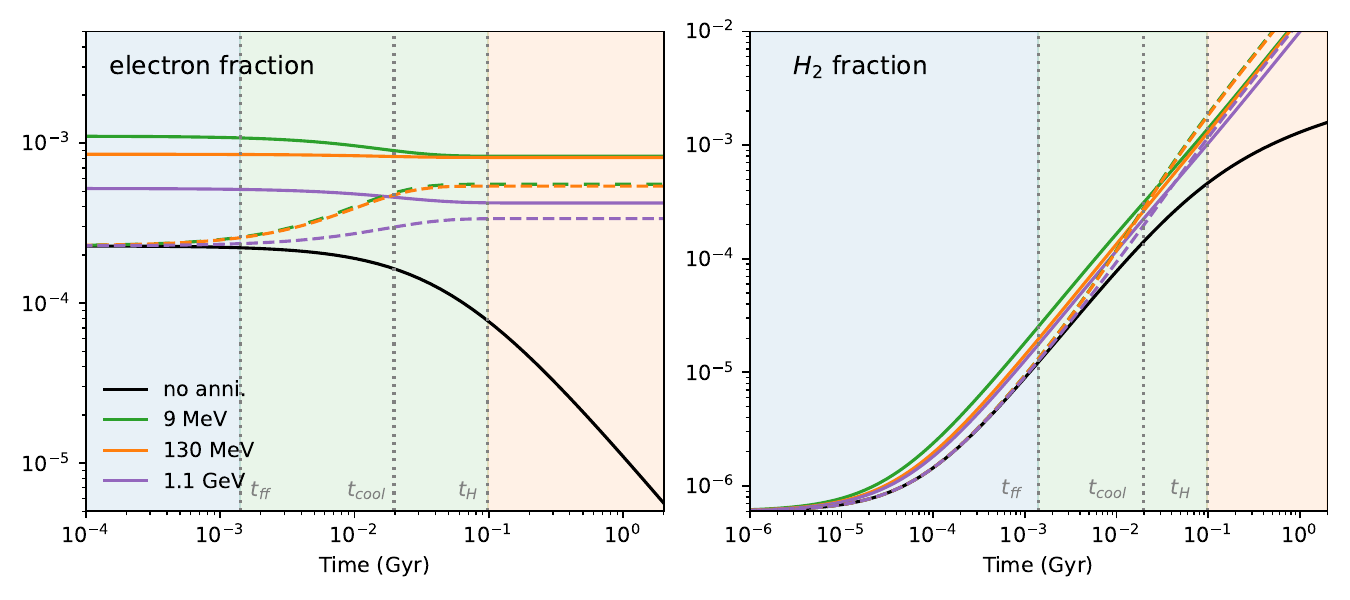}
    \caption{Fraction of electron and molecular hydrogen $\ce{H2}$ as a function of time during gas cooling. The halo has a mass $10^5 M_\odot$ at redshift $z=40$. We present the case without annihilation (black), as well as cases with dark matter annihilation for different particle masses. Initial electron fractions are determined by the background thermal evolution, with the effects of corresponding dark matter annihilation. The free-fall time $t_\mathrm{ff}$ and the cooling time criterion $t_\mathrm{cool} = 0.2 t_H$ are presented in the figure. Dashed lines represent scenarios where the same initial condition of gas at $t=0$ is assumed, with DM annihilation ionizing the gas thereafter.}
    \label{fig:cooling_frac}
\end{figure*}

The left panel of Fig.~\ref{fig:cooling_frac} shows the electron fraction calculated using Equation~\ref{eqn:xe_DM} in a $10^5 M_\odot$ halo at redshift 40. For each specific dark matter mass, the initial electron fraction is provided by the public cosmological recombination code \texttt{CosmoRec} \cite{Chluba2010}. The black line represents the scenario without dark matter annihilation. Dashed lines assume the same initial electron fraction at $t=0$ as the case without annihilation. The gas temperature is fixed at the virial temperature of the halo.

The production of molecular hydrogen is shown in the right panel of Fig.~\ref{fig:cooling_frac}. We set the initial molecular hydrogen fraction to $x_{\ce{H2}, 0} = 6\times 10^{-7}$ \cite{Galli2013}. Indirect effects of dark matter annihilation alter the initial gas number density and associated reaction rates. In small halos, dark matter annihilation reduces gas density, leading to a suppression of the reaction rates, as discussed in the context of the gas fraction. However, we found that this effect is minimal; the production of \ce{H2} is still dominated by the electron fraction. We found that the differences in molecular hydrogen fractions among the cases are primarily due to initial variations in the electron fraction. The molecular hydrogen fraction $x_{\ce{H2}, 0}$ has little impact on the results as long as it is small. Dashed lines represent scenarios in which the same initial fractions at $t = 0$ are assumed, with DM subsequently ionizing the gas.

\subsubsection{Cooling Criterion}

To find the minimum cooling mass inside halos via \ce{H2} cooling, we take the criterion that cooling time must be less than $20\%$ of the Hubble time, $t_\mathrm{cool} < 0.2 t_H$, as adopted in \cite{Tegmark1997, Machacek2001}, where the cooling time $t_\mathrm{cool}$ is given by 

\begin{equation}\label{eqn:t_cool}
    t_\mathrm{cool} = \frac{1}{\gamma - 1} \frac{n k_B T_\mathrm{vir}}{\Lambda_0 n_{\ce{H2}}}\;,
\end{equation}
where $\gamma = 5/3$ for primordial gas, $n$ is the total number density of gas, $k_B$ is the Boltzmann constant, $n_{\ce{H2}}$ is the number density of \ce{H2}, and $\Lambda_0$ is the total cooling rate of the gas per hydrogen molecule.

The total cooling rate per \ce{H2} molecule is given by

\begin{equation}
    \Lambda_{0} = \sum_k \Lambda_{\ce{H2},k} n_k\;,
\end{equation}
where $\Lambda_{\ce{H2},k}$ are the collisional excitation coefficients which are a function of temperature for each $k$, and $n_k$ is the number density of each species $k$, (such as \ce{H}, \ce{He}, \ce{H2}, \ce{H+}, and \ce{e-}).

The collisional excitation coefficients for \ce{H2} are reviewed in \cite{Glover2008}, where the cooling rate has been calculated in several collisional processes involving \ce{H}, \ce{H2}, \ce{He}, \ce{H+} and \ce{e-}. We adopt the cooling functions as calculated in that work, and maintain all the underlying assumptions of these functions. This includes adhering to the low-density environment and assuming an ortho-para ratio of 3:1. These assumptions are valid under the condition that the dark matter annihilation does not significantly alter the gas environment during the cooling process. 

Dark matter annihilation has been implemented into our halo evolution model in the following two ways:

1. We use the gas number density $n$ and ionization fraction $x_e$ from our evolution results, which account for the effects of dark matter annihilation. The gas density given by Equation~\ref{eqn:rho_core_LM} and \ref{eqn:rho_core_HM}, is modified by dark matter annihilation via filtering mass and IGM temperature. The initial ionization fraction of the gas, $x_e$, is determined using results from \texttt{CosmoRec}, which also include the impact of dark matter annihilation.

2. We modify the total cooling rate $\Lambda_0 n_{\ce{H2}}$ to $n_{\ce{H2}} \Lambda_{0} - n_{\ce{H}} \epsilon_\mathrm{heat}^\mathrm{DM}(z)$ to account for dark matter annihilation heating during gas cooling. Here, $\epsilon_\mathrm{heat}^\mathrm{DM}(z)$ represents the energy deposited into the heating channel per unit time per baryon, as defined by Equation~\ref{eq.epsilon_c}. We assume this value is solely a function of redshift.

\begin{figure}
    \centering
    \begin{subfigure}
        \centering
        \includegraphics[width=0.46\textwidth]{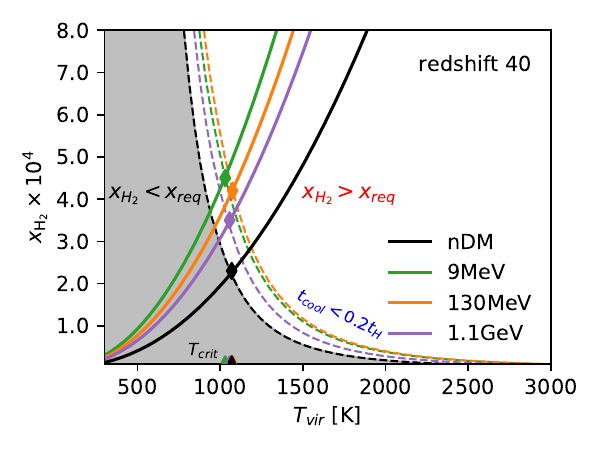}
    \end{subfigure}

    \begin{subfigure}
        \centering
        \includegraphics[width=0.46\textwidth]{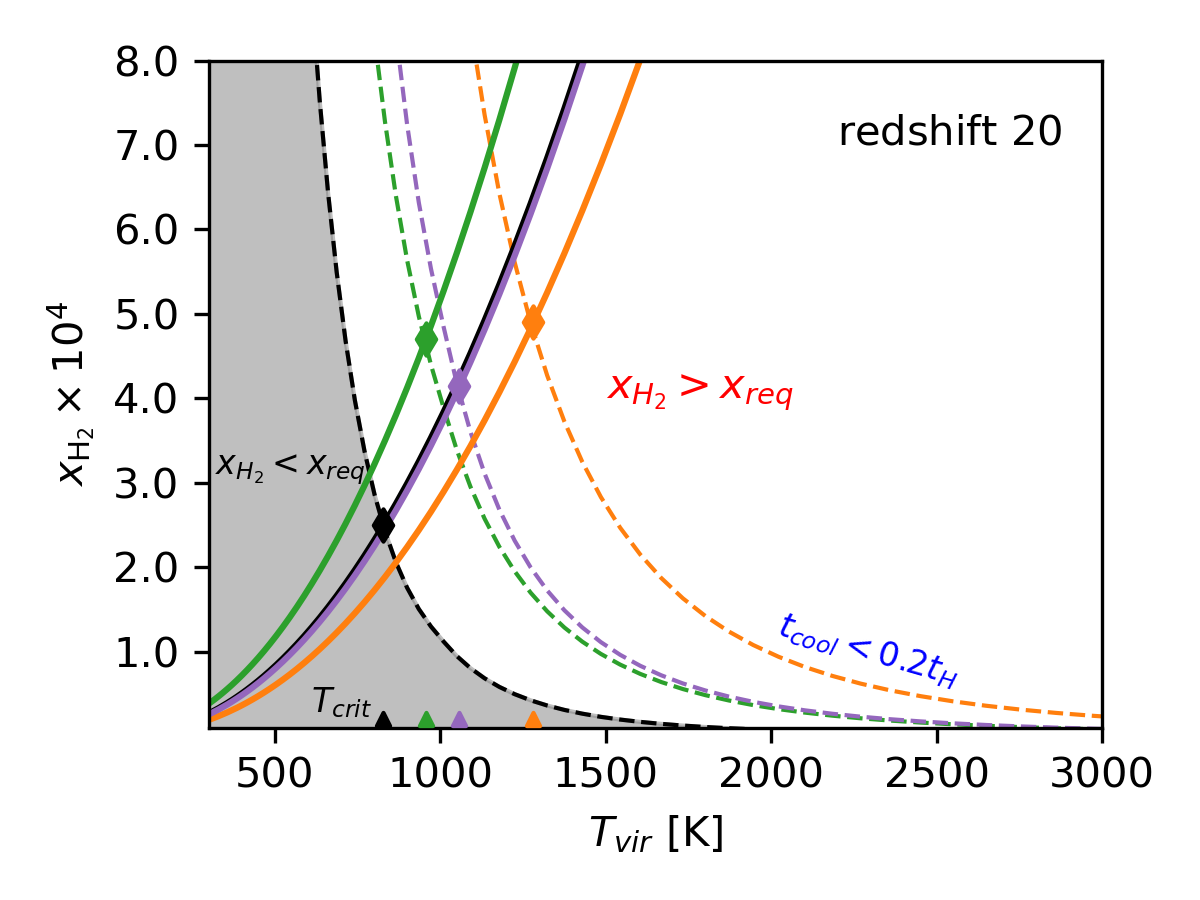}
    \end{subfigure}
    
    \caption{The molecular hydrogen fraction as a function of virial temperature at redshift $z=40$ and $z=20$. The dashed lines represent the required \ce{H2} fraction, determined by the criterion $t_\mathrm{cool} < 0.2 t_H$, while the solid lines correspond to the molecular hydrogen fraction at the same time. Cooling is not possible in the grey regions, where the produced \ce{H2} fraction is lower than the required fraction. Dark matter annihilation alters both $x_{\ce{H2}}$ and $x_{\mathrm{req}}$, as depicted by the different colored lines, each corresponding to a specific dark matter mass.}
    \label{fig:xH2}
\end{figure}

In Fig.~\ref{fig:xH2}, we plot the fraction of molecular hydrogen, \ce{H2}, in a halo in the presence of a dark matter annihilation background, as a function of virial temperature, $T_\mathrm{vir}$. The plot compares the required \ce{H2} fraction for cooling $x_{req}$ (dashed lines), determined by the criterion $t_\mathrm{cool} < 0.2 t_\mathrm{H}$, with the produced \ce{H2} fraction $x_{H_2}$ (solid lines), given by Equation~\ref{eqn:xH2_solution}. In the top panel, dark matter annihilation increases the produced \ce{H2} fraction. Dark matter annihilation ionizes the IGM in the early universe, thereby providing higher initial ionization fraction compared to the baseline. This increase varies with the dark matter mass, with 9 MeV and 130 MeV dark matter producing higher \ce{H2} fractions due to a larger energy deposition fraction in ionization. However, DM heating simultaneously suppresses the cooling rate by increasing the cooling time, $t_\mathrm{cool}$, which raises the required \ce{H2} fraction for cooling. The dashed lines in the plot shift upward in the dark matter annihilation case, reflecting this increase in the required \ce{H2} fraction.

The critical virial temperature, $T_\mathrm{crit}$, is defined as the point where the produced \ce{H2} fraction equals the required fraction for cooling to occur. Since dark matter annihilation can alter both $x_{\ce{H2}}$ and $x_{\mathrm{req}}$, the critical virial temperature is expected to change accordingly. However, in the top panel, both the produced and required \ce{H2} fractions increase due to dark matter annihilation, resulting in only minor changes to the critical virial temperature.

In the bottom panel, we plot the fraction of \ce{H2} at redshift $z=20$. At this redshift, dark matter annihilation has different effects on the produced \ce{H2} fraction depending on the dark matter mass. The 9 MeV dark matter increases the \ce{H2} fraction, while the 130 MeV dark matter decreases this fraction. The suppression of \ce{H2} fraction is primarily driven by the strong DM heating. The intensity of DM annihilation is stronger at lower redshift because of a stronger boost from DM structure formation. The resulting heating reduces the gas density and makes \ce{H2} production less efficient. This is also reflected in the dashed lines, where the required \ce{H2} fraction increases more significantly compared to the case at higher redshift. As a result, the critical virial temperature shifts to higher values, even for 9 MeV dark matter. At this redshift, dark matter annihilation has a primarily negative effect on molecular cooling.

\begin{figure}
    \centering
    \includegraphics[width=0.47\textwidth]{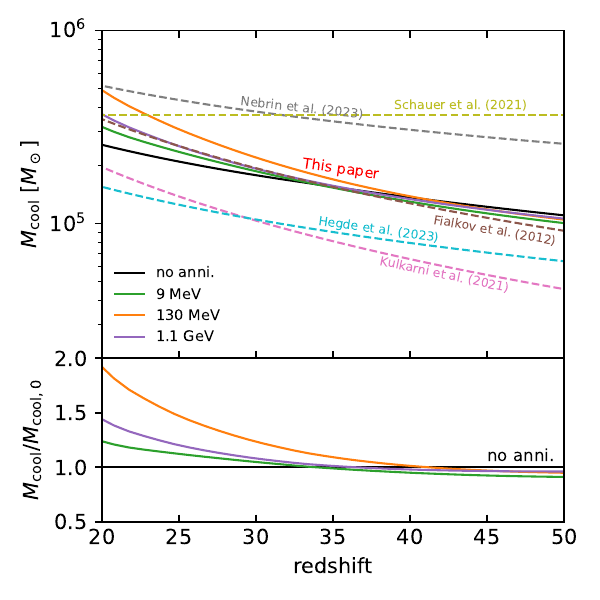}
    \caption{The minimum cooling mass, $M_\mathrm{cool}$, as a function of redshift with dark matter annihilation. In the top panel, the black line represents our baseline scenario without the effects of dark matter annihilation. The brown dashed line corresponds to the fit from simulations without a Lyman-Werner (LW) background and dark matter-baryon streaming, as detailed in \cite{Fialkov2012}. The pink dashed line is adopted from the fit in \cite{Kulkarni2021}, and the yellow-green line is from \cite{Schauer2021}. The grey dashed line shows the analytic model of \cite{Nebrin2023}, which assumes the criterion $t_\mathrm{cool} < 6 t_\mathrm{ff}$. In the bottom panel, different lines represent the relative ratio of the minimum cooling mass to the baseline $M_\mathrm{cool, 0}$ with different dark matter masses. The cyan line shows the analytic model of the minimum halo mass from \cite{Hegde2023}.}
    \label{fig:Mcool}
\end{figure}

In Fig.~\ref{fig:Mcool}, we predict the minimum cooling mass, $M_\mathrm{cool}$, in the presence of dark matter annihilation over redshifts ranging from $z = 20$ to $z = 50$. The solid black line in the top panel shows the minimum cooling mass in our baseline model without dark matter annihilation. We compare our results with those from several previous studies. The results from \cite{Fialkov2012, Kulkarni2021, Schauer2021, Nebrin2023} and \cite{Hegde2023} are shown as dashed lines. We set the Lyman-Werner (LW) background and streaming velocity to zero in these models to match our baseline model. Our cooling threshold is close to that of \cite{Fialkov2012}, but our slope is slightly lower. 

In \cite{Kulkarni2021}, the halo mass above which 50\% of halos host cool and dense gas is defined as $M_\mathrm{crit}$, while in \cite{Schauer2021}, they define a minimal ($M_\mathrm{min}$) and an average ($M_\mathrm{ave}$) halo mass at collapse. Our minimum cooling mass is generally greater than the value $M_\mathrm{crit}(z)$ as reported in \cite{Kulkarni2021} but lower than the redshift-independent value $M_\mathrm{min}$ (shown in the plot) and $M_\mathrm{ave}$ (not shown in the plot) from \cite{Schauer2021}. The grey dashed line represents the analytic model of \cite{Nebrin2023}, which is similar to our model but uses a stricter cooling criterion, $t_\mathrm{cool} < 6 t_\mathrm{ff}$, where $t_\mathrm{ff} = (3 \pi / 32 G \bar{\rho}_\mathrm{core})^{1/2}$ is the free-fall timescale, and $\bar{\rho}_\mathrm{core}$ is the mean core density of the dark matter halo. This stricter criterion results in larger minimum cooling mass estimates than in our work, which uses the criterion $t_\mathrm{cool} < 0.2 t_H$. Another key difference is that \cite{Nebrin2023} tailored the virial temperature $T_\mathrm{vir}$ by multiplying it by a factor of 0.75 relative to the formula given in Equation~\ref{eqn:Tvir}. This results in a lower cored density for a halo of same mass, and thereby suppresses the molecular cooling. These factors may explain why our results are lower. 

We acknowledge that a complete calculation would need to be more complex than our simplified model. For instance, we assumed that the gas density and temperature are fixed in order to obtain the cooling criterion for a halo. However, in reality, the gas temperature decreases as cooling proceeds, and the cooling rate, $\Lambda_0$, which depends on temperature, is not constant during this process. As a result, the calculation may slightly overestimate the cooling rate. We tested a method to incorporate the temperature evolution into our model and found that the resulting changes were small for times $t < t_\mathrm{cool}$. Therefore, our approximation is likely sufficient in most cases. 

Given the initial conditions of the gas, we calculate the cooling time-scale, $t_\mathrm{cool}$. The gas may undergo free-fall collapse if $t_\mathrm{cool} < t_\mathrm{ff}$ or experience collapse with an extended duration if $t_\mathrm{cool}$ is between $t_\mathrm{ff}$ and $t_H$. As shown in the green area of Fig.~\ref{fig:cooling_frac}, the fraction of \ce{H2} increases with time, while the required fraction decreases over time. This could lead to inaccuracies in the cooling mass estimation, as suggested by \cite{Nebrin2023}. 

The bottom panel of Fig.~\ref{fig:cooling_frac} shows the relative changes of minimum cooling mass with dark matter annihilation, $M_\mathrm{cool} / M_\mathrm{cool, 0}$, where $M_\mathrm{cool, 0}$ is the minimum cooling mass without dark matter annihilation, corresponding to the black line in the top panel. We found the dark matter annihilation impact increases with time. Dark matter annihilation slightly decreases the cooling mass at higher redshifts ($z>40$) but increases it at lower redshifts ($z<40$). At redshift $z=20$, the DM could increase the minimum cooling mass by factor of 2.

\subsubsection{LW Background}

In the presence of LW radiation, \ce{H2} can be photo-dissociated by photons. Although in this work we have not included the potential LW flux from dark matter annihilation, the LW flux from stellar feedback can lead to an increase in the minimum cooling mass~\cite{trenti2009formation, Fialkov2013, Visbal2014, Kulkarni2021, Schauer2021}. We will consider molecular cooling with dark matter annihilation in presence of a stellar LW background. Here, the formation of \ce{H2} in Equation~\ref{Equ:dxH2} becomes

\begin{equation}\label{eqn:xH2_LW}
    \frac{d x_{\ce{H2}}}{dt} = k_{1} x_e n_H \left( 1 + \frac{k_{3}}{k_{2}} x_e \right)^{-1} - k_\mathrm{LW} x_{\ce{H2}}\;,
\end{equation}
where $k_\mathrm{LW}$ is the photo-dissociation rate of \ce{H2}, given by~\cite{WolcottGreen2017},

\begin{equation}\label{eqn:k_LW}
    k_\mathrm{LW} = 1.38\times 10^{-12} f_{sh} J_\mathrm{LW}\;,
\end{equation}
where $J_\mathrm{LW}$ is the Lyman-Werner intensity in units of $10^{-21}\,\si{erg.s^{-1}.cm^{-2}.Hz^{-1}.sr^{-1}}$ and $f_{sh}$ is the self-shielding parameter taken from \cite{WolcottGreen2019}, as a function of the gas density and temperature. The provided fitting function is:

\begin{equation}
\begin{aligned}
f_{\text {sh }}(N_{\ce{H2}}, T) & =\frac{0.965}{\left(1+x / b_5\right)^{\alpha(n, T)}}+\frac{0.035}{(1+x)^{0.5}} \\
& \times \exp \left[-8.5 \times 10^{-4}(1+x)^{0.5}\right]\;,\\
\alpha(n, T) &=A_1(T) e^{-0.2856 \log \left(\mathrm{n} / \mathrm{cm}^{-3}\right)}+A_2(T) \;,\\
A_1(T) &= 0.8711 \log(T/K) - 1.928\;,\\
A_2(T) &=-0.9639\log(T/K)+3.892\;,
\end{aligned}
\end{equation}
where $b_5 = (2 k_B T / m_{\ce{H2}})^{1/2} / 10^{5}\ \si{cm^2.s^{-1}}$ and $x=N_{\ce{H2}} / 5\times 10^{14}\ \si{cm^2}$. 

\begin{figure}
    \centering
    \begin{subfigure}
        \centering
        \includegraphics[width=0.46\textwidth]{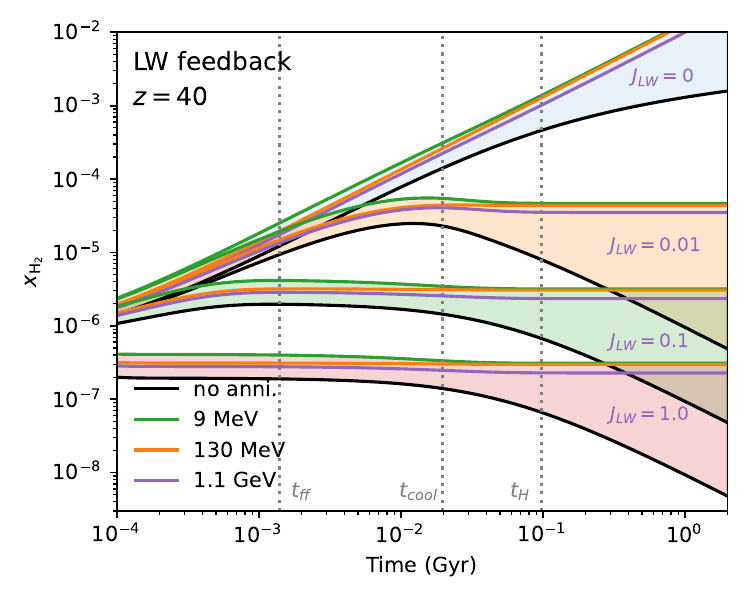}
    \end{subfigure}
    
    \begin{subfigure}
        \centering
        \includegraphics[width=0.46\textwidth]{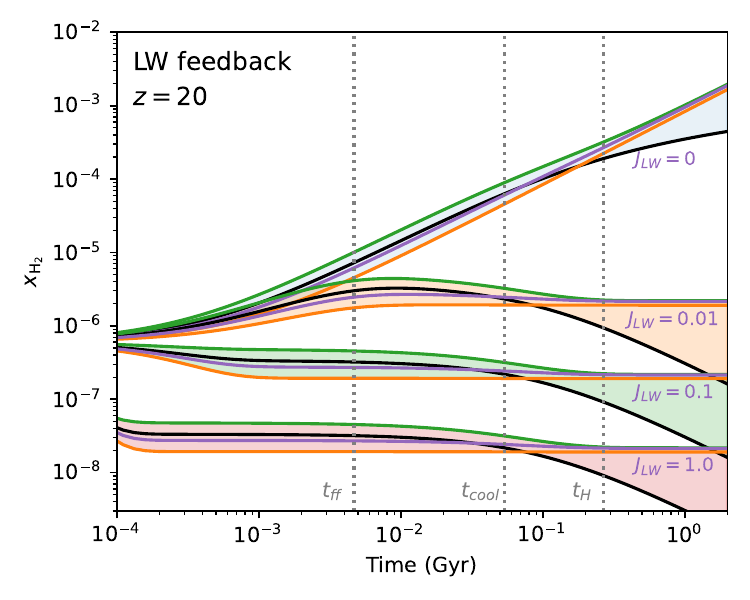}
    \end{subfigure}

    \caption{Molecular hydrogen fraction in a $10^5 \mathrm{M}_\odot$ halo with both DM annihilation and Lyman-Werner (LW) feedback. The \ce{H2} fractions at the redshift $z=40$ (top panel) and $z=20$ (bottom panel) are shown for different intensities of LW radiation (colored areas) and DM mass (colored lines). From top to bottom, the \ce{H2} fractions was calculated with LW intensity values $J_\mathrm{LW} = 0$, $0.01$, $0.1$, and $1.0$.}
    \label{fig:xH2_LW}
\end{figure}

Fig.~\ref{fig:xH2_LW} shows the production of the molecular hydrogen in a $10^5 \mathrm{M}_\odot$ dark matter halo, influenced by both Lyman-Werner feedback and DM annihilation. The \ce{H2} fraction is calculated using Equation~\ref{eqn:xH2_LW} for a range of LW intensities: $J_\mathrm{LW} = 0$, $0.01$, $0.1$,  and $1.0$. The initial fraction is $x_{\ce{H2}}=6\times10^{-7}$ for every case, as before. At each step, we estimate the \ce{H2} column density $N_{\ce{H2}} = 0.926 f_{\ce{H2}} n_\mathrm{core} R_\mathrm{core}$ according to \cite{Nebrin2023}, where $R_\mathrm{core}=0.1 R_\mathrm{vir}$ and $n_\mathrm{core}$ is the core gas density, and we calculate $k_\mathrm{LW}$ using Equation~\ref{eqn:k_LW}. 

In the top panel, at redshift $z=40$, although the \ce{H2} fraction has been suppressed by the LW photons, the effect of DM annihilation only slightly alters the \ce{H2} formation before the cooling time, $t<t_\mathrm{cool}$. While we have included a full range of LW values, the expected intensity at $z=40$ is very small, $J_\mathrm{LW} \approx 0.001$ \cite{Fialkov2013, Incatasciato2023}, so the effect of LW radiation is likely not significant. 
In the bottom panel at redshift $z=20$, where the LW intensity is expected to be stronger ($J_\mathrm{LW} \approx 1$), LW photons should efficiently dissociate \ce{H2} molecules. The \ce{H2} fraction decreases from $10^{-4}$ to $10^{-7}$ during the cooling phase, consistent with the findings of \cite{Kulkarni2021}. In the case with dark matter annihilation, the evolution is initially similar, but diverges from the baseline case after $t_\mathrm{cool}$, showing a larger \ce{H2} fraction compared to the case with LW radiation alone. 

\begin{figure}
    \centering
    \includegraphics[width=0.47\textwidth]{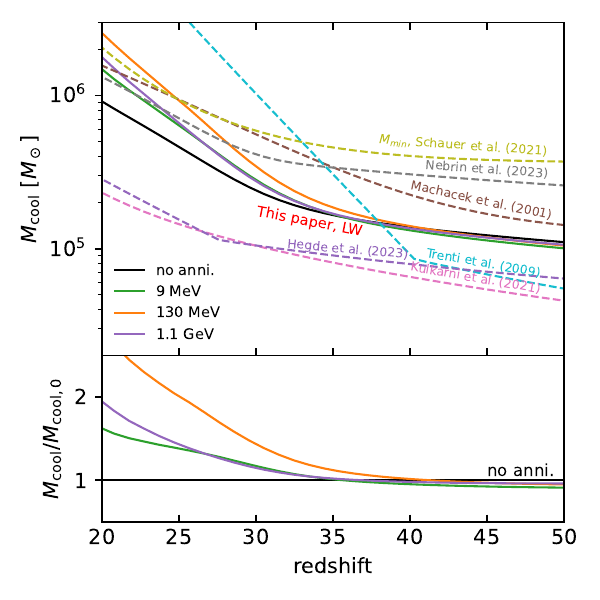}
    \caption{The minimum cooling mass $M_\mathrm{cool}$ as a function of redshift with both DM annihilation and LW feedback is included. Our results are compared to previous calculations from redshift $z=20$ to $50$. All cooling mass thresholds were calculated using the same LW background $J_\mathrm{LW}(z)$ as described in \cite{Incatasciato2023}. The bottom panel shows the relative effect of DM annihilation on the minimum cooling mass. At high redshift, where LW feedback is weak, the effect of DM annihilation is similar to the case before, but it becomes stronger at lower redshift as LW feedback becomes more significant. }
    \label{fig:Mcool_LW}
\end{figure}

To incorporate the stellar LW background in this study, we adopt the redshift-dependent LW intensity, $J_\mathrm{LW}(z)$, from \cite{Incatasciato2023}. This function is fitted for the range $6 < z < 23$, and we extrapolate it to higher redshifts. We acknowledge that the PopIII star formation model used in that work differs from our model, which may lead to inconsistencies in the LW feedback. However, it still provides a good approximation for evaluating the effect of DM annihilation under varying LW radiation intensities.

Fig.~\ref{fig:Mcool_LW} plots the minimum cooling mass with stellar LW feedback and dark matter annihilation. In the top panel, the solid black line represents the case with LW background alone. Compared to the case without LW radiation (as shown in Figure~\ref{fig:Mcool}), we find that the LW background causes minor changes at high redshift, but significantly increases the minimum cooling mass at lower redshift ($z<30$) for cases. We plot the result from previous studies in dashed lines. In our model, with LW feedback, the slope of $M_\mathrm{cool}$ at lower redshifts closely aligns with the analytic model of \cite{trenti2009formation}. We applied a self-shielding factor in Equation~\ref{eqn:xH2_LW}, resulting in a flatter slope that more closely matches simulation results at higher redshifts. The minimum cooling mass is sensitive to the choice of cooling time criterion (see Fig.~\ref{fig:xH2_LW}), which may explain why our slope is steeper than that of the analytic model with a shorter time criterion \cite{Nebrin2023} at low redshifts. We also include fitting functions from the simulations of \cite{Machacek2001} and \cite{Kulkarni2021}, as well as the redshift-independent result from \cite{Schauer2021}. Compared to our analytic results without DM annihilation (black line), the simulation results, particularly those of \cite{Kulkarni2021}, suggest that LW feedback has a less pronounced effect at lower redshifts.

In the bottom panel, we show how the influence of dark matter annihilation in the presence of a LW background varies with DM mass. Generally, DM annihilation very slightly decreases $M_\mathrm{cool}$ at high redshift, similar to the behavior seen in the case without LW feedback (as shown in Figure~\ref{fig:Mcool}). However, our result indicates that LW feedback magnifies the DM annihilation effect in molecular cooling, and the impact of DM annihilation increases over time, becoming slightly larger than in the case without LW feedback at lower redshifts.

We acknowledge that some radiative backgrounds, such as the cosmic X-ray background, are not included in this model. The cosmic X-ray background shares similar properties with DM annihilation, as both contribute to heating and ionizing primordial gas, which can either increase or decrease cooling. The transition between these two regimes occurs at gas densities between $n = 1$ and $100\ \si{cm^{-3}}$, depending on the strength of the X-ray background \cite{Hummel2015}. This effect has been included in recent molecular cooling studies and is expected to be important at lower redshifts ($z < 15$) \cite{Hegde2023}.

\section{\label{sec:simulation}Semi-Numerical Simulation}

We begin by calculating the corresponding energy injection rate from dark matter annihilation following Equation~\ref{eq.epsilon_c}. This model includes both smooth background annihilation and the enhanced annihilation resulting from collapsed structures (halos). To determine the thermal history of the IGM, we model the cosmic average thermal history in the public cosmological recombination code \texttt{CosmoRec}~\cite{Chluba2010} with energy injected from dark matter annihilation. We update the energy deposition fraction $f_c(z)$ of DM annihilation for different dark matter models using the pre-calculated table given by \cite{Slatyer2013} and calculate the gas temperature $T_g$ and ionization fraction $x_e$ from redshift $z=2700$ to $z=35$.

We calculate the star-forming baryon collapsed fraction $f_\mathrm{coll}$ in Equation~\ref{eq:fcoll}, as input for the public \texttt{21cmvFAST} code to estimate the star formation rate. In order to determine $f_\mathrm{coll}$, we calculate the Jeans mass $M_J$ using the IGM temperature given by \texttt{CosmoRec}. By considering the effects of dark matter annihilation on the thermal history, we can calculate the Jeans mass with dark matter annihilation. Then, we can obtain the gas fraction given by Equation~\ref{eq:fgas} as a function of redshift and halo mass in the presence of dark matter annihilation.

\begin{figure}
    \centering
    \includegraphics[width=0.47\textwidth]{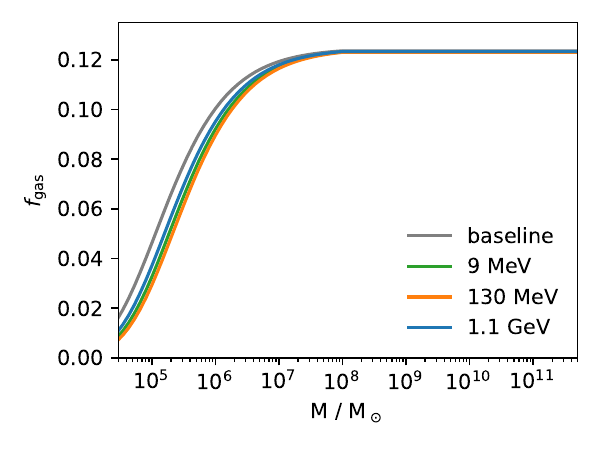}
    \caption{Gas fraction as a function of halo masses at redshift $z=20$, with the impact of dark matter (DM) annihilation for DM masses of $9$ MeV, $130$ MeV and $1.1$ GeV. The baseline (grey line) is given by Equation~\ref{eq:fgas} without DM annihilation. Dark matter annihilation leads to suppression in the gas fraction within low mass halos, but less in massive halos.}
    \label{fig:gas_frac_mass}
\end{figure}

Fig.~\ref{fig:gas_frac_mass} shows the gas fraction as a function of halo mass at redshift $z=20$, using the filtering mass defined in Equation~\ref{eq:fgas}. The gas thermal history is calculated in the recombination code \texttt{CosmoRec}. We observe a suppression in gas fraction, about 10\% to 40\%, at redshift $z=20$ for halo mass $M_h < 10^5 \mathrm{M}_\odot$, which varies with dark matter mass. The impact of dark matter annihilation on the gas fraction is less significant in halos of higher mass. For a halo of $10^6 \mathrm{M}_\odot$, the reduction in gas fraction is around $5\% \sim 10\%$. There is no significant difference for atomic cooling halos ($\mathrm{M}_h>10^8 \mathrm{M}_\odot$).

The differential brightness temperature of the high-redshift 21-cm signal is given by

\begin{equation}
    \delta T_b = \frac{T_S - T_\gamma}{1+z} (1-\exp^{-\tau_{\nu_0}})\;,
\end{equation}
where $T_\gamma$ is the CMB temperature, $\tau_{\nu_0}$ is the optical depth of the IGM for the 21cm signal, and $T_S$ is the spin temperature, given by

\begin{equation}
    T_S^{-1} = \frac{T_{\gamma}^{-1} + x_\alpha T_\alpha^{-1} + x_c T_k^{-1}}{1+x_c+x_\alpha}\;,
\end{equation}
where $T_k$ is the kinetic temperature of the gas, $T_\alpha$ is the color temperature, which is typically coupled to kinetic temperature, $T_\alpha \approx T_k$, $x_c$ is the collisional coupling coefficient, and $x_\alpha$ is the Wouthuysen-Field coupling coefficient. 

We use the public \texttt{21cmvFAST} semi-numerical code~\cite{Muoz2019} to calculate the spin temperature and sky-averaged 21cm brightness temperature. \texttt{21cmvFAST} modifies the \texttt{21cmFAST} code \cite{Mesinger2010}. It calculates star formation in molecular cooling halos and accounts for the effect of dark matter-baryon velocity offsets in the 21cm signal.

In this study, we modify \texttt{21cmvFAST} to implement dark matter annihilation at cosmic dawn. As described in \ref{sec:model}, we assume the dark matter particles annihilate 100\% into $\ce{e+e-}$, and the energy is deposited into heating, ionization and Lyman-$\alpha$ photons. The thermal evolution in the simulation is modified as described by Equations~\ref{Equ.evo_ion} and \ref{Equ.evo_lymana}. We employ the  energy deposition fractions $f_c(z)$ as used in \texttt{CosmoRec}. However, for redshifts $z < 25$, we rescale the deposition fraction using $f_c(z < 25) = \frac{f(z')}{f(z)} f_c(z')$, where $z'$ corresponds to the redshift where the free electron fraction at $x_e(z) = x_e(z')$. For star formation, we update the baryon collapsed fraction from Equation~\ref{eq:fcoll} to account the effect of dark matter annihilation on gas cooling and star formation. Following \citepalias{LopezHonorez2016}, we set the X-ray efficiency in \texttt{21cmvFAST} to be $\zeta_X = 10^{56}$, corresponding to the number of X-ray photons per solar mass in the simulation. We keep the default setting as described in \cite{Muoz2019} for all other options. For a discussion of the dark matter annihilation impact on the molecular cooling and streaming velocities, see Sec.~\ref{sec:H2cool} and \ref{sec:StreamingVelocity}. The initial conditions $T_e, X_e$ were taken from \texttt{CosmoRec}.

In this paper, all cosmological parameters in the calculation and simulations were adopted from Planck 2018, with $h=0.6766$, $\Omega_{m, 0}=0.3111$, and $\Omega_{b, 0} = 0.049$~\cite{aghanim2020planck}.

\section{\label{sec:discuss}Result}

\subsection{\label{sec:StreamingVelocity}DM-Baryon Velocity Offsets}

Dark matter-baryon velocity offsets play a significant role in early structure formation~\cite{tseliakhovich2010relative}. Recent studies have explored the impact of streaming velocity on the formation of dark matter structures, the gas content in small halos, the formation of the first stars, and the resulting implications for the 21cm signal during Cosmic Dawn~\cite{Dalal2010, mcquinn2012impact, Muoz2019}.

The influence of a nonzero streaming velocity on the 21cm signal, particularly during the Cosmic Dawn, manifests in three key ways:

\begin{enumerate}
    \item Suppression of the small-scale power spectrum, reducing the abundance of halos $dN/dM$ within the density field~\cite{tseliakhovich2010relative, Tseliakhovich2011}. 
    \item An increase of the filtering mass $M_\mathrm{F}$ and decrease of the gas fraction $f_\mathrm{gas}$ within small halos~\cite{Dalal2010, Fialkov2012, Naoz2012}.
    \item An increase of the cooling mass $M_\mathrm{cool}$, thereby influencing early star formation, eg.~\cite{Greif2011, Fialkov2012, Schauer2019, Kulkarni2021}.
\end{enumerate}

The equation for relative velocities can be written as \cite{tseliakhovich2010relative}

\begin{equation}
    v_{bc} = \frac{\hat{k}}{ik}[\theta_b-\theta_c]\;.
\end{equation}

where $\hat{k}$ is the unit vector in the direction k, and $\theta$ is the velocity divergence.

The variance of relative speed can be expressed as

\begin{equation}\label{vbc}
    \langle v^2_{bc} \rangle = \int\frac{dk}{k}\Delta^2_{vbc}(k)\;,
\end{equation}
where $\Delta^2_{vbc}(k)$ is a function of wavenumber $k$, which drops rapidly at $k>0.5 \si{Mpc^{-1}}$~\cite{tseliakhovich2010relative}.
At the recombination epoch ($z_{rec}=1020$), the relative speed is estimated to be $v_\mathrm{rms} = 30~\si{km/s}$; after baryons recombine and are no longer tied to the photons, their sound speed drops to $6~\si{km/s}$~\cite{Tseliakhovich2011}.

First, we calculate the dark matter annihilation power considering the impact of streaming. Equation~\ref{eq:dE_dVdt} shows that at lower redshifts, the average dark matter annihilation power is primarily due to contributions from collapsed structures. To account for the impact of streaming velocity, we use the modified power spectrum from perturbation theory with streaming included, as described in~\cite{tseliakhovich2010relative}, to calculate the halo mass function $dN/dM(M, z, v_{bc})$. Then, we apply the halo mass function in Equation~\ref{eq:boost_factor} to estimate the boost factor for dark matter annihilation. Because the streaming suppresses small-scale structure formation, this modification results in a reduction in the boost factor, especially at higher redshifts. Specifically, for halos following the NFW profile with a minimum mass of $10^{-9}\ \mathrm{M}_\odot$, with a streaming velocity of $v_{bc} = v_\mathrm{rms}(z)$, where $v_\mathrm{rms}(z)$ represents the expected root-mean-square of the streaming velocity at redshift $z$, the boost factor decreases by approximately 10\% at $z=40$ and by 4\% at $z=20$.

The gas fraction, considering the effect of streaming, can be computed using a Jeans analysis. The \textit{effective} sound speed, taking into account the streaming velocity $v_{bc}$, is given by ${c'}_{s}^2 = c_{s}^2 + v_{bc}^2$~\cite{Dalal2010, Fialkov2012}. This velocity term increases the filtering mass $M_\mathrm{F}$, thereby suppressing the gas fraction. In the presence of streaming, Equation~\ref{eq:fgas} becomes

\begin{equation}\label{eqn:f_gas_v}
    f_\mathrm{gas}(z, M, v_{bc})=f_{\mathrm{b}, 0}\left[1+\left(2^{\alpha / 3}-1\right)\left(\frac{M_\mathrm{F}(z, v_{bc})}{M_h}\right)^\alpha\right]^{-3 / \alpha}\;,
\end{equation}
where $f_{\mathrm{b}, 0}$ is the cosmic baryon fraction, and $M_\mathrm{F}(z, v_{bc})$ is the filtering mass as a function of streaming velocity $v_{bc}$ and redshift $z$.

\begin{figure}
    \centering
    \includegraphics[width=0.47\textwidth]{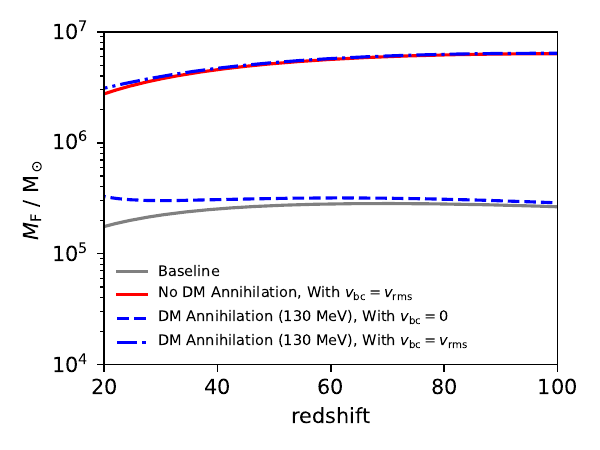}
    \caption{Filtering mass as a function of redshift with effects of dark matter (DM) annihilation and streaming velocity ($v_{bc}$). The baseline scenario, depicted by a solid grey line, assumes neither DM annihilation nor streaming velocity. The impact of streaming velocity alone is shown with a solid red line. The effects of DM annihilation of 130MeV DM particles without streaming velocity are represented by dashed lines. Dot-dashed lines indicate scenarios combining DM annihilation with streaming velocity $v_{bc} = v_\mathrm{rms}$.}
    \label{fig:FilteringMass_v}
\end{figure}

Fig.~\ref{fig:FilteringMass_v} shows the filtering mass $M_\mathrm{F}(z, v_{bc})$ with both dark matter annihilation and streaming. We set the streaming velocity $v_{bc} = v_\mathrm{rms}(z)$ and use the gas temperature from \texttt{CosmoRec} with 130 MeV dark matter. The result indicates that streaming velocity could significantly increase the filtering mass during Cosmic Dawn. We find that the increase in filtering mass due to gas heating from DM annihilation is overshadowed by the effects of streaming. The dot-dashed line shows that DM annihilation leads to less than a 10\% increase in the filtering mass at redshift $z=20$, compared to the 40\% increase observed in scenarios without streaming.

Streaming can prevent gas accretion into small halos, thereby increasing the cooling threshold for halo formation. Previous work by \cite{Fialkov2012} showed that gas cooling occurs only in halos with a circular velocity above $V_{\mathrm{cool}}$ in the presence of streaming, which is given by

\begin{equation}\label{eqn:V_cool}
    V_{\mathrm{cool}}(z)^2 = V_{\mathrm{cool}, 0}^2 + \left[\alpha v_{bc}(z)\right]^2\;,
\end{equation}

where $V_{\mathrm{cool}, 0}$ is the baseline cooling velocity when streaming is not considered ($v_{bc} = 0$), and $\alpha$ represents the scaling factor for the impact of streaming. According to the simulation fit, $V_{\mathrm{cool}, 0} = 3.714\ \si{km\ s^{-1}}$ and $\alpha = 4.015$.

The relation between $V_{\mathrm{cool}}$ and minimum cooling mass is given by~\cite{Barkana2001}

\begin{equation}\label{eqn:Mcool_v}
    M_\mathrm{cool}(z) \approx 7\times 10^6 \left(\frac{V_{\mathrm{cool}}(z)}{10\, \si{km.s^{-1}}}\right)^3 \left(\frac{1+z}{20} \right)^{-3/2}\ \mathrm{M}_\odot\;.
\end{equation}
This results in a minimum cooling mass of $M_\mathrm{cool} \approx 6 \times 10^5 \mathrm{M}_\odot$ at redshift $z=20$ when $v_{bc} = v_\mathrm{rms}$ compared to $M_\mathrm{cool} \approx 3.5 \times 10^5 \mathrm{M}_\odot$ when $v_{bc} = 0$.

Now, we can incorporate streaming into the molecular cooling model. First, we consider the effect of streaming on the gas density. Using Equation~\ref{eqn:f_gas_v}, we calculate the gas fraction, $f_\mathrm{gas}$, with streaming included. We modify the IGM temperature in the presence of streaming, following the prescription in~\cite{mcquinn2012impact}:

\begin{equation}
    T_\mathrm{IGM}(v_{bc}) =  T_{\mathrm{IGM}} (1 + 5 \mathcal{M}_{bc}^2 / 9)\;,
\end{equation}
where $\mathcal{M}_{bc} = v_{bc} / c_{s, \mathrm{IGM}}$ and $c_{s, \mathrm{IGM}}$ is the sound speed in the IGM. The increase in temperature results in a suppression of the gas density according to Equation~\ref{eqn:rho_core_LM}.

Subsequently, we recalculate the electron and molecular hydrogen fractions, and recalculate the cooling time as described in Sec.~\ref{sec:H2cool}.  The minimum cooling mass $M_\mathrm{cool}$ is now expressed as $M_\mathrm{cool}(z, v_{bc})$ as a function of streaming velocity $v_{bc}$. 

\begin{figure*}
    \centering
    \includegraphics[width=0.9\textwidth]{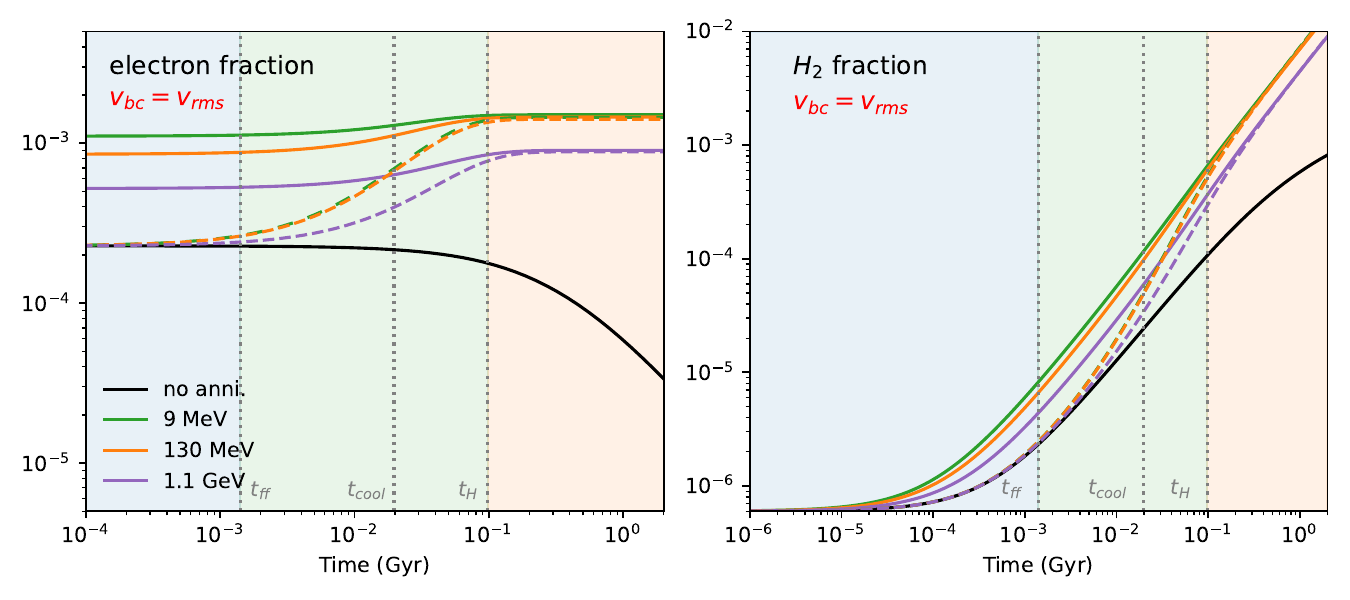}
    \caption{Fractions of electrons and molecular hydrogen as a function of time during gas cooling. The halo has a mass $10^5 M_\odot$ at redshift $z = 40$. This figure accounts for both streaming velocity and dark matter annihilation, with the streaming velocity $v_{bc}$ set to the root-mean-squared value $v_\mathrm{rms}$, which leads to a decrease in gas density. Dashed lines represent scenarios where the same initial condition of gas at $t=0$ is assumed, with DM annihilation ionizing the gas thereafter.}
    \label{fig:cooling_fraction_v}
\end{figure*}

Fig.~\ref{fig:cooling_fraction_v} depicts the fraction of free electrons $x_e$ and molecular hydrogen $x_{\ce{H2}}$ in the presence of a streaming velocity $v_{bc} = v_\mathrm{rms}$. We observe that the electron fraction decreases more slowly compared to the case with $v_{bc} = 0$ (Fig.~\ref{fig:cooling_frac}), since the recombination rate is higher in denser regions. Dark matter (DM) ionization becomes more efficient with streaming, resulting in the electron fraction being dominated by DM ionization. Consequently, $x_e$ becomes insensitive to the initial conditions after a sufficiently long time, $t > t_H$. The molecular hydrogen fraction, $x_{\ce{H2}}$, is relatively lower than that in the absence of streaming. However, we observe that DM annihilation in this scenario accelerates \ce{H2} production due to the increase in the electron fraction.

\begin{figure}
    \centering
    \includegraphics[width=0.47\textwidth]{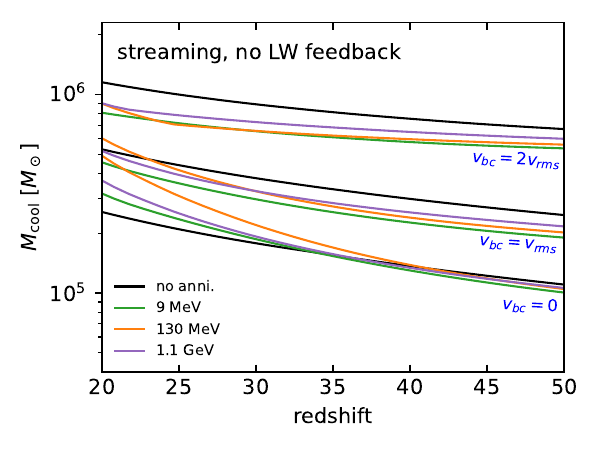}
    \caption{Minimum cooling mass as a function of redshift, considering the effects of dark matter annihilation and varying streaming velocities. The black lines (from top to bottom) represent the minimum cooling mass for streaming velocities of $v_{bc} = 2 v_\mathrm{rms}$, $v_{bc} = v_\mathrm{rms}$, and $v_{bc} = 0$ in our model. The colored lines indicate the results with dark matter annihilation of mass 9 MeV, 130 MeV and 1.1 GeV incorporated at different streaming velocities.}
    \label{fig:Mcool_v}
\end{figure}

Fig.~\ref{fig:Mcool_v} shows the minimum cooling mass $M_\mathrm{cool}(z)$ in the presence of both streaming and DM annihilation. The reduction in the gas fraction decreases the efficiency of molecular hydrogen production, leading to an increase in the minimum cooling mass by a factor of approximately 2. 
Dark matter annihilation shows a distinct effect in the case of streaming velocity. Without streaming, annihilation leads to a decrease in $M_\mathrm{cool}$ at high redshift, but an increase at low redshift. However, we found that annihilation primarily reduces $M_\mathrm{cool}$ in most cases when $v_{bc} = v_\mathrm{rms}$ and a reduction across nearly all redshifts at $v_{bc} = 2 v_\mathrm{rms}$. This is primarily due to streaming suppressing the gas density and molecular cooling, while the \ce{H2} production increases under DM-induced ionization. As a result, the effect of DM heating is not significant.

Without dark matter annihilation, our result indicates that the minimum cooling mass is increased by a factor of approximately $M_\mathrm{cool}(v_\mathrm{rms})/M_\mathrm{cool}(0) \approx 2.1$ when $v_{bc} = v_\mathrm{rms}$ at redshift $z=20$. This factor is slightly higher than the fitting result of $1.7$ given by~\cite{Fialkov2012} in Equation~\ref{eqn:Mcool_v}, but lower than the results from more recent simulations. For instance,~\cite{Kulkarni2021} reports a factor of $M_\mathrm{crit}(v_\mathrm{rms})/M_\mathrm{crit}(0) \approx 3.5$, and~\cite{Schauer2021} finds $M_\mathrm{min}(v_\mathrm{rms})/M_\mathrm{min}(0) \approx 4.1$.

\begin{figure}
    \centering
    \includegraphics[width=0.47\textwidth]{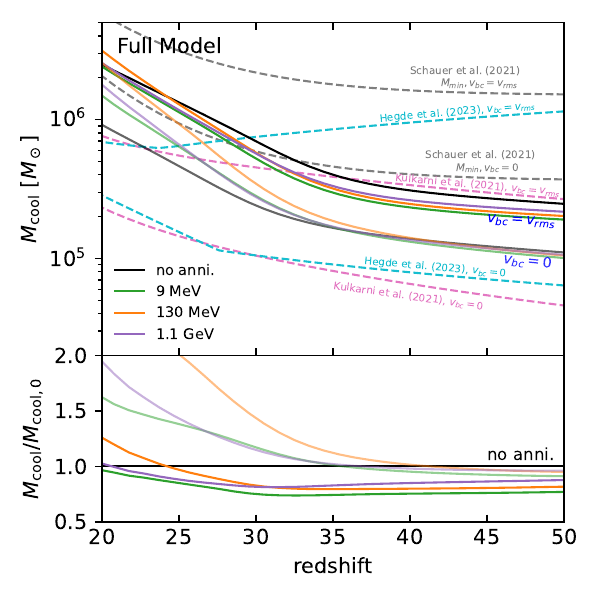}
    \caption{Minimum cooling mass in full scenario, include dark matter annihilation, Lyman-Werner feedback, and streaming velocity. The top panel plots $M_\mathrm{cool}$ as a function of redshift. The black line represents the scenario without dark matter annihilation. Different colors represent dark matter annihilation for various dark matter masses (9 MeV, 130 MeV, and 1.1 GeV). Two distinct scenarios are presented in the figure: one with streaming velocity ($v_{bc} = v_\mathrm{rms}$) and one without ($v_{bc}=0$). These results are compared with those from previous simulations~\cite{Kulkarni2021, Schauer2021}. The bottom panel plots the relative cooling mass compared to the scenario without dark matter annihilation as a function of redshift. Dark matter annihilation has a relatively greater impact in regions without streaming velocity than in those with streaming velocity.}
    \label{fig:Mcool_compare_full}
\end{figure}

\begin{figure}
    \centering
    \begin{subfigure}
        \centering
        \includegraphics[width=0.47\textwidth]{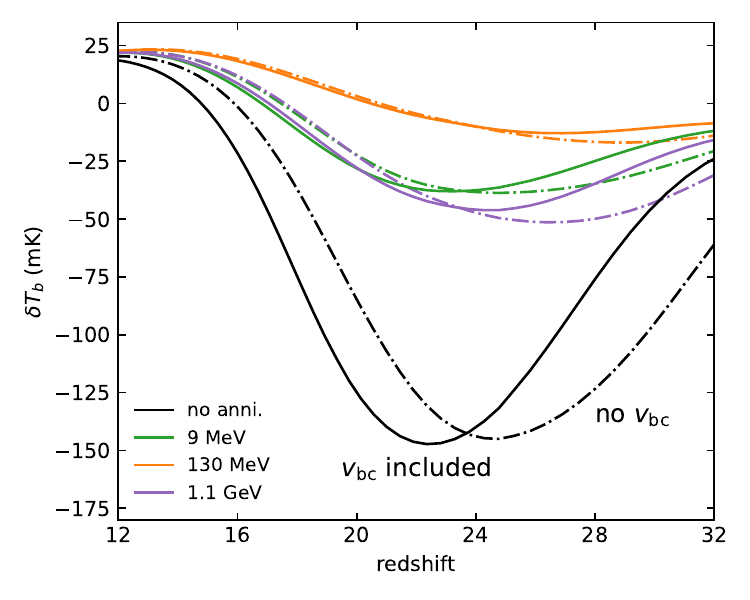}
    \end{subfigure}

    \begin{subfigure}
        \centering
        \includegraphics[width=0.47\textwidth]{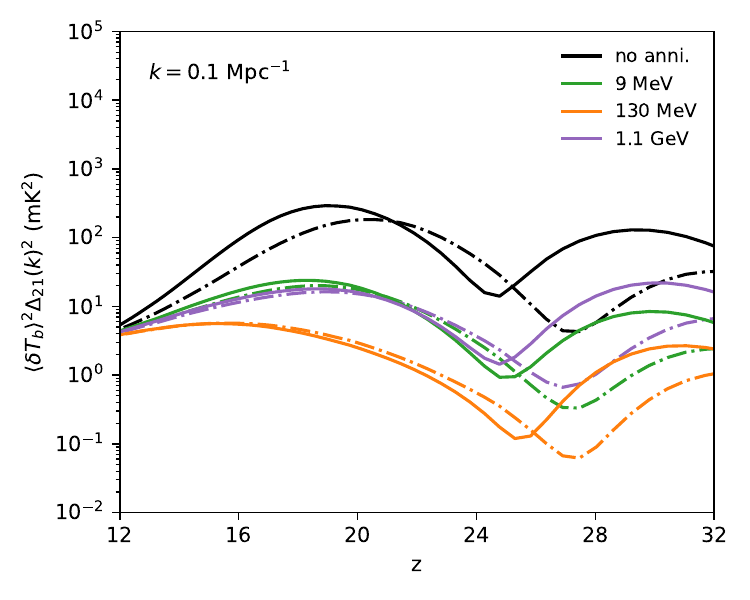}
    \end{subfigure}
    
    \caption{The simulated 21cm signal as a function of redshift for the no-annihilation case and dark matter masses of 9 MeV, 130 MeV, and 1.1 GeV. The top panel shows the global 21cm signal. The bottom panel shows the 21cm power spectrum at a scale of $k=0.1$ Mpc$^{-1}$. Solid lines show our results with both dark matter annihilation and streaming velocity effects, while dot-dashed lines represent cases without the streaming velocity effect.}
    \label{fig:dTb_vbc}
\end{figure}

Fig.~\ref{fig:Mcool_compare_full} illustrates the complete scenario for the minimum cooling mass, including dark matter annihilation, stellar Lyman-Werner feedback, and streaming. Standard Lyman-Werner feedback, as described by~\cite{Incatasciato2023}, was applied in the calculation. We acknowledge significant discrepancies exist in the predicted minimum cooling masses across recent studies as shown in Fig.~\ref{fig:dTb_vbc}. For instance, the simulation by \cite{Schauer2021} reports a significantly larger mass threshold compared to \cite{Kulkarni2021}. The reason for this divergence remains unclear, but may result from different definitions or differences in the calculation of self-shielding factor, as noted in \cite{Hegde2023}. We use the recent self-shielding function described by~\cite{WolcottGreen2019}, also used in \cite{Kulkarni2021, Hegde2023}, which leads our masses to a similar slope. We also present the analytic model from \cite{Hegde2023} in the figure, which takes a similar approach to that in our model. One key distinction arises in \cite{Hegde2023}, which imposes the requirement of minimum cooling mass $M_\mathrm{min} \geq M_\mathrm{F}$, where $M_\mathrm{F}$ is the filtering mass. Our model relaxes this constraint, by instead linking the filtering mass to the gas fraction. This leads us to a minimum cooling mass that is  significantly smaller at high redshifts with streaming, compared to \cite{Hegde2023}. Other discrepancies  could also arise from our choice of cooling time, the use of a simplified molecular hydrogen chemistry model,
or the incomplete treatment of the gas density profile during cooling in the analytic model. These limitations highlight the need for future studies with more physically complete models.

Regarding streaming, we point out that our model reveals two distinct impacts of dark matter annihilation, depending on whether streaming is present or absent. This is because the effects of dark matter heating and ionization depend on gas density, and the core gas density is significantly suppressed by streaming velocity. As shown in the bottom panel, dark matter annihilation has a suppression on molecular cooling, thereby increase the cooling mass without streaming at redshift $z\lesssim 30$, but this suppression is reduced with streaming, making the cooling mass remain lower than the case without annihilation. In our model, streaming velocity alters gas density by affecting the effective sound speed and filtering mass, a method also used in previous analyses such as~\cite{Naoz2012}. However, instead of estimating the impact of streaming on the minimum cooling mass through the halo of circular velocity $V_\mathrm{cool}$ fitted from simulations~\cite{Fialkov2012}, we directly derive the minimum cooling mass with streaming in our analytic model by adjusting the gas density. This approach allows us to account for potential interactions between DM annihilation, LW feedback and streaming within a consistent framework.

The top panel of Fig.~\ref{fig:dTb_vbc} shows the global 21cm signal as a function of redshift. The solid lines represent the full result with both dark matter annihilation and streaming velocity effects. The black line represents the no-annihilation case from our model. Dark matter annihilation significantly shallow the temperature trough, while streaming causes a slight redshift offset in the brightness temperature by $\Delta z \sim 2$ across all annihilation scenarios due to suppressed star formation.

\color{black} The bottom panel of Fig.~\ref{fig:dTb_vbc} shows the 21cm power spectrum as a function of redshift at a scale $k=0.1 \si{Mpc^{-1}}$ with the same settings as in the global 21cm signal. We identified two peaks due to Lyman-$\alpha$ coupling and X-ray heating. As expected, including the streaming velocity delays the formation of the structure, thereby postponing the  onset of both Lyman-$\alpha$ coupling and X-ray heating. Moreover, the spatial fluctuations induced by the streaming velocity enhance the overall amplitude of the 21cm signal fluctuations, which is also shown in previous results \cite{Muoz2019, Muoz2022}.

\subsection{\label{sec:disscuss.compare}Comparison to previous works}

As discussed in Sec.~\ref{sec:previous}, the primary aim of this work is to estimate the effect of dark matter annihilation on the 21cm signal, considering realistic expectations for dark matter-baryon velocity offsets. We study the annihilation $\chi \chi \to e^{-} e^{+}$ with three DM masses: 9 MeV, 130 MeV and 1.1 GeV. In our analytic cooling model, we discuss the effect of LW feedback and streaming.

The study by \citetalias{LopezHonorez2016} investigates the 21cm signal in the context of background dark matter annihilation using \texttt{21cmFAST}. Both \citetalias{LopezHonorez2016}'s study and ours calculate the thermal history of the IGM and the effect on 21cm signal in presence of the DM annihilation. However, \citetalias{LopezHonorez2016} does not account for molecular cooling, instead, setting the minimum virial temperature between \(10^4\) and \(10^5\) K. The atomic cooling-only model results in a significant delay in early galaxy formation relative to a model with molecular cooling, as evident in the position of the absorption trough in the 21cm signal. 

\begin{figure}
    \centering
    \includegraphics[width=0.46\textwidth]{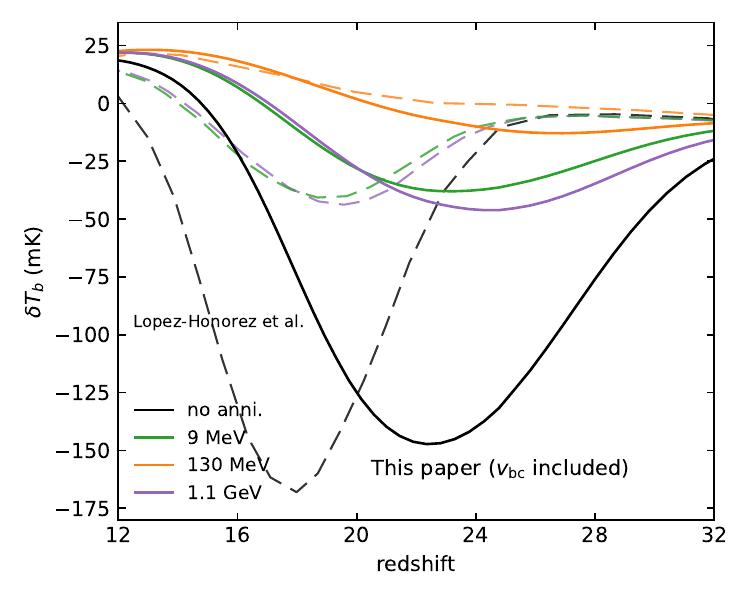}
    \caption{Comparison of the 21cm brightness temperature between our study and the previous results from \citetalias{LopezHonorez2016}. Long dashed lines plot the previous results without molecular cooling and streaming velocity. Our results are depicted with solid lines, incorporating the effects of dark matter annihilation and streaming velocity.}
    \label{fig:dTb_Lopez}
\end{figure}

Figure \ref{fig:dTb_Lopez} presents our complete results with streaming velocity (solid lines) compared to those from \citetalias{LopezHonorez2016} (dashed lines) with same parameter settings. For DM mass 9 MeV, 130 MeV and 1.1 GeV, our works find similar absorption troughs, primarily due to the same thermal history of IGM. We also study the DM annihilation in molecular cooling halos, and obtain a mass-dependent cooling mass. The increase of the minimum cooling mass due to dark matter annihilation results in a delay in star formation and therefore the 21cm signal. However, the effect is not significant in the global 21cm signal, as it is overshadowed by the effect in the IGM, which greatly boosts the differential brightness temperature. Additionally, the effect of streaming results in a slight offset in the brightness temperature as discussed. When both effects are combined, the redshift shifts by $\Delta z \approx 5$ for all dark matter masses compared to the previous results.

\begin{figure}
    \centering
    \includegraphics[width=0.45\textwidth]{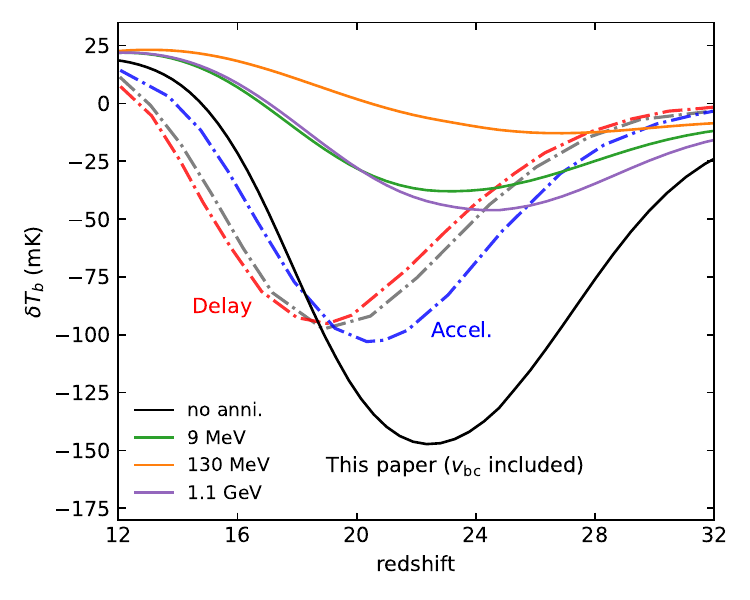}
    \caption{Comparison of the 21cm brightness temperature between our study and~\cite{Qin2023}. Solid lines represent our result with both dark matter annihilation and streaming velocity effects. Dot-dashed lines from \cite{Qin2023} show the baseline case without exotic injections (gray) and illustrate how the exotic energy accelerate (blue) and delay (red) the star formation under different models.}
    \label{fig:21cm_compare_Qin}
\end{figure}

The study by~\cite{Qin2023} examined the formation of the first stars under the influence of exotic energy injection from dark matter decay or annihilation, using the \texttt{DarkHistory} code~\cite{Liu2020}. They employed the \texttt{Zeus21} code~\cite{Muoz2023} with a modified star formation model in molecular cooling halos to calculate the 21cm signal. A comparison of our results is presented in Fig.~\ref{fig:21cm_compare_Qin}. Our findings indicate that, as in the study by~\cite{Qin2023}, molecular cooling in small halos shifts the 21cm absorption signal towards higher redshift. The authors of \cite{Qin2023} studied two dark matter models, each characterized by a particle mass of 185 MeV, decaying to $e^+ e^-$ pairs. The model with lifetime $\tau=25.6$ raises the mass threshold and the 21cm signal is slightly delayed, while the model with lifetime $\tau=26.4$ accelerates the 21cm signal. Their work focused on the timing of Cosmic Dawn through the enhancement or suppression of \ce{H2} from exotic energy injection. As such, they did not model other sources of feedback, e.g. stellar LW emission, DM-baryon relative velocities, or their combination, or IGM  heating. In contrast, our study includes the effects of IGM heating and ionization, while considering regular stellar LW feedback (indirectly influenced by dark matter annihilation and streaming). However, we do not account for direct LW energy deposition from dark matter annihilation. Another notable difference in our work is the star formation efficiency $f_\star$ for molecular cooling halos. While their work adopts the model from~\cite{Muoz2023}, which assumes star-formation efficiency $f_\star=10^{-2.5}$; our choices follows \citepalias{LopezHonorez2016}, where a high star-formation efficiency $f_\star=10^{-1}$ was used. This selection allows for a direct comparison with \citepalias{LopezHonorez2016} and ensures consistency within our modeling framework. We acknowledge that this value could lead to a pronounced earlier onset due to star formation in molecular cooling halos. Therefore, even with LW feedback and streaming considered, our 21cm signal still appears earlier than in their results.

We note that our predicted 21cm power spectrum appears at earlier redshifts compared to previous studies (e.g., \citepalias{LopezHonorez2016}). This earlier onset is consistent with the trend seen in our global signal analysis. The differences can be primarily attributed to the star formation in the molecular cooling halos.

\section{\label{sec:observable}Observational Prospects}

The observation of the global 21cm signal is challenging due to a Galacticforeground thousands of times stronger than the weak cosmological background signal. However, several observational programs have been developed to detect this signal.

One is the Experiment to Detect the Global EoR Signature(EDGES,~\cite{Monsalve2017}), an instrument located in Western Australia. Recently, the project reported an unexpectedly large absorption trough~\cite{Barkana2018} around 78 MHz, which, if validated, could provide strong constraints on dark matter annihilation during Cosmic Dawn~\cite{Liu2018}. However, this signal anomaly is still under debate and needs confirmation. Other experiments include the Probing Radio Intensity at high-Z from Marion(PRI$^\mathrm{Z}$M,~\cite{Philip2019}), Shaped Antenna measurement of the background RAdio Spectrum(SARAS,~\cite{Singh2017}), and Radio Experiment for the Analysis of Cosmic Hydrogen(REACH,~\cite{deLeraAcedo2019}). So far, these projects have not confirmed the EDGES signal.

Probes of high-redshift galaxies such as the Hubble Space Telescope (HST) and James Webb Space Telescope (JWST) also provide essential information in studying the high redshift universe near Cosmic Dawn. The detection of the high-z UV radiation background could have significant implications on the global 21cm signal~\cite{Mirocha2016}. Moreover, the molecular hydrogen \ce{H2} lines from the PopIII star formation in primordial gas could potentially be observable by JWST~\cite{Mizusawa2005}, providing insights into early star formation.

\section{\label{sec:conclusion}Discussion}

This study investigates the impact of dark matter (DM) annihilation during the Cosmic Dawn, focusing on sub-GeV dark matter particles annihilating into $e^+ e^-$. Our analysis shows that energy from DM annihilation impacts heat, ionization, and excitation in the cosmic gas, leading to an increase in both gas temperature and ionization fraction during this epoch.

We identified a strong dependence of baryon collapse on the thermal history, via the filtering mass. In mini-halos, where gravitational forces are weaker, heating from DM annihilation can increases the gas temperature and suppress gas collapse. Though the filtering mass is less affected due to weak annihilation effects at earlier times, we found a DM mass-dependent gas fraction of mini-halos at Cosmic Dawn.

We developed an analytic cooling model that incorporates DM annihilation. We calculate the production and the required fraction of molecular hydrogen for different redshifts and DM masses, showing both positive and negative impacts on molecular cooling. At high redshift $z>30$, DM ionization increases the electron fraction, slightly lowering the minimum cooling mass. However, at lower redshifts ($z < 30$), DM heating dominates, raising the minimum cooling mass. The impact on the cooling mass with redshift is influenced by the DM mass, which determines the fraction of energy deposited in each channel.

We also account the influence of photo-dissociation from Lyman-Werner (LW) photons. In the presence of stellar LW feedback, the molecular hydrogen fraction is significantly reduced, which increases the minimum cooling mass at lower redshifts across all considered DM masses.

We extend our analysis to account for the effects of dark matter-baryon velocity offsets (streaming) driven by baryon acoustic oscillations to better understand the 21cm signal and early star formation during Cosmic Dawn. Streaming suppresses small-scale halo formation, reducing the abundance of halos -- particularly those near the Jeans scale. This reduction leads to a slight decrease in the DM annihilation boost factor.

Our study also examined the combined effect of DM annihilation and streaming on Population III star formation. The streaming significantly reduces the gas content in small halos, directly inhibiting star formation and altering the environment for molecular cooling. In our model, the minimum cooling mass increases as a function of the streaming velocity, and DM annihilation exhibits distinct features for different cases of streaming velocities. It shows the strongest suppression of molecular cooling in the absence of streaming, but may lead to an acceleration of cooling in the case of a large streaming velocity at the same redshift.

We calculated the 21cm signal using simulations from the \texttt{21cmvFAST} code, incorporating the cooling mass results from our molecular cooling model. The inclusion of star formation in molecular cooling halos shifts the global 21cm signal to earlier redshifts compared to previous studies. However, the precise shift in redshift is highly sensitive to the choice of parameters, such as the cooling criterion in the analytic model and the star formation efficiency of molecular cooling halos. When we include streaming  in our calculations, we observe a slight delay in the global 21cm signal, due to the suppression of the gas fraction and molecular cooling.

\section{\label{sec:future}Future Work}

The primary aim of this work is to investigate the interactions between dark matter annihilation, dark matter-baryon velocity offsets, and molecular cooling, and their impact on early structure formation and the 21cm signal. This work attempts to bridge the gap between previous studies by incorporating multiple effects, offering a more comprehensive model of how these factors interplay.

We acknowledge that the modeling presented here is complex; however, it is essential to consider these interacting processes to achieve a more complete understanding of early cosmology. Dark matter annihilation can either enhance or suppress molecular cooling, leading to either an increase or decrease in the formation of the first stars. This effect can be further amplified by Lyman-Werner (LW) feedback and modified by the intensity of streaming velocities. Additionally, incorporating molecular cooling into the calculation of the 21cm signal can significantly shift the signal to earlier redshifts. The interplay among these effects is non-trivial, and understanding them together is crucial for making more accurate predictions.

One important avenue for future work is improving the estimates for energy deposition from dark matter annihilation. On small scales, dark matter annihilation within halos can locally deposit energy into the gas, raising its temperature and increasing the ionization fraction. In the Appendix, we outline initial steps toward estimating these local effects. The gas temperature and ionization, which are raised by background dark matter annihilation power, can be further enhanced by local energy deposition within halos. Accurate estimates of the deposition fraction from specific annihilation products could be obtained through numerical simulations, which would include a detailed treatment of particle interactions with the primordial gas in profiled halos. We are currently developing such a program, which we expect will enable us to draw robust conclusions about the impact of dark matter annihilation on structure formation.

Looking ahead, a key goal will be to trace the influence of dark matter annihilation and other small-scale factors through to observable galaxy formation. By doing so, we can apply redshift- and scale-dependent modifications to models of galaxy formation. Future studies should focus on connecting the detailed physics of small scales with observable large-scale phenomena, such as galaxy formation and evolution and the evolution of the IGM, which can provide testable predictions for current and near-future observations and potentially constrain the nature of dark matter.

\begin{acknowledgments}
We thank James Taylor, Wenzer Qin, James Gurian, Aaron Vincent, Julian Munoz, Sarah Schon, and Neal Dalal for helpful discussions. We thank Weikang Lin for the discussions at the early stages of this work. This material is based upon work supported by the National Science Foundation under Grant No. 2108931. The authors acknowledge the use of high-performance computing (HPC) resources provided by North Carolina State University. We also thank the Digital Research Alliance of Canada for additional computational support.

\end{acknowledgments}

\bibliography{main}

\begin{thebibliography}{86}%
\makeatletter
\providecommand \@ifxundefined [1]{%
 \@ifx{#1\undefined}
}%
\providecommand \@ifnum [1]{%
 \ifnum #1\expandafter \@firstoftwo
 \else \expandafter \@secondoftwo
 \fi
}%
\providecommand \@ifx [1]{%
 \ifx #1\expandafter \@firstoftwo
 \else \expandafter \@secondoftwo
 \fi
}%
\providecommand \natexlab [1]{#1}%
\providecommand \enquote  [1]{``#1''}%
\providecommand \bibnamefont  [1]{#1}%
\providecommand \bibfnamefont [1]{#1}%
\providecommand \citenamefont [1]{#1}%
\providecommand \href@noop [0]{\@secondoftwo}%
\providecommand \href [0]{\begingroup \@sanitize@url \@href}%
\providecommand \@href[1]{\@@startlink{#1}\@@href}%
\providecommand \@@href[1]{\endgroup#1\@@endlink}%
\providecommand \@sanitize@url [0]{\catcode `\\12\catcode `\$12\catcode `\&12\catcode `\#12\catcode `\^12\catcode `\_12\catcode `\%12\relax}%
\providecommand \@@startlink[1]{}%
\providecommand \@@endlink[0]{}%
\providecommand \url  [0]{\begingroup\@sanitize@url \@url }%
\providecommand \@url [1]{\endgroup\@href {#1}{\urlprefix }}%
\providecommand \urlprefix  [0]{URL }%
\providecommand \Eprint [0]{\href }%
\providecommand \doibase [0]{https://doi.org/}%
\providecommand \selectlanguage [0]{\@gobble}%
\providecommand \bibinfo  [0]{\@secondoftwo}%
\providecommand \bibfield  [0]{\@secondoftwo}%
\providecommand \translation [1]{[#1]}%
\providecommand \BibitemOpen [0]{}%
\providecommand \bibitemStop [0]{}%
\providecommand \bibitemNoStop [0]{.\EOS\space}%
\providecommand \EOS [0]{\spacefactor3000\relax}%
\providecommand \BibitemShut  [1]{\csname bibitem#1\endcsname}%
\let\auto@bib@innerbib\@empty
\bibitem [{\citenamefont {Hooper}\ \emph {et~al.}(2013)\citenamefont {Hooper}, \citenamefont {Kelso},\ and\ \citenamefont {Queiroz}}]{Hooper2013}%
  \BibitemOpen
  \bibfield  {author} {\bibinfo {author} {\bibfnamefont {D.}~\bibnamefont {Hooper}}, \bibinfo {author} {\bibfnamefont {C.}~\bibnamefont {Kelso}},\ and\ \bibinfo {author} {\bibfnamefont {F.~S.}\ \bibnamefont {Queiroz}},\ }\bibfield  {title} {\bibinfo {title} {Stringent constraints on the dark matter annihilation cross section from the region of the galactic center},\ }\href {https://doi.org/10.1016/j.astropartphys.2013.04.007} {\bibfield  {journal} {\bibinfo  {journal} {Astroparticle Physics}\ }\textbf {\bibinfo {volume} {46}},\ \bibinfo {pages} {55} (\bibinfo {year} {2013})}\BibitemShut {NoStop}%
\bibitem [{\citenamefont {collaboration}(2015)}]{collaboration2015}%
  \BibitemOpen
  \bibfield  {author} {\bibinfo {author} {\bibfnamefont {T.~F.~L.}\ \bibnamefont {collaboration}},\ }\bibfield  {title} {\bibinfo {title} {Limits on dark matter annihilation signals from the fermi {LAT} 4-year measurement of the isotropic gamma-ray background},\ }\href {https://doi.org/10.1088/1475-7516/2015/09/008} {\bibfield  {journal} {\bibinfo  {journal} {Journal of Cosmology and Astroparticle Physics}\ }\textbf {\bibinfo {volume} {2015}}\bibinfo  {number} { (09)},\ \bibinfo {pages} {008}}\BibitemShut {NoStop}%
\bibitem [{\citenamefont {Daylan}\ \emph {et~al.}(2016)\citenamefont {Daylan}, \citenamefont {Finkbeiner}, \citenamefont {Hooper}, \citenamefont {Linden}, \citenamefont {Portillo}, \citenamefont {Rodd},\ and\ \citenamefont {Slatyer}}]{daylan2016characterization}%
  \BibitemOpen
\bibfield  {number} {  }\bibfield  {author} {\bibinfo {author} {\bibfnamefont {T.}~\bibnamefont {Daylan}}, \bibinfo {author} {\bibfnamefont {D.~P.}\ \bibnamefont {Finkbeiner}}, \bibinfo {author} {\bibfnamefont {D.}~\bibnamefont {Hooper}}, \bibinfo {author} {\bibfnamefont {T.}~\bibnamefont {Linden}}, \bibinfo {author} {\bibfnamefont {S.~K.}\ \bibnamefont {Portillo}}, \bibinfo {author} {\bibfnamefont {N.~L.}\ \bibnamefont {Rodd}},\ and\ \bibinfo {author} {\bibfnamefont {T.~R.}\ \bibnamefont {Slatyer}},\ }\bibfield  {title} {\bibinfo {title} {The characterization of the gamma-ray signal from the central milky way: A case for annihilating dark matter},\ }\href@noop {} {\bibfield  {journal} {\bibinfo  {journal} {Physics of the Dark Universe}\ }\textbf {\bibinfo {volume} {12}},\ \bibinfo {pages} {1} (\bibinfo {year} {2016})}\BibitemShut {NoStop}%
\bibitem [{\citenamefont {Padmanabhan}\ and\ \citenamefont {Finkbeiner}(2005)}]{Padmanabhan2005}%
  \BibitemOpen
  \bibfield  {author} {\bibinfo {author} {\bibfnamefont {N.}~\bibnamefont {Padmanabhan}}\ and\ \bibinfo {author} {\bibfnamefont {D.~P.}\ \bibnamefont {Finkbeiner}},\ }\bibfield  {title} {\bibinfo {title} {Detecting dark matter annihilation with {CMB} polarization: Signatures and experimental prospects},\ }\bibfield  {journal} {\bibinfo  {journal} {Physical Review D}\ }\textbf {\bibinfo {volume} {72}},\ \href {https://doi.org/10.1103/physrevd.72.023508} {10.1103/physrevd.72.023508} (\bibinfo {year} {2005})\BibitemShut {NoStop}%
\bibitem [{\citenamefont {Galli}\ \emph {et~al.}(2013)\citenamefont {Galli}, \citenamefont {Slatyer}, \citenamefont {Valdes},\ and\ \citenamefont {Iocco}}]{Galli2013}%
  \BibitemOpen
  \bibfield  {author} {\bibinfo {author} {\bibfnamefont {S.}~\bibnamefont {Galli}}, \bibinfo {author} {\bibfnamefont {T.~R.}\ \bibnamefont {Slatyer}}, \bibinfo {author} {\bibfnamefont {M.}~\bibnamefont {Valdes}},\ and\ \bibinfo {author} {\bibfnamefont {F.}~\bibnamefont {Iocco}},\ }\bibfield  {title} {\bibinfo {title} {Systematic uncertainties in constraining dark matter annihilation from the cosmic microwave background},\ }\bibfield  {journal} {\bibinfo  {journal} {Physical Review D}\ }\textbf {\bibinfo {volume} {88}},\ \href {https://doi.org/10.1103/physrevd.88.063502} {10.1103/physrevd.88.063502} (\bibinfo {year} {2013})\BibitemShut {NoStop}%
\bibitem [{\citenamefont {Ghara}\ \emph {et~al.}(2015)\citenamefont {Ghara}, \citenamefont {Choudhury},\ and\ \citenamefont {Datta}}]{Ghara2015}%
  \BibitemOpen
  \bibfield  {author} {\bibinfo {author} {\bibfnamefont {R.}~\bibnamefont {Ghara}}, \bibinfo {author} {\bibfnamefont {T.~R.}\ \bibnamefont {Choudhury}},\ and\ \bibinfo {author} {\bibfnamefont {K.~K.}\ \bibnamefont {Datta}},\ }\bibfield  {title} {\bibinfo {title} {21~cm signal from cosmic dawn: imprints of spin temperature fluctuations and peculiar velocities},\ }\href {https://doi.org/10.1093/mnras/stu2512} {\bibfield  {journal} {\bibinfo  {journal} {Monthly Notices of the Royal Astronomical Society}\ }\textbf {\bibinfo {volume} {447}},\ \bibinfo {pages} {1806} (\bibinfo {year} {2015})}\BibitemShut {NoStop}%
\bibitem [{\citenamefont {Vald{\'{e}}s}\ \emph {et~al.}(2012)\citenamefont {Vald{\'{e}}s}, \citenamefont {Evoli}, \citenamefont {Mesinger}, \citenamefont {Ferrara},\ and\ \citenamefont {Yoshida}}]{Valds2012}%
  \BibitemOpen
  \bibfield  {author} {\bibinfo {author} {\bibfnamefont {M.}~\bibnamefont {Vald{\'{e}}s}}, \bibinfo {author} {\bibfnamefont {C.}~\bibnamefont {Evoli}}, \bibinfo {author} {\bibfnamefont {A.}~\bibnamefont {Mesinger}}, \bibinfo {author} {\bibfnamefont {A.}~\bibnamefont {Ferrara}},\ and\ \bibinfo {author} {\bibfnamefont {N.}~\bibnamefont {Yoshida}},\ }\bibfield  {title} {\bibinfo {title} {The nature of dark matter from the global high-redshift h{\hspace{0.167em}}i 21{\hspace{0.167em}}cm signal},\ }\href {https://doi.org/10.1093/mnras/sts458} {\bibfield  {journal} {\bibinfo  {journal} {Monthly Notices of the Royal Astronomical Society}\ }\textbf {\bibinfo {volume} {429}},\ \bibinfo {pages} {1705} (\bibinfo {year} {2012})}\BibitemShut {NoStop}%
\bibitem [{\citenamefont {Evoli}\ \emph {et~al.}(2014)\citenamefont {Evoli}, \citenamefont {Mesinger},\ and\ \citenamefont {Ferrara}}]{Evoli2014}%
  \BibitemOpen
  \bibfield  {author} {\bibinfo {author} {\bibfnamefont {C.}~\bibnamefont {Evoli}}, \bibinfo {author} {\bibfnamefont {A.}~\bibnamefont {Mesinger}},\ and\ \bibinfo {author} {\bibfnamefont {A.}~\bibnamefont {Ferrara}},\ }\bibfield  {title} {\bibinfo {title} {Unveiling the nature of dark matter with high redshift 21 cm line experiments},\ }\href {https://doi.org/10.1088/1475-7516/2014/11/024} {\bibfield  {journal} {\bibinfo  {journal} {Journal of Cosmology and Astroparticle Physics}\ }\textbf {\bibinfo {volume} {2014}}\bibinfo  {number} { (11)},\ \bibinfo {pages} {024}}\BibitemShut {NoStop}%
\bibitem [{\citenamefont {Furlanetto}\ \emph {et~al.}(2006)\citenamefont {Furlanetto}, \citenamefont {Oh},\ and\ \citenamefont {Pierpaoli}}]{Furlanetto2006}%
  \BibitemOpen
\bibfield  {number} {  }\bibfield  {author} {\bibinfo {author} {\bibfnamefont {S.~R.}\ \bibnamefont {Furlanetto}}, \bibinfo {author} {\bibfnamefont {S.~P.}\ \bibnamefont {Oh}},\ and\ \bibinfo {author} {\bibfnamefont {E.}~\bibnamefont {Pierpaoli}},\ }\bibfield  {title} {\bibinfo {title} {Effects of dark matter decay and annihilation on the high-redshift 21~cm background},\ }\bibfield  {journal} {\bibinfo  {journal} {Physical Review D}\ }\textbf {\bibinfo {volume} {74}},\ \href {https://doi.org/10.1103/physrevd.74.103502} {10.1103/physrevd.74.103502} (\bibinfo {year} {2006})\BibitemShut {NoStop}%
\bibitem [{\citenamefont {Pritchard}\ and\ \citenamefont {Loeb}(2008)}]{Pritchard2008}%
  \BibitemOpen
  \bibfield  {author} {\bibinfo {author} {\bibfnamefont {J.~R.}\ \bibnamefont {Pritchard}}\ and\ \bibinfo {author} {\bibfnamefont {A.}~\bibnamefont {Loeb}},\ }\bibfield  {title} {\bibinfo {title} {Evolution of the 21~cm signal throughout cosmic history},\ }\bibfield  {journal} {\bibinfo  {journal} {Physical Review D}\ }\textbf {\bibinfo {volume} {78}},\ \href {https://doi.org/10.1103/physrevd.78.103511} {10.1103/physrevd.78.103511} (\bibinfo {year} {2008})\BibitemShut {NoStop}%
\bibitem [{\citenamefont {Pritchard}\ and\ \citenamefont {Loeb}(2012)}]{pritchard2012}%
  \BibitemOpen
  \bibfield  {author} {\bibinfo {author} {\bibfnamefont {J.~R.}\ \bibnamefont {Pritchard}}\ and\ \bibinfo {author} {\bibfnamefont {A.}~\bibnamefont {Loeb}},\ }\bibfield  {title} {\bibinfo {title} {21 cm cosmology in the 21st century},\ }\href@noop {} {\bibfield  {journal} {\bibinfo  {journal} {Reports on Progress in Physics}\ }\textbf {\bibinfo {volume} {75}},\ \bibinfo {pages} {086901} (\bibinfo {year} {2012})}\BibitemShut {NoStop}%
\bibitem [{\citenamefont {Cirelli}\ \emph {et~al.}(2009)\citenamefont {Cirelli}, \citenamefont {Iocco},\ and\ \citenamefont {Panci}}]{Cirelli2009}%
  \BibitemOpen
  \bibfield  {author} {\bibinfo {author} {\bibfnamefont {M.}~\bibnamefont {Cirelli}}, \bibinfo {author} {\bibfnamefont {F.}~\bibnamefont {Iocco}},\ and\ \bibinfo {author} {\bibfnamefont {P.}~\bibnamefont {Panci}},\ }\bibfield  {title} {\bibinfo {title} {Constraints on dark matter annihilations from reionization and heating of the intergalactic gas},\ }\href {https://doi.org/10.1088/1475-7516/2009/10/009} {\bibfield  {journal} {\bibinfo  {journal} {Journal of Cosmology and Astroparticle Physics}\ }\textbf {\bibinfo {volume} {2009}}\bibinfo  {number} { (10)},\ \bibinfo {pages} {009}}\BibitemShut {NoStop}%
\bibitem [{\citenamefont {Ripamonti}\ \emph {et~al.}(2010)\citenamefont {Ripamonti}, \citenamefont {Iocco}, \citenamefont {Ferrara}, \citenamefont {Schneider}, \citenamefont {Bressan},\ and\ \citenamefont {Marigo}}]{Ripamonti2010}%
  \BibitemOpen
\bibfield  {number} {  }\bibfield  {author} {\bibinfo {author} {\bibfnamefont {E.}~\bibnamefont {Ripamonti}}, \bibinfo {author} {\bibfnamefont {F.}~\bibnamefont {Iocco}}, \bibinfo {author} {\bibfnamefont {A.}~\bibnamefont {Ferrara}}, \bibinfo {author} {\bibfnamefont {R.}~\bibnamefont {Schneider}}, \bibinfo {author} {\bibfnamefont {A.}~\bibnamefont {Bressan}},\ and\ \bibinfo {author} {\bibfnamefont {P.}~\bibnamefont {Marigo}},\ }\bibfield  {title} {\bibinfo {title} {First star formation with dark matter annihilation},\ }\href {https://doi.org/10.1111/j.1365-2966.2010.16854.x} {\bibfield  {journal} {\bibinfo  {journal} {Monthly Notices of the Royal Astronomical Society}\ }\textbf {\bibinfo {volume} {406}},\ \bibinfo {pages} {2605} (\bibinfo {year} {2010})}\BibitemShut {NoStop}%
\bibitem [{\citenamefont {Bromm}\ \emph {et~al.}(2002)\citenamefont {Bromm}, \citenamefont {Coppi},\ and\ \citenamefont {Larson}}]{Bromm2002}%
  \BibitemOpen
  \bibfield  {author} {\bibinfo {author} {\bibfnamefont {V.}~\bibnamefont {Bromm}}, \bibinfo {author} {\bibfnamefont {P.~S.}\ \bibnamefont {Coppi}},\ and\ \bibinfo {author} {\bibfnamefont {R.~B.}\ \bibnamefont {Larson}},\ }\bibfield  {title} {\bibinfo {title} {The formation of the first stars. i. the primordial star-forming cloud},\ }\href {https://doi.org/10.1086/323947} {\bibfield  {journal} {\bibinfo  {journal} {The Astrophysical Journal}\ }\textbf {\bibinfo {volume} {564}},\ \bibinfo {pages} {23} (\bibinfo {year} {2002})}\BibitemShut {NoStop}%
\bibitem [{\citenamefont {Bromm}(2013)}]{Bromm2013}%
  \BibitemOpen
  \bibfield  {author} {\bibinfo {author} {\bibfnamefont {V.}~\bibnamefont {Bromm}},\ }\bibfield  {title} {\bibinfo {title} {Formation of the first stars},\ }\href {https://doi.org/10.1088/0034-4885/76/11/112901} {\bibfield  {journal} {\bibinfo  {journal} {Reports on Progress in Physics}\ }\textbf {\bibinfo {volume} {76}},\ \bibinfo {pages} {112901} (\bibinfo {year} {2013})}\BibitemShut {NoStop}%
\bibitem [{\citenamefont {Hirano}\ \emph {et~al.}(2015)\citenamefont {Hirano}, \citenamefont {Hosokawa}, \citenamefont {Yoshida}, \citenamefont {Omukai},\ and\ \citenamefont {Yorke}}]{Hirano2015}%
  \BibitemOpen
  \bibfield  {author} {\bibinfo {author} {\bibfnamefont {S.}~\bibnamefont {Hirano}}, \bibinfo {author} {\bibfnamefont {T.}~\bibnamefont {Hosokawa}}, \bibinfo {author} {\bibfnamefont {N.}~\bibnamefont {Yoshida}}, \bibinfo {author} {\bibfnamefont {K.}~\bibnamefont {Omukai}},\ and\ \bibinfo {author} {\bibfnamefont {H.~W.}\ \bibnamefont {Yorke}},\ }\bibfield  {title} {\bibinfo {title} {Primordial star formation under the influence of far ultraviolet radiation: 1540 cosmological haloes and the stellar mass distribution},\ }\href {https://doi.org/10.1093/mnras/stv044} {\bibfield  {journal} {\bibinfo  {journal} {Monthly Notices of the Royal Astronomical Society}\ }\textbf {\bibinfo {volume} {448}},\ \bibinfo {pages} {568} (\bibinfo {year} {2015})}\BibitemShut {NoStop}%
\bibitem [{\citenamefont {Lopez-Honorez}\ \emph {et~al.}(2016)\citenamefont {Lopez-Honorez}, \citenamefont {Mena}, \citenamefont {Molin{\'{e}}}, \citenamefont {Palomares-Ruiz},\ and\ \citenamefont {Vincent}}]{LopezHonorez2016}%
  \BibitemOpen
  \bibfield  {author} {\bibinfo {author} {\bibfnamefont {L.}~\bibnamefont {Lopez-Honorez}}, \bibinfo {author} {\bibfnamefont {O.}~\bibnamefont {Mena}}, \bibinfo {author} {\bibfnamefont {{\'{A}}.}~\bibnamefont {Molin{\'{e}}}}, \bibinfo {author} {\bibfnamefont {S.}~\bibnamefont {Palomares-Ruiz}},\ and\ \bibinfo {author} {\bibfnamefont {A.~C.}\ \bibnamefont {Vincent}},\ }\bibfield  {title} {\bibinfo {title} {The 21 cm signal and the interplay between dark matter annihilations and astrophysical processes},\ }\href {https://doi.org/10.1088/1475-7516/2016/08/004} {\bibfield  {journal} {\bibinfo  {journal} {Journal of Cosmology and Astroparticle Physics}\ }\textbf {\bibinfo {volume} {2016}}\bibinfo  {number} { (08)},\ \bibinfo {pages} {004}}\BibitemShut {NoStop}%
\bibitem [{\citenamefont {Slatyer}(2016)}]{Slatyer2016}%
  \BibitemOpen
\bibfield  {number} {  }\bibfield  {author} {\bibinfo {author} {\bibfnamefont {T.~R.}\ \bibnamefont {Slatyer}},\ }\bibfield  {title} {\bibinfo {title} {Indirect dark matter signatures in the cosmic dark ages. i. generalizing the bound on s-wave dark matter annihilation from planck results},\ }\href@noop {} {\bibfield  {journal} {\bibinfo  {journal} {Physical Review D}\ }\textbf {\bibinfo {volume} {93}},\ \bibinfo {pages} {023527} (\bibinfo {year} {2016})}\BibitemShut {NoStop}%
\bibitem [{\citenamefont {Tseliakhovich}\ and\ \citenamefont {Hirata}(2010)}]{tseliakhovich2010relative}%
  \BibitemOpen
  \bibfield  {author} {\bibinfo {author} {\bibfnamefont {D.}~\bibnamefont {Tseliakhovich}}\ and\ \bibinfo {author} {\bibfnamefont {C.}~\bibnamefont {Hirata}},\ }\bibfield  {title} {\bibinfo {title} {Relative velocity of dark matter and baryonic fluids and the formation of the first structures},\ }\href@noop {} {\bibfield  {journal} {\bibinfo  {journal} {Physical Review D}\ }\textbf {\bibinfo {volume} {82}},\ \bibinfo {pages} {083520} (\bibinfo {year} {2010})}\BibitemShut {NoStop}%
\bibitem [{\citenamefont {Tseliakhovich}\ \emph {et~al.}(2011)\citenamefont {Tseliakhovich}, \citenamefont {Barkana},\ and\ \citenamefont {Hirata}}]{Tseliakhovich2011}%
  \BibitemOpen
  \bibfield  {author} {\bibinfo {author} {\bibfnamefont {D.}~\bibnamefont {Tseliakhovich}}, \bibinfo {author} {\bibfnamefont {R.}~\bibnamefont {Barkana}},\ and\ \bibinfo {author} {\bibfnamefont {C.~M.}\ \bibnamefont {Hirata}},\ }\bibfield  {title} {\bibinfo {title} {Suppression and spatial variation of early galaxies and minihaloes},\ }\href {https://doi.org/10.1111/j.1365-2966.2011.19541.x} {\bibfield  {journal} {\bibinfo  {journal} {Monthly Notices of the Royal Astronomical Society}\ }\textbf {\bibinfo {volume} {418}},\ \bibinfo {pages} {906} (\bibinfo {year} {2011})}\BibitemShut {NoStop}%
\bibitem [{\citenamefont {Mu{\~{n}}oz}(2019)}]{Muoz2019}%
  \BibitemOpen
  \bibfield  {author} {\bibinfo {author} {\bibfnamefont {J.~B.}\ \bibnamefont {Mu{\~{n}}oz}},\ }\bibfield  {title} {\bibinfo {title} {Robust velocity-induced acoustic oscillations at cosmic dawn},\ }\bibfield  {journal} {\bibinfo  {journal} {Physical Review D}\ }\textbf {\bibinfo {volume} {100}},\ \href {https://doi.org/10.1103/physrevd.100.063538} {10.1103/physrevd.100.063538} (\bibinfo {year} {2019})\BibitemShut {NoStop}%
\bibitem [{\citenamefont {Schauer}\ \emph {et~al.}(2022)\citenamefont {Schauer}, \citenamefont {Boylan-Kolchin}, \citenamefont {Colston}, \citenamefont {Sameie}, \citenamefont {Bromm}, \citenamefont {Bullock},\ and\ \citenamefont {Wetzel}}]{Schauer2022Dwarf}%
  \BibitemOpen
  \bibfield  {author} {\bibinfo {author} {\bibfnamefont {A.~T.}\ \bibnamefont {Schauer}}, \bibinfo {author} {\bibfnamefont {M.}~\bibnamefont {Boylan-Kolchin}}, \bibinfo {author} {\bibfnamefont {K.}~\bibnamefont {Colston}}, \bibinfo {author} {\bibfnamefont {O.}~\bibnamefont {Sameie}}, \bibinfo {author} {\bibfnamefont {V.}~\bibnamefont {Bromm}}, \bibinfo {author} {\bibfnamefont {J.~S.}\ \bibnamefont {Bullock}},\ and\ \bibinfo {author} {\bibfnamefont {A.}~\bibnamefont {Wetzel}},\ }\bibfield  {title} {\bibinfo {title} {Dwarf galaxy formation with and without dark matter-baryon streaming velocities},\ }\href@noop {} {\bibfield  {journal} {\bibinfo  {journal} {arXiv preprint arXiv:2210.12815}\ } (\bibinfo {year} {2022})}\BibitemShut {NoStop}%
\bibitem [{\citenamefont {Natarajan}\ and\ \citenamefont {Schwarz}(2009)}]{Natarajan2009}%
  \BibitemOpen
  \bibfield  {author} {\bibinfo {author} {\bibfnamefont {A.}~\bibnamefont {Natarajan}}\ and\ \bibinfo {author} {\bibfnamefont {D.~J.}\ \bibnamefont {Schwarz}},\ }\bibfield  {title} {\bibinfo {title} {Dark matter annihilation and its effect on cmb and hydrogen 21 cm observations},\ }\bibfield  {journal} {\bibinfo  {journal} {Physical Review D}\ }\textbf {\bibinfo {volume} {80}},\ \href {https://doi.org/10.1103/physrevd.80.043529} {10.1103/physrevd.80.043529} (\bibinfo {year} {2009})\BibitemShut {NoStop}%
\bibitem [{\citenamefont {Cheung}\ \emph {et~al.}(2019)\citenamefont {Cheung}, \citenamefont {Kuo}, \citenamefont {Ng},\ and\ \citenamefont {Tsai}}]{Cheung2019}%
  \BibitemOpen
  \bibfield  {author} {\bibinfo {author} {\bibfnamefont {K.}~\bibnamefont {Cheung}}, \bibinfo {author} {\bibfnamefont {J.-L.}\ \bibnamefont {Kuo}}, \bibinfo {author} {\bibfnamefont {K.-W.}\ \bibnamefont {Ng}},\ and\ \bibinfo {author} {\bibfnamefont {Y.-L.~S.}\ \bibnamefont {Tsai}},\ }\bibfield  {title} {\bibinfo {title} {The impact of edges 21-cm data on dark matter interactions},\ }\href {https://doi.org/10.1016/j.physletb.2018.11.058} {\bibfield  {journal} {\bibinfo  {journal} {Physics Letters B}\ }\textbf {\bibinfo {volume} {789}},\ \bibinfo {pages} {137–144} (\bibinfo {year} {2019})}\BibitemShut {NoStop}%
\bibitem [{\citenamefont {Basu}\ \emph {et~al.}(2020)\citenamefont {Basu}, \citenamefont {Banerjee}, \citenamefont {Pandey},\ and\ \citenamefont {Majumdar}}]{basu2020lower}%
  \BibitemOpen
  \bibfield  {author} {\bibinfo {author} {\bibfnamefont {R.}~\bibnamefont {Basu}}, \bibinfo {author} {\bibfnamefont {S.}~\bibnamefont {Banerjee}}, \bibinfo {author} {\bibfnamefont {M.}~\bibnamefont {Pandey}},\ and\ \bibinfo {author} {\bibfnamefont {D.}~\bibnamefont {Majumdar}},\ }\bibfield  {title} {\bibinfo {title} {Lower bounds on dark matter annihilation cross-sections by studying the fluctuations of 21-cm line with dark matter candidate in inert doublet model (idm) with the combined effects of dark matter scattering and annihilation},\ }\href@noop {} {\bibfield  {journal} {\bibinfo  {journal} {arXiv preprint arXiv:2010.11007}\ } (\bibinfo {year} {2020})}\BibitemShut {NoStop}%
\bibitem [{\citenamefont {Cang}\ \emph {et~al.}(2023)\citenamefont {Cang}, \citenamefont {Gao},\ and\ \citenamefont {Ma}}]{Cang2023}%
  \BibitemOpen
  \bibfield  {author} {\bibinfo {author} {\bibfnamefont {J.}~\bibnamefont {Cang}}, \bibinfo {author} {\bibfnamefont {Y.}~\bibnamefont {Gao}},\ and\ \bibinfo {author} {\bibfnamefont {Y.-Z.}\ \bibnamefont {Ma}},\ }\href {https://doi.org/10.48550/ARXIV.2312.17499} {\bibinfo {title} {Signatures of inhomogeneous dark matter annihilation on 21-cm}} (\bibinfo {year} {2023})\BibitemShut {NoStop}%
\bibitem [{\citenamefont {Qin}\ \emph {et~al.}(2023)\citenamefont {Qin}, \citenamefont {Munoz}, \citenamefont {Liu},\ and\ \citenamefont {Slatyer}}]{Qin2023}%
  \BibitemOpen
  \bibfield  {author} {\bibinfo {author} {\bibfnamefont {W.}~\bibnamefont {Qin}}, \bibinfo {author} {\bibfnamefont {J.~B.}\ \bibnamefont {Munoz}}, \bibinfo {author} {\bibfnamefont {H.}~\bibnamefont {Liu}},\ and\ \bibinfo {author} {\bibfnamefont {T.~R.}\ \bibnamefont {Slatyer}},\ }\href {https://doi.org/10.48550/ARXIV.2308.12992} {\bibinfo {title} {Birth of the first stars amidst decaying and annihilating dark matter}} (\bibinfo {year} {2023})\BibitemShut {NoStop}%
\bibitem [{\citenamefont {Liu}\ \emph {et~al.}(2016)\citenamefont {Liu}, \citenamefont {Slatyer},\ and\ \citenamefont {Zavala}}]{Liu2016}%
  \BibitemOpen
  \bibfield  {author} {\bibinfo {author} {\bibfnamefont {H.}~\bibnamefont {Liu}}, \bibinfo {author} {\bibfnamefont {T.~R.}\ \bibnamefont {Slatyer}},\ and\ \bibinfo {author} {\bibfnamefont {J.}~\bibnamefont {Zavala}},\ }\bibfield  {title} {\bibinfo {title} {Contributions to cosmic reionization from dark matter annihilation and decay},\ }\bibfield  {journal} {\bibinfo  {journal} {Physical Review D}\ }\textbf {\bibinfo {volume} {94}},\ \href {https://doi.org/10.1103/physrevd.94.063507} {10.1103/physrevd.94.063507} (\bibinfo {year} {2016})\BibitemShut {NoStop}%
\bibitem [{\citenamefont {Shull}\ and\ \citenamefont {Van~Steenberg}(1985)}]{shull1985}%
  \BibitemOpen
  \bibfield  {author} {\bibinfo {author} {\bibfnamefont {J.~M.}\ \bibnamefont {Shull}}\ and\ \bibinfo {author} {\bibfnamefont {M.~E.}\ \bibnamefont {Van~Steenberg}},\ }\bibfield  {title} {\bibinfo {title} {X-ray secondary heating and ionization in quasar emission-line clouds},\ }\href@noop {} {\bibfield  {journal} {\bibinfo  {journal} {Astrophysical Journal, Part 1 (ISSN 0004-637X), vol. 298, Nov. 1, 1985, p. 268-274.}\ }\textbf {\bibinfo {volume} {298}},\ \bibinfo {pages} {268} (\bibinfo {year} {1985})}\BibitemShut {NoStop}%
\bibitem [{\citenamefont {Chen}\ and\ \citenamefont {Kamionkowski}(2004)}]{chen2004particle}%
  \BibitemOpen
  \bibfield  {author} {\bibinfo {author} {\bibfnamefont {X.}~\bibnamefont {Chen}}\ and\ \bibinfo {author} {\bibfnamefont {M.}~\bibnamefont {Kamionkowski}},\ }\bibfield  {title} {\bibinfo {title} {Particle decays during the cosmic dark ages},\ }\href@noop {} {\bibfield  {journal} {\bibinfo  {journal} {Physical Review D}\ }\textbf {\bibinfo {volume} {70}},\ \bibinfo {pages} {043502} (\bibinfo {year} {2004})}\BibitemShut {NoStop}%
\bibitem [{\citenamefont {Gnedin}(2000)}]{Gnedin2000}%
  \BibitemOpen
  \bibfield  {author} {\bibinfo {author} {\bibfnamefont {N.~Y.}\ \bibnamefont {Gnedin}},\ }\bibfield  {title} {\bibinfo {title} {Effect of reionization on structure formation in the universe},\ }\href {https://doi.org/10.1086/317042} {\bibfield  {journal} {\bibinfo  {journal} {The Astrophysical Journal}\ }\textbf {\bibinfo {volume} {542}},\ \bibinfo {pages} {535} (\bibinfo {year} {2000})}\BibitemShut {NoStop}%
\bibitem [{\citenamefont {Naoz}\ and\ \citenamefont {Barkana}(2007)}]{Naoz2007}%
  \BibitemOpen
  \bibfield  {author} {\bibinfo {author} {\bibfnamefont {S.}~\bibnamefont {Naoz}}\ and\ \bibinfo {author} {\bibfnamefont {R.}~\bibnamefont {Barkana}},\ }\bibfield  {title} {\bibinfo {title} {The formation and gas content of high-redshift galaxies and minihaloes},\ }\href {https://doi.org/10.1111/j.1365-2966.2007.11636.x} {\bibfield  {journal} {\bibinfo  {journal} {Monthly Notices of the Royal Astronomical Society}\ }\textbf {\bibinfo {volume} {377}},\ \bibinfo {pages} {667} (\bibinfo {year} {2007})}\BibitemShut {NoStop}%
\bibitem [{\citenamefont {Barkana}\ and\ \citenamefont {Loeb}(2011)}]{barkana2011scale}%
  \BibitemOpen
  \bibfield  {author} {\bibinfo {author} {\bibfnamefont {R.}~\bibnamefont {Barkana}}\ and\ \bibinfo {author} {\bibfnamefont {A.}~\bibnamefont {Loeb}},\ }\bibfield  {title} {\bibinfo {title} {Scale-dependent bias of galaxies from baryonic acoustic oscillations},\ }\href@noop {} {\bibfield  {journal} {\bibinfo  {journal} {Monthly Notices of the Royal Astronomical Society}\ }\textbf {\bibinfo {volume} {415}},\ \bibinfo {pages} {3113} (\bibinfo {year} {2011})}\BibitemShut {NoStop}%
\bibitem [{\citenamefont {Barkana}\ and\ \citenamefont {Loeb}(2001)}]{Barkana2001}%
  \BibitemOpen
  \bibfield  {author} {\bibinfo {author} {\bibfnamefont {R.}~\bibnamefont {Barkana}}\ and\ \bibinfo {author} {\bibfnamefont {A.}~\bibnamefont {Loeb}},\ }\bibfield  {title} {\bibinfo {title} {In the beginning: the first sources of light and the reionization of the universe},\ }\href {https://doi.org/10.1016/s0370-1573(01)00019-9} {\bibfield  {journal} {\bibinfo  {journal} {Physics Reports}\ }\textbf {\bibinfo {volume} {349}},\ \bibinfo {pages} {125} (\bibinfo {year} {2001})}\BibitemShut {NoStop}%
\bibitem [{\citenamefont {Fialkov}\ \emph {et~al.}(2013)\citenamefont {Fialkov}, \citenamefont {Barkana}, \citenamefont {Visbal}, \citenamefont {Tseliakhovich},\ and\ \citenamefont {Hirata}}]{Fialkov2013}%
  \BibitemOpen
  \bibfield  {author} {\bibinfo {author} {\bibfnamefont {A.}~\bibnamefont {Fialkov}}, \bibinfo {author} {\bibfnamefont {R.}~\bibnamefont {Barkana}}, \bibinfo {author} {\bibfnamefont {E.}~\bibnamefont {Visbal}}, \bibinfo {author} {\bibfnamefont {D.}~\bibnamefont {Tseliakhovich}},\ and\ \bibinfo {author} {\bibfnamefont {C.~M.}\ \bibnamefont {Hirata}},\ }\bibfield  {title} {\bibinfo {title} {The 21-cm signature of the first stars during the lyman–werner feedback era},\ }\href {https://doi.org/10.1093/mnras/stt650} {\bibfield  {journal} {\bibinfo  {journal} {Monthly Notices of the Royal Astronomical Society}\ }\textbf {\bibinfo {volume} {432}},\ \bibinfo {pages} {2909–2916} (\bibinfo {year} {2013})}\BibitemShut {NoStop}%
\bibitem [{\citenamefont {Collaboration}\ \emph {et~al.}(2014)\citenamefont {Collaboration}, \citenamefont {Ade}, \citenamefont {Aghanim}, \citenamefont {Armitage-Caplan}, \citenamefont {Arnaud}, \citenamefont {Ashdown}, \citenamefont {Atrio-Barandela}, \citenamefont {Aumont}, \citenamefont {Baccigalupi}, \citenamefont {Banday} \emph {et~al.}}]{collaboration2014planck}%
  \BibitemOpen
  \bibfield  {author} {\bibinfo {author} {\bibfnamefont {P.}~\bibnamefont {Collaboration}}, \bibinfo {author} {\bibfnamefont {P.}~\bibnamefont {Ade}}, \bibinfo {author} {\bibfnamefont {N.}~\bibnamefont {Aghanim}}, \bibinfo {author} {\bibfnamefont {C.}~\bibnamefont {Armitage-Caplan}}, \bibinfo {author} {\bibfnamefont {M.}~\bibnamefont {Arnaud}}, \bibinfo {author} {\bibfnamefont {M.}~\bibnamefont {Ashdown}}, \bibinfo {author} {\bibfnamefont {F.}~\bibnamefont {Atrio-Barandela}}, \bibinfo {author} {\bibfnamefont {J.}~\bibnamefont {Aumont}}, \bibinfo {author} {\bibfnamefont {C.}~\bibnamefont {Baccigalupi}}, \bibinfo {author} {\bibfnamefont {A.}~\bibnamefont {Banday}}, \emph {et~al.},\ }\bibfield  {title} {\bibinfo {title} {Planck 2013 results. xvi. cosmological parameters},\ }\href@noop {} {\bibfield  {journal} {\bibinfo  {journal} {A\&A}\ }\textbf {\bibinfo {volume} {571}},\ \bibinfo {pages} {A16} (\bibinfo {year} {2014})}\BibitemShut {NoStop}%
\bibitem [{\citenamefont {Mack}(2014)}]{Mack2014}%
  \BibitemOpen
  \bibfield  {author} {\bibinfo {author} {\bibfnamefont {K.~J.}\ \bibnamefont {Mack}},\ }\bibfield  {title} {\bibinfo {title} {Known unknowns of dark matter annihilation over cosmic time},\ }\href {https://doi.org/10.1093/mnras/stu129} {\bibfield  {journal} {\bibinfo  {journal} {Monthly Notices of the Royal Astronomical Society}\ }\textbf {\bibinfo {volume} {439}},\ \bibinfo {pages} {2728–2735} (\bibinfo {year} {2014})}\BibitemShut {NoStop}%
\bibitem [{\citenamefont {Sheth}\ \emph {et~al.}(2001)\citenamefont {Sheth}, \citenamefont {Mo},\ and\ \citenamefont {Tormen}}]{Sheth2001}%
  \BibitemOpen
  \bibfield  {author} {\bibinfo {author} {\bibfnamefont {R.~K.}\ \bibnamefont {Sheth}}, \bibinfo {author} {\bibfnamefont {H.~J.}\ \bibnamefont {Mo}},\ and\ \bibinfo {author} {\bibfnamefont {G.}~\bibnamefont {Tormen}},\ }\bibfield  {title} {\bibinfo {title} {Ellipsoidal collapse and an improved model for the number and spatial distribution of dark matter haloes},\ }\href {https://doi.org/10.1046/j.1365-8711.2001.04006.x} {\bibfield  {journal} {\bibinfo  {journal} {Monthly Notices of the Royal Astronomical Society}\ }\textbf {\bibinfo {volume} {323}},\ \bibinfo {pages} {1} (\bibinfo {year} {2001})}\BibitemShut {NoStop}%
\bibitem [{\citenamefont {Diemer}\ and\ \citenamefont {Joyce}(2019)}]{diemer2019accurate}%
  \BibitemOpen
  \bibfield  {author} {\bibinfo {author} {\bibfnamefont {B.}~\bibnamefont {Diemer}}\ and\ \bibinfo {author} {\bibfnamefont {M.}~\bibnamefont {Joyce}},\ }\bibfield  {title} {\bibinfo {title} {An accurate physical model for halo concentrations},\ }\href@noop {} {\bibfield  {journal} {\bibinfo  {journal} {The Astrophysical Journal}\ }\textbf {\bibinfo {volume} {871}},\ \bibinfo {pages} {168} (\bibinfo {year} {2019})}\BibitemShut {NoStop}%
\bibitem [{\citenamefont {Schon}\ \emph {et~al.}(2014)\citenamefont {Schon}, \citenamefont {Mack}, \citenamefont {Avram}, \citenamefont {Wyithe},\ and\ \citenamefont {Barberio}}]{schon2014dark}%
  \BibitemOpen
  \bibfield  {author} {\bibinfo {author} {\bibfnamefont {S.}~\bibnamefont {Schon}}, \bibinfo {author} {\bibfnamefont {K.~J.}\ \bibnamefont {Mack}}, \bibinfo {author} {\bibfnamefont {C.~A.}\ \bibnamefont {Avram}}, \bibinfo {author} {\bibfnamefont {J.~S.~B.}\ \bibnamefont {Wyithe}},\ and\ \bibinfo {author} {\bibfnamefont {E.}~\bibnamefont {Barberio}},\ }\bibfield  {title} {\bibinfo {title} {Dark matter annihilation in the first galaxy halos},\ }\href@noop {} {\bibfield  {journal} {\bibinfo  {journal} {arXiv preprint arXiv:1411.3783}\ } (\bibinfo {year} {2014})}\BibitemShut {NoStop}%
\bibitem [{\citenamefont {Sch\"{o}n}\ \emph {et~al.}(2017)\citenamefont {Sch\"{o}n}, \citenamefont {Mack},\ and\ \citenamefont {Wyithe}}]{Schn2017}%
  \BibitemOpen
  \bibfield  {author} {\bibinfo {author} {\bibfnamefont {S.}~\bibnamefont {Sch\"{o}n}}, \bibinfo {author} {\bibfnamefont {K.~J.}\ \bibnamefont {Mack}},\ and\ \bibinfo {author} {\bibfnamefont {J.~S.~B.}\ \bibnamefont {Wyithe}},\ }\bibfield  {title} {\bibinfo {title} {Dark matter annihilation in the circumgalactic medium at high redshifts},\ }\href {https://doi.org/10.1093/mnras/stx2968} {\bibfield  {journal} {\bibinfo  {journal} {Monthly Notices of the Royal Astronomical Society}\ }\textbf {\bibinfo {volume} {474}},\ \bibinfo {pages} {3067} (\bibinfo {year} {2017})}\BibitemShut {NoStop}%
\bibitem [{\citenamefont {Clark}\ \emph {et~al.}(2017)\citenamefont {Clark}, \citenamefont {Iwanus}, \citenamefont {Elahi}, \citenamefont {Lewis},\ and\ \citenamefont {Scott}}]{Clark2017}%
  \BibitemOpen
  \bibfield  {author} {\bibinfo {author} {\bibfnamefont {H.~A.}\ \bibnamefont {Clark}}, \bibinfo {author} {\bibfnamefont {N.}~\bibnamefont {Iwanus}}, \bibinfo {author} {\bibfnamefont {P.~J.}\ \bibnamefont {Elahi}}, \bibinfo {author} {\bibfnamefont {G.~F.}\ \bibnamefont {Lewis}},\ and\ \bibinfo {author} {\bibfnamefont {P.}~\bibnamefont {Scott}},\ }\bibfield  {title} {\bibinfo {title} {Heating of galactic gas by dark matter annihilation in ultracompact minihalos},\ }\href {https://doi.org/10.1088/1475-7516/2017/05/048} {\bibfield  {journal} {\bibinfo  {journal} {Journal of Cosmology and Astroparticle Physics}\ }\textbf {\bibinfo {volume} {2017}}\bibinfo  {number} { (05)},\ \bibinfo {pages} {048–048}}\BibitemShut {NoStop}%
\bibitem [{\citenamefont {Galli}\ and\ \citenamefont {Palla}(1998)}]{Galli1998}%
  \BibitemOpen
\bibfield  {number} {  }\bibfield  {author} {\bibinfo {author} {\bibfnamefont {D.}~\bibnamefont {Galli}}\ and\ \bibinfo {author} {\bibfnamefont {F.}~\bibnamefont {Palla}},\ }\bibfield  {title} {\bibinfo {title} {The chemistry of the early universe},\ }\bibfield  {journal} {\bibinfo  {journal} {arXiv preprint astro-ph/9803315}\ }\href {https://doi.org/10.48550/ARXIV.ASTRO-PH/9803315} {10.48550/ARXIV.ASTRO-PH/9803315} (\bibinfo {year} {1998})\BibitemShut {NoStop}%
\bibitem [{\citenamefont {Glover}\ and\ \citenamefont {Savin}(2006)}]{Glover2006}%
  \BibitemOpen
  \bibfield  {author} {\bibinfo {author} {\bibfnamefont {S.}~\bibnamefont {Glover}}\ and\ \bibinfo {author} {\bibfnamefont {D.~W.}\ \bibnamefont {Savin}},\ }\bibfield  {title} {\bibinfo {title} {cooling in primordial gas},\ }\href {https://doi.org/10.1098/rsta.2006.1867} {\bibfield  {journal} {\bibinfo  {journal} {Philosophical Transactions of the Royal Society A: Mathematical, Physical and Engineering Sciences}\ }\textbf {\bibinfo {volume} {364}},\ \bibinfo {pages} {3107} (\bibinfo {year} {2006})}\BibitemShut {NoStop}%
\bibitem [{\citenamefont {Yoshida}\ \emph {et~al.}(2007)\citenamefont {Yoshida}, \citenamefont {Oh}, \citenamefont {Kitayama},\ and\ \citenamefont {Hernquist}}]{Yoshida2007}%
  \BibitemOpen
  \bibfield  {author} {\bibinfo {author} {\bibfnamefont {N.}~\bibnamefont {Yoshida}}, \bibinfo {author} {\bibfnamefont {S.~P.}\ \bibnamefont {Oh}}, \bibinfo {author} {\bibfnamefont {T.}~\bibnamefont {Kitayama}},\ and\ \bibinfo {author} {\bibfnamefont {L.}~\bibnamefont {Hernquist}},\ }\bibfield  {title} {\bibinfo {title} {Early cosmological h<scp>ii</scp>/he<scp>iii</scp>regions and their impact on second‐generation star formation},\ }\href {https://doi.org/10.1086/518227} {\bibfield  {journal} {\bibinfo  {journal} {The Astrophysical Journal}\ }\textbf {\bibinfo {volume} {663}},\ \bibinfo {pages} {687–707} (\bibinfo {year} {2007})}\BibitemShut {NoStop}%
\bibitem [{\citenamefont {Glover}\ and\ \citenamefont {Abel}(2008)}]{Glover2008}%
  \BibitemOpen
  \bibfield  {author} {\bibinfo {author} {\bibfnamefont {S.~C.~O.}\ \bibnamefont {Glover}}\ and\ \bibinfo {author} {\bibfnamefont {T.}~\bibnamefont {Abel}},\ }\bibfield  {title} {\bibinfo {title} {Uncertainties in h2and hd chemistry and cooling and their role in early structure formation},\ }\href {https://doi.org/10.1111/j.1365-2966.2008.13224.x} {\bibfield  {journal} {\bibinfo  {journal} {Monthly Notices of the Royal Astronomical Society}\ }\textbf {\bibinfo {volume} {388}},\ \bibinfo {pages} {1627–1651} (\bibinfo {year} {2008})}\BibitemShut {NoStop}%
\bibitem [{\citenamefont {Glover}\ and\ \citenamefont {Savin}(2009)}]{Glover2009}%
  \BibitemOpen
  \bibfield  {author} {\bibinfo {author} {\bibfnamefont {S.~C.~O.}\ \bibnamefont {Glover}}\ and\ \bibinfo {author} {\bibfnamefont {D.~W.}\ \bibnamefont {Savin}},\ }\bibfield  {title} {\bibinfo {title} {Is h+3cooling ever important in primordial gas?},\ }\href {https://doi.org/10.1111/j.1365-2966.2008.14156.x} {\bibfield  {journal} {\bibinfo  {journal} {Monthly Notices of the Royal Astronomical Society}\ }\textbf {\bibinfo {volume} {393}},\ \bibinfo {pages} {911–948} (\bibinfo {year} {2009})}\BibitemShut {NoStop}%
\bibitem [{\citenamefont {Machacek}\ \emph {et~al.}(2001)\citenamefont {Machacek}, \citenamefont {Bryan},\ and\ \citenamefont {Abel}}]{Machacek2001}%
  \BibitemOpen
  \bibfield  {author} {\bibinfo {author} {\bibfnamefont {M.~E.}\ \bibnamefont {Machacek}}, \bibinfo {author} {\bibfnamefont {G.~L.}\ \bibnamefont {Bryan}},\ and\ \bibinfo {author} {\bibfnamefont {T.}~\bibnamefont {Abel}},\ }\bibfield  {title} {\bibinfo {title} {Simulations of pregalactic structure formation with radiative feedback},\ }\href {https://doi.org/10.1086/319014} {\bibfield  {journal} {\bibinfo  {journal} {The Astrophysical Journal}\ }\textbf {\bibinfo {volume} {548}},\ \bibinfo {pages} {509} (\bibinfo {year} {2001})}\BibitemShut {NoStop}%
\bibitem [{\citenamefont {Visbal}\ \emph {et~al.}(2014)\citenamefont {Visbal}, \citenamefont {Haiman}, \citenamefont {Terrazas}, \citenamefont {Bryan},\ and\ \citenamefont {Barkana}}]{Visbal2014}%
  \BibitemOpen
  \bibfield  {author} {\bibinfo {author} {\bibfnamefont {E.}~\bibnamefont {Visbal}}, \bibinfo {author} {\bibfnamefont {Z.}~\bibnamefont {Haiman}}, \bibinfo {author} {\bibfnamefont {B.}~\bibnamefont {Terrazas}}, \bibinfo {author} {\bibfnamefont {G.~L.}\ \bibnamefont {Bryan}},\ and\ \bibinfo {author} {\bibfnamefont {R.}~\bibnamefont {Barkana}},\ }\bibfield  {title} {\bibinfo {title} {High-redshift star formation in a time-dependent lyman{\textendash}werner background},\ }\href {https://doi.org/10.1093/mnras/stu1710} {\bibfield  {journal} {\bibinfo  {journal} {Monthly Notices of the Royal Astronomical Society}\ }\textbf {\bibinfo {volume} {445}},\ \bibinfo {pages} {107} (\bibinfo {year} {2014})}\BibitemShut {NoStop}%
\bibitem [{\citenamefont {Tegmark}\ \emph {et~al.}(1997)\citenamefont {Tegmark}, \citenamefont {Silk}, \citenamefont {Rees}, \citenamefont {Blanchard}, \citenamefont {Abel},\ and\ \citenamefont {Palla}}]{Tegmark1997}%
  \BibitemOpen
  \bibfield  {author} {\bibinfo {author} {\bibfnamefont {M.}~\bibnamefont {Tegmark}}, \bibinfo {author} {\bibfnamefont {J.}~\bibnamefont {Silk}}, \bibinfo {author} {\bibfnamefont {M.~J.}\ \bibnamefont {Rees}}, \bibinfo {author} {\bibfnamefont {A.}~\bibnamefont {Blanchard}}, \bibinfo {author} {\bibfnamefont {T.}~\bibnamefont {Abel}},\ and\ \bibinfo {author} {\bibfnamefont {F.}~\bibnamefont {Palla}},\ }\bibfield  {title} {\bibinfo {title} {How small were the first cosmological objects?},\ }\href {https://doi.org/10.1086/303434} {\bibfield  {journal} {\bibinfo  {journal} {The Astrophysical Journal}\ }\textbf {\bibinfo {volume} {474}},\ \bibinfo {pages} {1} (\bibinfo {year} {1997})}\BibitemShut {NoStop}%
\bibitem [{\citenamefont {Hutchins}(1976)}]{hutchins1976thermal}%
  \BibitemOpen
  \bibfield  {author} {\bibinfo {author} {\bibfnamefont {J.~B.}\ \bibnamefont {Hutchins}},\ }\bibfield  {title} {\bibinfo {title} {The thermal effects of h2 molecules in rotating and collapsing spheroidal gas clouds},\ }\href@noop {} {\bibfield  {journal} {\bibinfo  {journal} {Astrophysical Journal, vol. 205, Apr. 1, 1976, pt. 1, p. 103-121.}\ }\textbf {\bibinfo {volume} {205}},\ \bibinfo {pages} {103} (\bibinfo {year} {1976})}\BibitemShut {NoStop}%
\bibitem [{\citenamefont {Kreckel}\ \emph {et~al.}(2010)\citenamefont {Kreckel}, \citenamefont {Bruhns}, \citenamefont {{\v{C}}{\'\i}{\v{z}}ek}, \citenamefont {Glover}, \citenamefont {Miller}, \citenamefont {Urbain},\ and\ \citenamefont {Savin}}]{kreckel2010experimental}%
  \BibitemOpen
  \bibfield  {author} {\bibinfo {author} {\bibfnamefont {H.}~\bibnamefont {Kreckel}}, \bibinfo {author} {\bibfnamefont {H.}~\bibnamefont {Bruhns}}, \bibinfo {author} {\bibfnamefont {M.}~\bibnamefont {{\v{C}}{\'\i}{\v{z}}ek}}, \bibinfo {author} {\bibfnamefont {S.}~\bibnamefont {Glover}}, \bibinfo {author} {\bibfnamefont {K.}~\bibnamefont {Miller}}, \bibinfo {author} {\bibfnamefont {X.}~\bibnamefont {Urbain}},\ and\ \bibinfo {author} {\bibfnamefont {D.~W.}\ \bibnamefont {Savin}},\ }\bibfield  {title} {\bibinfo {title} {Experimental results for h2 formation from h- and h and implications for first star formation},\ }\href@noop {} {\bibfield  {journal} {\bibinfo  {journal} {Science}\ }\textbf {\bibinfo {volume} {329}},\ \bibinfo {pages} {69} (\bibinfo {year} {2010})}\BibitemShut {NoStop}%
\bibitem [{\citenamefont {Stenrup}\ \emph {et~al.}(2009)\citenamefont {Stenrup}, \citenamefont {Larson},\ and\ \citenamefont {Elander}}]{stenrup2009mutual}%
  \BibitemOpen
  \bibfield  {author} {\bibinfo {author} {\bibfnamefont {M.}~\bibnamefont {Stenrup}}, \bibinfo {author} {\bibfnamefont {{\AA}.}~\bibnamefont {Larson}},\ and\ \bibinfo {author} {\bibfnamefont {N.}~\bibnamefont {Elander}},\ }\bibfield  {title} {\bibinfo {title} {Mutual neutralization in low-energy h++ h- collisions: A quantum ab initio study},\ }\href@noop {} {\bibfield  {journal} {\bibinfo  {journal} {Physical Review A—Atomic, Molecular, and Optical Physics}\ }\textbf {\bibinfo {volume} {79}},\ \bibinfo {pages} {012713} (\bibinfo {year} {2009})}\BibitemShut {NoStop}%
\bibitem [{\citenamefont {Nebrin}\ \emph {et~al.}(2023)\citenamefont {Nebrin}, \citenamefont {Giri},\ and\ \citenamefont {Mellema}}]{Nebrin2023}%
  \BibitemOpen
  \bibfield  {author} {\bibinfo {author} {\bibfnamefont {O.}~\bibnamefont {Nebrin}}, \bibinfo {author} {\bibfnamefont {S.~K.}\ \bibnamefont {Giri}},\ and\ \bibinfo {author} {\bibfnamefont {G.}~\bibnamefont {Mellema}},\ }\bibfield  {title} {\bibinfo {title} {Starbursts in low-mass haloes at cosmic dawn. i. the critical halo mass for star formation},\ }\href {https://doi.org/10.1093/mnras/stad1852} {\bibfield  {journal} {\bibinfo  {journal} {Monthly Notices of the Royal Astronomical Society}\ }\textbf {\bibinfo {volume} {524}},\ \bibinfo {pages} {2290–2311} (\bibinfo {year} {2023})}\BibitemShut {NoStop}%
\bibitem [{\citenamefont {Chluba}\ and\ \citenamefont {Thomas}(2010)}]{Chluba2010}%
  \BibitemOpen
  \bibfield  {author} {\bibinfo {author} {\bibfnamefont {J.}~\bibnamefont {Chluba}}\ and\ \bibinfo {author} {\bibfnamefont {R.~M.}\ \bibnamefont {Thomas}},\ }\bibfield  {title} {\bibinfo {title} {Towards a complete treatment of the cosmological recombination problem},\ }\href {https://doi.org/10.1111/j.1365-2966.2010.17940.x} {\bibfield  {journal} {\bibinfo  {journal} {Monthly Notices of the Royal Astronomical Society}\ ,\ \bibinfo {pages} {no}} (\bibinfo {year} {2010})}\BibitemShut {NoStop}%
\bibitem [{\citenamefont {Fialkov}\ \emph {et~al.}(2012)\citenamefont {Fialkov}, \citenamefont {Barkana}, \citenamefont {Tseliakhovich},\ and\ \citenamefont {Hirata}}]{Fialkov2012}%
  \BibitemOpen
  \bibfield  {author} {\bibinfo {author} {\bibfnamefont {A.}~\bibnamefont {Fialkov}}, \bibinfo {author} {\bibfnamefont {R.}~\bibnamefont {Barkana}}, \bibinfo {author} {\bibfnamefont {D.}~\bibnamefont {Tseliakhovich}},\ and\ \bibinfo {author} {\bibfnamefont {C.~M.}\ \bibnamefont {Hirata}},\ }\bibfield  {title} {\bibinfo {title} {Impact of the relative motion between the dark matter and baryons on the first stars: semi-analytical modelling},\ }\href {https://doi.org/10.1111/j.1365-2966.2012.21318.x} {\bibfield  {journal} {\bibinfo  {journal} {Monthly Notices of the Royal Astronomical Society}\ }\textbf {\bibinfo {volume} {424}},\ \bibinfo {pages} {1335} (\bibinfo {year} {2012})}\BibitemShut {NoStop}%
\bibitem [{\citenamefont {Kulkarni}\ \emph {et~al.}(2021)\citenamefont {Kulkarni}, \citenamefont {Visbal},\ and\ \citenamefont {Bryan}}]{Kulkarni2021}%
  \BibitemOpen
  \bibfield  {author} {\bibinfo {author} {\bibfnamefont {M.}~\bibnamefont {Kulkarni}}, \bibinfo {author} {\bibfnamefont {E.}~\bibnamefont {Visbal}},\ and\ \bibinfo {author} {\bibfnamefont {G.~L.}\ \bibnamefont {Bryan}},\ }\bibfield  {title} {\bibinfo {title} {The critical dark matter halo mass for population {III} star formation: Dependence on lyman{\textendash}werner radiation, baryon-dark matter streaming velocity, and redshift},\ }\href {https://doi.org/10.3847/1538-4357/ac08a3} {\bibfield  {journal} {\bibinfo  {journal} {The Astrophysical Journal}\ }\textbf {\bibinfo {volume} {917}},\ \bibinfo {pages} {40} (\bibinfo {year} {2021})}\BibitemShut {NoStop}%
\bibitem [{\citenamefont {Schauer}\ \emph {et~al.}(2021)\citenamefont {Schauer}, \citenamefont {Glover}, \citenamefont {Klessen},\ and\ \citenamefont {Clark}}]{Schauer2021}%
  \BibitemOpen
  \bibfield  {author} {\bibinfo {author} {\bibfnamefont {A.~T.~P.}\ \bibnamefont {Schauer}}, \bibinfo {author} {\bibfnamefont {S.~C.~O.}\ \bibnamefont {Glover}}, \bibinfo {author} {\bibfnamefont {R.~S.}\ \bibnamefont {Klessen}},\ and\ \bibinfo {author} {\bibfnamefont {P.}~\bibnamefont {Clark}},\ }\bibfield  {title} {\bibinfo {title} {The influence of streaming velocities and lyman{\textendash}werner radiation on the formation of the first stars},\ }\href {https://doi.org/10.1093/mnras/stab1953} {\bibfield  {journal} {\bibinfo  {journal} {Monthly Notices of the Royal Astronomical Society}\ }\textbf {\bibinfo {volume} {507}},\ \bibinfo {pages} {1775} (\bibinfo {year} {2021})}\BibitemShut {NoStop}%
\bibitem [{\citenamefont {Hegde}\ and\ \citenamefont {Furlanetto}(2023)}]{Hegde2023}%
  \BibitemOpen
  \bibfield  {author} {\bibinfo {author} {\bibfnamefont {S.}~\bibnamefont {Hegde}}\ and\ \bibinfo {author} {\bibfnamefont {S.~R.}\ \bibnamefont {Furlanetto}},\ }\bibfield  {title} {\bibinfo {title} {A self-consistent semi-analytic model for population iii star formation in minihaloes},\ }\href {https://doi.org/10.1093/mnras/stad2308} {\bibfield  {journal} {\bibinfo  {journal} {Monthly Notices of the Royal Astronomical Society}\ }\textbf {\bibinfo {volume} {525}},\ \bibinfo {pages} {428–447} (\bibinfo {year} {2023})}\BibitemShut {NoStop}%
\bibitem [{\citenamefont {Trenti}\ and\ \citenamefont {Stiavelli}(2009)}]{trenti2009formation}%
  \BibitemOpen
  \bibfield  {author} {\bibinfo {author} {\bibfnamefont {M.}~\bibnamefont {Trenti}}\ and\ \bibinfo {author} {\bibfnamefont {M.}~\bibnamefont {Stiavelli}},\ }\bibfield  {title} {\bibinfo {title} {Formation rates of population iii stars and chemical enrichment of halos during the reionization era},\ }\href@noop {} {\bibfield  {journal} {\bibinfo  {journal} {The Astrophysical Journal}\ }\textbf {\bibinfo {volume} {694}},\ \bibinfo {pages} {879} (\bibinfo {year} {2009})}\BibitemShut {NoStop}%
\bibitem [{\citenamefont {Wolcott-Green}\ \emph {et~al.}(2017)\citenamefont {Wolcott-Green}, \citenamefont {Haiman},\ and\ \citenamefont {Bryan}}]{WolcottGreen2017}%
  \BibitemOpen
  \bibfield  {author} {\bibinfo {author} {\bibfnamefont {J.}~\bibnamefont {Wolcott-Green}}, \bibinfo {author} {\bibfnamefont {Z.}~\bibnamefont {Haiman}},\ and\ \bibinfo {author} {\bibfnamefont {G.}~\bibnamefont {Bryan}},\ }\bibfield  {title} {\bibinfo {title} {Beyond j crit: a critical curve for suppression of h2-cooling in protogalaxies},\ }\href@noop {} {\bibfield  {journal} {\bibinfo  {journal} {Monthly Notices of the Royal Astronomical Society}\ }\textbf {\bibinfo {volume} {469}},\ \bibinfo {pages} {3329} (\bibinfo {year} {2017})}\BibitemShut {NoStop}%
\bibitem [{\citenamefont {Wolcott-Green}\ and\ \citenamefont {Haiman}(2019)}]{WolcottGreen2019}%
  \BibitemOpen
  \bibfield  {author} {\bibinfo {author} {\bibfnamefont {J.}~\bibnamefont {Wolcott-Green}}\ and\ \bibinfo {author} {\bibfnamefont {Z.}~\bibnamefont {Haiman}},\ }\bibfield  {title} {\bibinfo {title} {H2 self-shielding with non-lte rovibrational populations: implications for cooling in protogalaxies},\ }\href@noop {} {\bibfield  {journal} {\bibinfo  {journal} {Monthly Notices of the Royal Astronomical Society}\ }\textbf {\bibinfo {volume} {484}},\ \bibinfo {pages} {2467} (\bibinfo {year} {2019})}\BibitemShut {NoStop}%
\bibitem [{\citenamefont {Incatasciato}\ \emph {et~al.}(2023)\citenamefont {Incatasciato}, \citenamefont {Khochfar},\ and\ \citenamefont {O{\~{n}}orbe}}]{Incatasciato2023}%
  \BibitemOpen
  \bibfield  {author} {\bibinfo {author} {\bibfnamefont {A.}~\bibnamefont {Incatasciato}}, \bibinfo {author} {\bibfnamefont {S.}~\bibnamefont {Khochfar}},\ and\ \bibinfo {author} {\bibfnamefont {J.}~\bibnamefont {O{\~{n}}orbe}},\ }\bibfield  {title} {\bibinfo {title} {Modelling the cosmological lyman{\textendash}werner background radiation field in the early universe},\ }\href {https://doi.org/10.1093/mnras/stad1008} {\bibfield  {journal} {\bibinfo  {journal} {Monthly Notices of the Royal Astronomical Society}\ }\textbf {\bibinfo {volume} {522}},\ \bibinfo {pages} {330} (\bibinfo {year} {2023})}\BibitemShut {NoStop}%
\bibitem [{\citenamefont {Hummel}\ \emph {et~al.}(2015)\citenamefont {Hummel}, \citenamefont {Stacy}, \citenamefont {Jeon}, \citenamefont {Oliveri},\ and\ \citenamefont {Bromm}}]{Hummel2015}%
  \BibitemOpen
  \bibfield  {author} {\bibinfo {author} {\bibfnamefont {J.~A.}\ \bibnamefont {Hummel}}, \bibinfo {author} {\bibfnamefont {A.}~\bibnamefont {Stacy}}, \bibinfo {author} {\bibfnamefont {M.}~\bibnamefont {Jeon}}, \bibinfo {author} {\bibfnamefont {A.}~\bibnamefont {Oliveri}},\ and\ \bibinfo {author} {\bibfnamefont {V.}~\bibnamefont {Bromm}},\ }\bibfield  {title} {\bibinfo {title} {The first stars: formation under x-ray feedback},\ }\href {https://doi.org/10.1093/mnras/stv1902} {\bibfield  {journal} {\bibinfo  {journal} {Monthly Notices of the Royal Astronomical Society}\ }\textbf {\bibinfo {volume} {453}},\ \bibinfo {pages} {4137–4148} (\bibinfo {year} {2015})}\BibitemShut {NoStop}%
\bibitem [{\citenamefont {Slatyer}(2013)}]{Slatyer2013}%
  \BibitemOpen
  \bibfield  {author} {\bibinfo {author} {\bibfnamefont {T.~R.}\ \bibnamefont {Slatyer}},\ }\bibfield  {title} {\bibinfo {title} {Energy injection and absorption in the cosmic dark ages},\ }\bibfield  {journal} {\bibinfo  {journal} {Physical Review D}\ }\textbf {\bibinfo {volume} {87}},\ \href {https://doi.org/10.1103/physrevd.87.123513} {10.1103/physrevd.87.123513} (\bibinfo {year} {2013})\BibitemShut {NoStop}%
\bibitem [{\citenamefont {Mesinger}\ \emph {et~al.}(2010)\citenamefont {Mesinger}, \citenamefont {Furlanetto},\ and\ \citenamefont {Cen}}]{Mesinger2010}%
  \BibitemOpen
  \bibfield  {author} {\bibinfo {author} {\bibfnamefont {A.}~\bibnamefont {Mesinger}}, \bibinfo {author} {\bibfnamefont {S.}~\bibnamefont {Furlanetto}},\ and\ \bibinfo {author} {\bibfnamefont {R.}~\bibnamefont {Cen}},\ }\bibfield  {title} {\bibinfo {title} {21cmfast: a fast, seminumerical simulation of the high-redshift 21-cm signal},\ }\href {https://doi.org/10.1111/j.1365-2966.2010.17731.x} {\bibfield  {journal} {\bibinfo  {journal} {Monthly Notices of the Royal Astronomical Society}\ }\textbf {\bibinfo {volume} {411}},\ \bibinfo {pages} {955} (\bibinfo {year} {2010})}\BibitemShut {NoStop}%
\bibitem [{\citenamefont {Aghanim}\ \emph {et~al.}(2020)\citenamefont {Aghanim}, \citenamefont {Akrami}, \citenamefont {Ashdown}, \citenamefont {Aumont}, \citenamefont {Baccigalupi}, \citenamefont {Ballardini}, \citenamefont {Banday}, \citenamefont {Barreiro}, \citenamefont {Bartolo}, \citenamefont {Basak} \emph {et~al.}}]{aghanim2020planck}%
  \BibitemOpen
  \bibfield  {author} {\bibinfo {author} {\bibfnamefont {N.}~\bibnamefont {Aghanim}}, \bibinfo {author} {\bibfnamefont {Y.}~\bibnamefont {Akrami}}, \bibinfo {author} {\bibfnamefont {M.}~\bibnamefont {Ashdown}}, \bibinfo {author} {\bibfnamefont {J.}~\bibnamefont {Aumont}}, \bibinfo {author} {\bibfnamefont {C.}~\bibnamefont {Baccigalupi}}, \bibinfo {author} {\bibfnamefont {M.}~\bibnamefont {Ballardini}}, \bibinfo {author} {\bibfnamefont {A.}~\bibnamefont {Banday}}, \bibinfo {author} {\bibfnamefont {R.}~\bibnamefont {Barreiro}}, \bibinfo {author} {\bibfnamefont {N.}~\bibnamefont {Bartolo}}, \bibinfo {author} {\bibfnamefont {S.}~\bibnamefont {Basak}}, \emph {et~al.},\ }\bibfield  {title} {\bibinfo {title} {Planck 2018 results-vi. cosmological parameters},\ }\href@noop {} {\bibfield  {journal} {\bibinfo  {journal} {Astronomy \& Astrophysics}\ }\textbf {\bibinfo {volume} {641}},\ \bibinfo {pages} {A6} (\bibinfo {year} {2020})}\BibitemShut {NoStop}%
\bibitem [{\citenamefont {Dalal}\ \emph {et~al.}(2010)\citenamefont {Dalal}, \citenamefont {Pen},\ and\ \citenamefont {Seljak}}]{Dalal2010}%
  \BibitemOpen
  \bibfield  {author} {\bibinfo {author} {\bibfnamefont {N.}~\bibnamefont {Dalal}}, \bibinfo {author} {\bibfnamefont {U.-L.}\ \bibnamefont {Pen}},\ and\ \bibinfo {author} {\bibfnamefont {U.}~\bibnamefont {Seljak}},\ }\bibfield  {title} {\bibinfo {title} {Large-scale {BAO} signatures of the smallest galaxies},\ }\href {https://doi.org/10.1088/1475-7516/2010/11/007} {\bibfield  {journal} {\bibinfo  {journal} {Journal of Cosmology and Astroparticle Physics}\ }\textbf {\bibinfo {volume} {2010}}\bibinfo  {number} { (11)},\ \bibinfo {pages} {007}}\BibitemShut {NoStop}%
\bibitem [{\citenamefont {McQuinn}\ and\ \citenamefont {O'Leary}(2012)}]{mcquinn2012impact}%
  \BibitemOpen
\bibfield  {number} {  }\bibfield  {author} {\bibinfo {author} {\bibfnamefont {M.}~\bibnamefont {McQuinn}}\ and\ \bibinfo {author} {\bibfnamefont {R.~M.}\ \bibnamefont {O'Leary}},\ }\bibfield  {title} {\bibinfo {title} {The impact of the supersonic baryon--dark matter velocity difference on the z~ 20 21 cm background},\ }\href@noop {} {\bibfield  {journal} {\bibinfo  {journal} {The Astrophysical Journal}\ }\textbf {\bibinfo {volume} {760}},\ \bibinfo {pages} {3} (\bibinfo {year} {2012})}\BibitemShut {NoStop}%
\bibitem [{\citenamefont {Naoz}\ \emph {et~al.}(2012)\citenamefont {Naoz}, \citenamefont {Yoshida},\ and\ \citenamefont {Gnedin}}]{Naoz2012}%
  \BibitemOpen
  \bibfield  {author} {\bibinfo {author} {\bibfnamefont {S.}~\bibnamefont {Naoz}}, \bibinfo {author} {\bibfnamefont {N.}~\bibnamefont {Yoshida}},\ and\ \bibinfo {author} {\bibfnamefont {N.~Y.}\ \bibnamefont {Gnedin}},\ }\bibfield  {title} {\bibinfo {title} {Simulations of early baryonic structure formation with stream velocity. i. halo abundance},\ }\href {https://doi.org/10.1088/0004-637x/747/2/128} {\bibfield  {journal} {\bibinfo  {journal} {The Astrophysical Journal}\ }\textbf {\bibinfo {volume} {747}},\ \bibinfo {pages} {128} (\bibinfo {year} {2012})}\BibitemShut {NoStop}%
\bibitem [{\citenamefont {Greif}\ \emph {et~al.}(2011)\citenamefont {Greif}, \citenamefont {White}, \citenamefont {Klessen},\ and\ \citenamefont {Springel}}]{Greif2011}%
  \BibitemOpen
  \bibfield  {author} {\bibinfo {author} {\bibfnamefont {T.~H.}\ \bibnamefont {Greif}}, \bibinfo {author} {\bibfnamefont {S.~D.~M.}\ \bibnamefont {White}}, \bibinfo {author} {\bibfnamefont {R.~S.}\ \bibnamefont {Klessen}},\ and\ \bibinfo {author} {\bibfnamefont {V.}~\bibnamefont {Springel}},\ }\bibfield  {title} {\bibinfo {title} {The delay of population iii star formation by supersonic streaming velocities},\ }\href {https://doi.org/10.1088/0004-637x/736/2/147} {\bibfield  {journal} {\bibinfo  {journal} {The Astrophysical Journal}\ }\textbf {\bibinfo {volume} {736}},\ \bibinfo {pages} {147} (\bibinfo {year} {2011})}\BibitemShut {NoStop}%
\bibitem [{\citenamefont {Schauer}\ \emph {et~al.}(2019)\citenamefont {Schauer}, \citenamefont {Glover}, \citenamefont {Klessen},\ and\ \citenamefont {Ceverino}}]{Schauer2019}%
  \BibitemOpen
  \bibfield  {author} {\bibinfo {author} {\bibfnamefont {A.~T.~P.}\ \bibnamefont {Schauer}}, \bibinfo {author} {\bibfnamefont {S.~C.~O.}\ \bibnamefont {Glover}}, \bibinfo {author} {\bibfnamefont {R.~S.}\ \bibnamefont {Klessen}},\ and\ \bibinfo {author} {\bibfnamefont {D.}~\bibnamefont {Ceverino}},\ }\bibfield  {title} {\bibinfo {title} {The influence of streaming velocities on the formation of the first stars},\ }\href {https://doi.org/10.1093/mnras/stz013} {\bibfield  {journal} {\bibinfo  {journal} {Monthly Notices of the Royal Astronomical Society}\ }\textbf {\bibinfo {volume} {484}},\ \bibinfo {pages} {3510} (\bibinfo {year} {2019})}\BibitemShut {NoStop}%
\bibitem [{\citenamefont {Muñoz}\ \emph {et~al.}(2022)\citenamefont {Muñoz}, \citenamefont {Qin}, \citenamefont {Mesinger}, \citenamefont {Murray}, \citenamefont {Greig},\ and\ \citenamefont {Mason}}]{Muoz2022}%
  \BibitemOpen
  \bibfield  {author} {\bibinfo {author} {\bibfnamefont {J.~B.}\ \bibnamefont {Muñoz}}, \bibinfo {author} {\bibfnamefont {Y.}~\bibnamefont {Qin}}, \bibinfo {author} {\bibfnamefont {A.}~\bibnamefont {Mesinger}}, \bibinfo {author} {\bibfnamefont {S.~G.}\ \bibnamefont {Murray}}, \bibinfo {author} {\bibfnamefont {B.}~\bibnamefont {Greig}},\ and\ \bibinfo {author} {\bibfnamefont {C.}~\bibnamefont {Mason}},\ }\bibfield  {title} {\bibinfo {title} {The impact of the first galaxies on cosmic dawn and reionization},\ }\href {https://doi.org/10.1093/mnras/stac185} {\bibfield  {journal} {\bibinfo  {journal} {Monthly Notices of the Royal Astronomical Society}\ }\textbf {\bibinfo {volume} {511}},\ \bibinfo {pages} {3657–3681} (\bibinfo {year} {2022})}\BibitemShut {NoStop}%
\bibitem [{\citenamefont {Liu}\ \emph {et~al.}(2020)\citenamefont {Liu}, \citenamefont {Ridgway},\ and\ \citenamefont {Slatyer}}]{Liu2020}%
  \BibitemOpen
  \bibfield  {author} {\bibinfo {author} {\bibfnamefont {H.}~\bibnamefont {Liu}}, \bibinfo {author} {\bibfnamefont {G.~W.}\ \bibnamefont {Ridgway}},\ and\ \bibinfo {author} {\bibfnamefont {T.~R.}\ \bibnamefont {Slatyer}},\ }\bibfield  {title} {\bibinfo {title} {Code package for calculating modified cosmic ionization and thermal histories with dark matter and other exotic energy injections},\ }\bibfield  {journal} {\bibinfo  {journal} {Physical Review D}\ }\textbf {\bibinfo {volume} {101}},\ \href {https://doi.org/10.1103/physrevd.101.023530} {10.1103/physrevd.101.023530} (\bibinfo {year} {2020})\BibitemShut {NoStop}%
\bibitem [{\citenamefont {Mu{\~{n}}oz}(2023)}]{Muoz2023}%
  \BibitemOpen
  \bibfield  {author} {\bibinfo {author} {\bibfnamefont {J.~B.}\ \bibnamefont {Mu{\~{n}}oz}},\ }\bibfield  {title} {\bibinfo {title} {An effective model for the cosmic-dawn 21-cm signal},\ }\href {https://doi.org/10.1093/mnras/stad1512} {\bibfield  {journal} {\bibinfo  {journal} {Monthly Notices of the Royal Astronomical Society}\ }\textbf {\bibinfo {volume} {523}},\ \bibinfo {pages} {2587} (\bibinfo {year} {2023})}\BibitemShut {NoStop}%
\bibitem [{\citenamefont {Monsalve}\ \emph {et~al.}(2017)\citenamefont {Monsalve}, \citenamefont {Rogers}, \citenamefont {Bowman},\ and\ \citenamefont {Mozdzen}}]{Monsalve2017}%
  \BibitemOpen
  \bibfield  {author} {\bibinfo {author} {\bibfnamefont {R.~A.}\ \bibnamefont {Monsalve}}, \bibinfo {author} {\bibfnamefont {A.~E.~E.}\ \bibnamefont {Rogers}}, \bibinfo {author} {\bibfnamefont {J.~D.}\ \bibnamefont {Bowman}},\ and\ \bibinfo {author} {\bibfnamefont {T.~J.}\ \bibnamefont {Mozdzen}},\ }\bibfield  {title} {\bibinfo {title} {Results from {EDGES} high-band. i. constraints on phenomenological models for the global 21 cm signal},\ }\href {https://doi.org/10.3847/1538-4357/aa88d1} {\bibfield  {journal} {\bibinfo  {journal} {The Astrophysical Journal}\ }\textbf {\bibinfo {volume} {847}},\ \bibinfo {pages} {64} (\bibinfo {year} {2017})}\BibitemShut {NoStop}%
\bibitem [{\citenamefont {Barkana}(2018)}]{Barkana2018}%
  \BibitemOpen
  \bibfield  {author} {\bibinfo {author} {\bibfnamefont {R.}~\bibnamefont {Barkana}},\ }\bibfield  {title} {\bibinfo {title} {Possible interaction between baryons and dark-matter particles revealed by the first stars},\ }\href {https://doi.org/10.1038/nature25791} {\bibfield  {journal} {\bibinfo  {journal} {Nature}\ }\textbf {\bibinfo {volume} {555}},\ \bibinfo {pages} {71} (\bibinfo {year} {2018})}\BibitemShut {NoStop}%
\bibitem [{\citenamefont {Liu}\ and\ \citenamefont {Slatyer}(2018)}]{Liu2018}%
  \BibitemOpen
  \bibfield  {author} {\bibinfo {author} {\bibfnamefont {H.}~\bibnamefont {Liu}}\ and\ \bibinfo {author} {\bibfnamefont {T.~R.}\ \bibnamefont {Slatyer}},\ }\bibfield  {title} {\bibinfo {title} {Implications of a 21-cm signal for dark matter annihilation and decay},\ }\bibfield  {journal} {\bibinfo  {journal} {Physical Review D}\ }\textbf {\bibinfo {volume} {98}},\ \href {https://doi.org/10.1103/physrevd.98.023501} {10.1103/physrevd.98.023501} (\bibinfo {year} {2018})\BibitemShut {NoStop}%
\bibitem [{\citenamefont {Philip}\ \emph {et~al.}(2019)\citenamefont {Philip}, \citenamefont {Abdurashidova}, \citenamefont {Chiang}, \citenamefont {Ghazi}, \citenamefont {Gumba}, \citenamefont {Heilgendorff}, \citenamefont {J{\'{a}}uregui-Garc{\'{\i}}a}, \citenamefont {Malepe}, \citenamefont {Nunhokee}, \citenamefont {Peterson}, \citenamefont {Sievers}, \citenamefont {Simes},\ and\ \citenamefont {Spann}}]{Philip2019}%
  \BibitemOpen
  \bibfield  {author} {\bibinfo {author} {\bibfnamefont {L.}~\bibnamefont {Philip}}, \bibinfo {author} {\bibfnamefont {Z.}~\bibnamefont {Abdurashidova}}, \bibinfo {author} {\bibfnamefont {H.~C.}\ \bibnamefont {Chiang}}, \bibinfo {author} {\bibfnamefont {N.}~\bibnamefont {Ghazi}}, \bibinfo {author} {\bibfnamefont {A.}~\bibnamefont {Gumba}}, \bibinfo {author} {\bibfnamefont {H.~M.}\ \bibnamefont {Heilgendorff}}, \bibinfo {author} {\bibfnamefont {J.~M.}\ \bibnamefont {J{\'{a}}uregui-Garc{\'{\i}}a}}, \bibinfo {author} {\bibfnamefont {K.}~\bibnamefont {Malepe}}, \bibinfo {author} {\bibfnamefont {C.~D.}\ \bibnamefont {Nunhokee}}, \bibinfo {author} {\bibfnamefont {J.}~\bibnamefont {Peterson}}, \bibinfo {author} {\bibfnamefont {J.~L.}\ \bibnamefont {Sievers}}, \bibinfo {author} {\bibfnamefont {V.}~\bibnamefont {Simes}},\ and\ \bibinfo {author} {\bibfnamefont {R.}~\bibnamefont {Spann}},\ }\bibfield  {title} {\bibinfo {title} {Probing radio intensity at high-z from marion: 2017 instrument},\ }\bibfield  {journal}
  {\bibinfo  {journal} {Journal of Astronomical Instrumentation}\ }\textbf {\bibinfo {volume} {08}},\ \href {https://doi.org/10.1142/s2251171719500041} {10.1142/s2251171719500041} (\bibinfo {year} {2019})\BibitemShut {NoStop}%
\bibitem [{\citenamefont {Singh}\ \emph {et~al.}(2017)\citenamefont {Singh}, \citenamefont {Subrahmanyan}, \citenamefont {Shankar}, \citenamefont {Rao}, \citenamefont {Fialkov}, \citenamefont {Cohen}, \citenamefont {Barkana}, \citenamefont {Girish}, \citenamefont {Raghunathan}, \citenamefont {Somashekar},\ and\ \citenamefont {Srivani}}]{Singh2017}%
  \BibitemOpen
  \bibfield  {author} {\bibinfo {author} {\bibfnamefont {S.}~\bibnamefont {Singh}}, \bibinfo {author} {\bibfnamefont {R.}~\bibnamefont {Subrahmanyan}}, \bibinfo {author} {\bibfnamefont {N.~U.}\ \bibnamefont {Shankar}}, \bibinfo {author} {\bibfnamefont {M.~S.}\ \bibnamefont {Rao}}, \bibinfo {author} {\bibfnamefont {A.}~\bibnamefont {Fialkov}}, \bibinfo {author} {\bibfnamefont {A.}~\bibnamefont {Cohen}}, \bibinfo {author} {\bibfnamefont {R.}~\bibnamefont {Barkana}}, \bibinfo {author} {\bibfnamefont {B.~S.}\ \bibnamefont {Girish}}, \bibinfo {author} {\bibfnamefont {A.}~\bibnamefont {Raghunathan}}, \bibinfo {author} {\bibfnamefont {R.}~\bibnamefont {Somashekar}},\ and\ \bibinfo {author} {\bibfnamefont {K.~S.}\ \bibnamefont {Srivani}},\ }\bibfield  {title} {\bibinfo {title} {First results on the epoch of reionization from first light with {SARAS} 2},\ }\href {https://doi.org/10.3847/2041-8213/aa831b} {\bibfield  {journal} {\bibinfo  {journal} {The Astrophysical Journal}\ }\textbf {\bibinfo {volume} {845}},\
  \bibinfo {pages} {L12} (\bibinfo {year} {2017})}\BibitemShut {NoStop}%
\bibitem [{\citenamefont {de~Lera~Acedo}(2019)}]{deLeraAcedo2019}%
  \BibitemOpen
  \bibfield  {author} {\bibinfo {author} {\bibfnamefont {E.}~\bibnamefont {de~Lera~Acedo}},\ }\bibfield  {title} {\bibinfo {title} {{REACH}: Radio experiment for the analysis of cosmic hydrogen},\ }in\ \href {https://doi.org/10.1109/iceaa.2019.8879199} {\emph {\bibinfo {booktitle} {2019 International Conference on Electromagnetics in Advanced Applications ({ICEAA})}}}\ (\bibinfo  {publisher} {{IEEE}},\ \bibinfo {year} {2019})\BibitemShut {NoStop}%
\bibitem [{\citenamefont {Mirocha}\ \emph {et~al.}(2016)\citenamefont {Mirocha}, \citenamefont {Furlanetto},\ and\ \citenamefont {Sun}}]{Mirocha2016}%
  \BibitemOpen
  \bibfield  {author} {\bibinfo {author} {\bibfnamefont {J.}~\bibnamefont {Mirocha}}, \bibinfo {author} {\bibfnamefont {S.~R.}\ \bibnamefont {Furlanetto}},\ and\ \bibinfo {author} {\bibfnamefont {G.}~\bibnamefont {Sun}},\ }\bibfield  {title} {\bibinfo {title} {The global 21-cm signal in the context of the high-zgalaxy luminosity function},\ }\href {https://doi.org/10.1093/mnras/stw2412} {\bibfield  {journal} {\bibinfo  {journal} {Monthly Notices of the Royal Astronomical Society}\ }\textbf {\bibinfo {volume} {464}},\ \bibinfo {pages} {1365–1379} (\bibinfo {year} {2016})}\BibitemShut {NoStop}%
\bibitem [{\citenamefont {Mizusawa}\ \emph {et~al.}(2005)\citenamefont {Mizusawa}, \citenamefont {Omukai},\ and\ \citenamefont {Nishi}}]{Mizusawa2005}%
  \BibitemOpen
  \bibfield  {author} {\bibinfo {author} {\bibfnamefont {H.}~\bibnamefont {Mizusawa}}, \bibinfo {author} {\bibfnamefont {K.}~\bibnamefont {Omukai}},\ and\ \bibinfo {author} {\bibfnamefont {R.}~\bibnamefont {Nishi}},\ }\bibfield  {title} {\bibinfo {title} {Primordial molecular emission in population {III} galaxies},\ }\href {https://doi.org/10.1093/pasj/57.6.951} {\bibfield  {journal} {\bibinfo  {journal} {Publications of the Astronomical Society of Japan}\ }\textbf {\bibinfo {volume} {57}},\ \bibinfo {pages} {951} (\bibinfo {year} {2005})}\BibitemShut {NoStop}%
\bibitem [{\citenamefont {Dijkstra}\ \emph {et~al.}(2007)\citenamefont {Dijkstra}, \citenamefont {Lidz},\ and\ \citenamefont {Wyithe}}]{dijkstra2007impact}%
  \BibitemOpen
  \bibfield  {author} {\bibinfo {author} {\bibfnamefont {M.}~\bibnamefont {Dijkstra}}, \bibinfo {author} {\bibfnamefont {A.}~\bibnamefont {Lidz}},\ and\ \bibinfo {author} {\bibfnamefont {J.~S.~B.}\ \bibnamefont {Wyithe}},\ }\bibfield  {title} {\bibinfo {title} {The impact of the igm on high-redshift ly$\alpha$ emission lines},\ }\href@noop {} {\bibfield  {journal} {\bibinfo  {journal} {Monthly Notices of the Royal Astronomical Society}\ }\textbf {\bibinfo {volume} {377}},\ \bibinfo {pages} {1175} (\bibinfo {year} {2007})}\BibitemShut {NoStop}%
\bibitem [{\citenamefont {Bruns~Jr}\ \emph {et~al.}(2012)\citenamefont {Bruns~Jr}, \citenamefont {Wyithe}, \citenamefont {Bland-Hawthorn},\ and\ \citenamefont {Dijkstra}}]{bruns2012clustering}%
  \BibitemOpen
  \bibfield  {author} {\bibinfo {author} {\bibfnamefont {L.~R.}\ \bibnamefont {Bruns~Jr}}, \bibinfo {author} {\bibfnamefont {J.~S.~B.}\ \bibnamefont {Wyithe}}, \bibinfo {author} {\bibfnamefont {J.}~\bibnamefont {Bland-Hawthorn}},\ and\ \bibinfo {author} {\bibfnamefont {M.}~\bibnamefont {Dijkstra}},\ }\bibfield  {title} {\bibinfo {title} {Clustering of ly$\alpha$ emitters around luminous quasars at z= 2--3: an alternative probe of reionization on galaxy formation},\ }\href@noop {} {\bibfield  {journal} {\bibinfo  {journal} {Monthly Notices of the Royal Astronomical Society}\ }\textbf {\bibinfo {volume} {421}},\ \bibinfo {pages} {2543} (\bibinfo {year} {2012})}\BibitemShut {NoStop}%
\bibitem [{\citenamefont {Correa}\ \emph {et~al.}(2015)\citenamefont {Correa}, \citenamefont {Wyithe}, \citenamefont {Schaye},\ and\ \citenamefont {Duffy}}]{Correa2015}%
  \BibitemOpen
  \bibfield  {author} {\bibinfo {author} {\bibfnamefont {C.~A.}\ \bibnamefont {Correa}}, \bibinfo {author} {\bibfnamefont {J.~S.~B.}\ \bibnamefont {Wyithe}}, \bibinfo {author} {\bibfnamefont {J.}~\bibnamefont {Schaye}},\ and\ \bibinfo {author} {\bibfnamefont {A.~R.}\ \bibnamefont {Duffy}},\ }\bibfield  {title} {\bibinfo {title} {The accretion history of dark matter haloes – ii. the connections with the mass power spectrum and the density profile},\ }\href {https://doi.org/10.1093/mnras/stv697} {\bibfield  {journal} {\bibinfo  {journal} {Monthly Notices of the Royal Astronomical Society}\ }\textbf {\bibinfo {volume} {450}},\ \bibinfo {pages} {1521–1537} (\bibinfo {year} {2015})}\BibitemShut {NoStop}%
\end{thebibliography}%

\newpage
\appendix
\section{\label{sec:model.local}Local Effect of Dark Matter Annihilation}

While we have neglected the effects of dark matter (DM) annihilation local to DM halos in this work, the local DM has the potential to significantly influence the surrounding gas by depositing energy into it. This process can play a crucial role in various astrophysical phenomena, impacting the thermal and ionization states of the circumgalactic medium of early galaxies. Understanding the local effects of dark matter annihilation is essential for comprehending the behavior of gas, as well as the first stars and galaxies in the early universe. This section briefly summarizes the mechanisms through which dark matter annihilation contributes to energy deposition in the surrounding gas to estimate its importance in the context of gas cooling.

For a small halo, although a dense DM core can inject significant DM annihilation power, a large fraction, $f_\mathrm{esc}$, of dark matter annihilation energy is expected to escape from the local environment and boost the global DM annihilation background. The total deposited fraction in surrounding gas is $f_\mathrm{local} = 1 - f_\mathrm{esc}$. The energy deposition of dark matter annihilation products is determined by the interactions between the annihilation products from dark matter and the baryons or photons in the local environment and the Cosmic Microwave Background (CMB), via the process such as inverse Compton scattering, photo-ionization, etc. This has been studied in energy transfer simulations on large scales~\cite{Evoli2014, schon2014dark}, but only roughly on mini-halo scales~\cite{Schn2017}. Therefore, we apply a parameterized semi-analytic model and provide a rough estimate of the local effects here.

First, we aim to estimate the intensity of local dark matter annihilation $\epsilon^\mathrm{DM}_\mathrm{c, local}$, and compare it with the global DM annihilation background $\epsilon^\mathrm{DM}_\mathrm{c, bg}$. The latter can be calculated using Equation~\ref{eq.epsilon_c}. The parameter $\epsilon^\mathrm{DM}_\mathrm{c, local}$ reflects the energy transfer from the dark matter annihilation products produced by the local halo to the local baryonic environment. The calculation depends on the DM annihilation model and the gas profile, and requires detailed simulations for a fully rigorous treatment.


Recent small-scale energy transfer simulations suggest that an electron with an energy of $10^8$ eV injected into a $10^6\, \mathrm{M}_\odot$ halo at redshift $z = 40$ deposits approximately $0.1\, \si{eV / pc^3}$ in the circumgalactic medium (CGM) \cite{Schn2017}. We adopt the CGM gas density profile $\rho \approx 20 \bar \rho (r / R_\mathrm{vir})^{-1}$ \cite{dijkstra2007impact, bruns2012clustering} and use it to calculate the total energy deposition $\epsilon^\mathrm{DM}_\mathrm{local}$, where $\bar \rho$ is the mean gas density of IGM. This results in $\log (\epsilon^\mathrm{DM}_\mathrm{local} / E_h) \sim -64$, where $E_h$ is the total dark matter annihilation power of the halo. 
For simplicity, we parameterize this dimensionless factor as $h = \log (\epsilon^\mathrm{DM}_\mathrm{local} / E_h)$, setting a baseline value of $h = -65$ and assuming it is independent of both halo mass and redshift. The local energy deposition in channel $c$ is given by $\epsilon^\mathrm{DM}_\mathrm{c, local} = f_c \epsilon^\mathrm{DM}_\mathrm{local}$, where $f_c$ represents the fraction of total energy deposited in channel $c$. We assign deposition fractions for heating and for the ionization of H and He as $f_\mathrm{heat} = [1 + 2x_e + f_{\ce{He}}(1 +
2 Z_{\ce{HeII}})]/3[1 + f_{\ce{He}}]$, $f_\mathrm{ion, \ce{H}} = (1 - x_e) / 3$, and $f_\mathrm{ion, \ce{He}} = (1 - Z_{\ce{HeII}}) / 3$, respectively\cite{chen2004particle, Chluba2010}, where $Z_{\ce{HeII}}$ is the
fraction of singly ionized helium atoms relative to the total number
of helium nuclei.

To make a comparison between this effect and the DM background, we can write the local energy deposition in terms of the local boost factor $B_\mathrm{local}(z)$ and the DM annihilation power of a smooth background,

\begin{equation}\label{eqn:epsilon_local_boost}
    \epsilon^\mathrm{DM}_\mathrm{c, local} = \frac{1}{n_\mathrm{B}} f_c B_\mathrm{local}(z) \frac{dE}{dV dt}\bigg|_\mathrm{smooth}\;,
\end{equation}
where $n_\mathrm{B}$ is the local gas density, and $B_\mathrm{local}(z)$ is the local annihilation power as a fraction of the smooth background DM annihilation power, which is given by Equation~\ref{eq:dE_dVdt}. 

\begin{figure}
    \centering
    \includegraphics[width=0.46\textwidth]{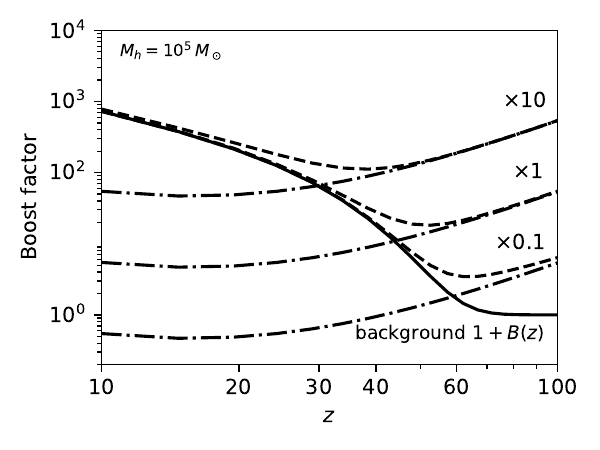}
    \caption{The local boost factor in dark matter (DM) annihilation around a $10^5 \mathrm{M}_\odot$ dark matter halo at different redshift. The solid line represents the global boost factor $1+B(z)$, the dot-dashed line indicates the contribution from the local $10^5 \mathrm{M}_\odot$ halo with different local deposit efficiencies, and the dashed line shows the total boost factor.}
    \label{fig:boost_factor}
\end{figure}

The local boost factor, $B_\mathrm{local}(z)$, can be calculated by \ref{eqn:epsilon_local_boost} with given value of efficiency $h$. 
Figure~\ref{fig:boost_factor} illustrates the boost factor of dark matter annihilation near a $10^5 \mathrm{M}_\odot$ dark matter halo. The structure boost factor from external annihilation, following Equation~\ref{eq:boost_factor}, is shown with a solid line. 
The boost factor $B(z)$ in the global DM annihilation (as inEquation~\ref{eq:boost_factor}) assumes that dark matter annihilation products completely escape from the local environment and deposit all their energy into the IGM. To account for local energy deposition, the escape fraction $f_\mathrm{esc}$ should be considered in Equation~\ref{eq:boost_factor}. In this work, we set $f_\mathrm{esc}\approx 1$. The local boost factors, calculated at different values of the parameter $0.1$, $1$ and $10$ times of base value $h = -65$, are represented by dot-dashed lines in Figure~\ref{fig:boost_factor}. The dashed lines indicate the total boost factor as a function of redshift. The structure boost factor from the background is very low at high redshifts due to the low abundance of halos. In contrast, the local boost factor for the halo is dominant at higher redshifts but decreases over time. After redshift $z=40$, the structure boost factor from the  background becomes more significant.

For the circumgalactic medium (CGM) gas, the gas temperature is represented by the IGM temperature, $T_\mathrm{IGM}$, and an additional component, $\Delta T$, from local DM heating. This heating leads to a scale-dependent collapse mass, because local DM energy deposition is a function of mass scale $M_h$. For massive halos, the DM heating is expected to be stronger, causing the surrounding gas temperature to increase. Therefore the Jeans mass is a function of mass scale, $M_\mathrm{J}(z, M_h)$. This can be approximated by following approach. As we fixed the local deposition efficiency, $h=\log (\epsilon^\mathrm{DM}_\mathrm{local} / E_h)$, the energy deposition per baryon is proportional to the total dark matter annihilation energy of the halo, $E_h$. As a result, the gas temperature raised by $\Delta T \propto E_h$, and the Jeans mass, as a function of gas temperature, follows the relation $M_\mathrm{J} \propto ({T_\mathrm{IGM} + \Delta T})^{3/2}$. 

To calculate the DM heating and ionization in local gas, we apply the local DM annihilation rate $\epsilon^\mathrm{DM}_\mathrm{c, local}$ into Equation~\ref{Equ.evo_T}, \ref{Equ.evo_ion} and \ref{Equ.evo_lymana} to obtain the thermal evolution of local gas surrounding the halo.

\begin{figure*}
    \centering
    \includegraphics[width=0.90\textwidth]{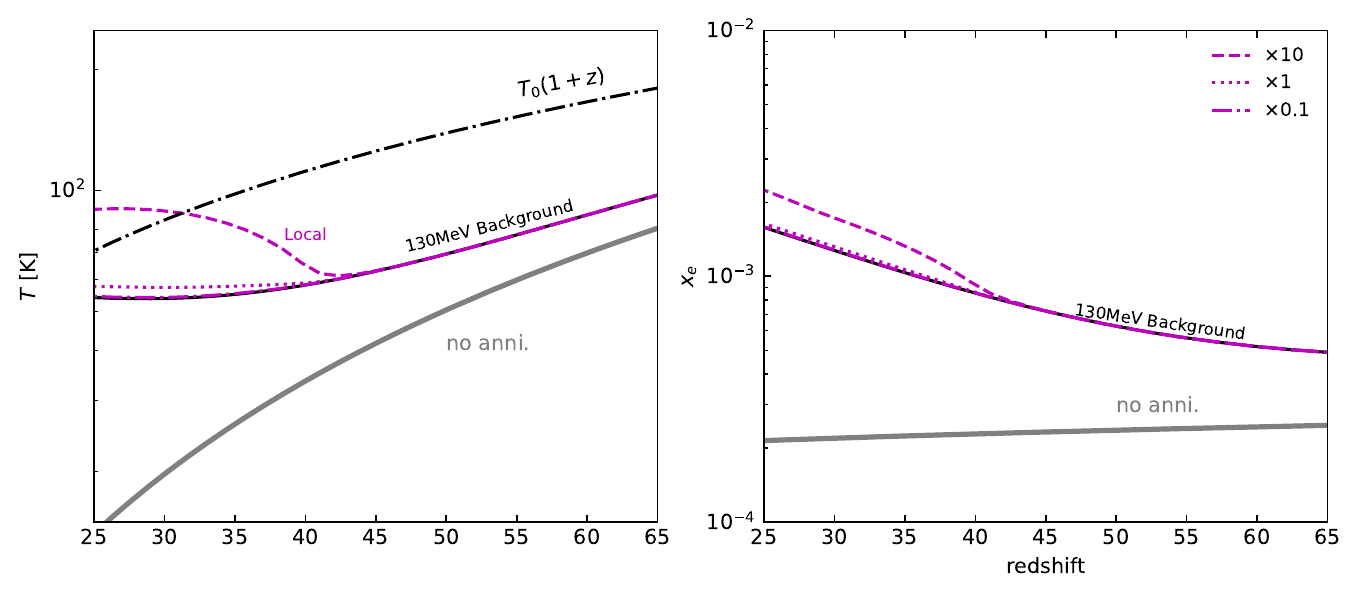}
    \caption{Kinetic temperature and ionization fraction of gas surrounding a $10^6 \mathrm{M}_\odot$ dark matter halo, calculated using \texttt{CosmoRec}. The halo formation was assumed to occur at redshift $z_\mathrm{form}=40$. The dotted-dashed line indicates the Cosmic Microwave Background (CMB) temperature. The grey solid line represents the baseline scenario without dark matter annihilation, and the solid black line represents the scenario with only 130 MeV background annihilation. The dotted and dashed lines corresponding to the gas properity with both 130 MeV DM background and local annihilation power with value of $0.1$, $1$ and $10$ times of base value.}
    \label{fig:DeltaT}
\end{figure*}

Fig.~\ref{fig:DeltaT} shows the gas temperature and ionized fraction surrounding a single $10^6 \mathrm{M}_\odot$ dark matter halo. We assume the halo formed at $z \approx 40$, with mass accretion following the power law $M = M_0 (1+z)^\alpha e^{\beta z}$\cite{Correa2015}, reaching $M = 10^6\ \mathrm{M}_\odot$ at redshift $z = 40$. Initially, the thermal history of the primordial gas is influenced solely by background dark matter annihilation (denoted by the black solid line). The impact of the local halo begins near the time of its formation. In the case of a low value of local deposit efficiency $h = -66$ and $-65$, the gas temperature and ionization fraction have almost no detectable change. However, once the energy deposition becomes significant ($h \gtrsim -64$), it can substantially increase the temperature and ionization fraction of the gas.

To summarize, dark matter annihilation from the halo can deposit energy into the surrounding gas before escaping the local environment. This energy deposition may enhance the total annihilation effects in small scale, such as increasing the gas temperature and ionization fraction. The impact of the local effect highly depends on the deposition efficiency. In our scenario, the local boost most significantly affects higher mass halos. While still preliminary, our calculation shows the importance of this avenue for probing the characteristics of dark matter annihilation, particularly in small halos, via its influence on gas and star formation. Further constraints require a more refined model for deposition efficiencies, for more robust galaxy formation simulations and comparison with observational data.

\end{document}